\documentclass[journal=jctcce,manuscript=article]{achemso}

\usepackage{gnuplottex}
\usepackage{mathtools}
\usepackage{float}
\usepackage{array}
\usepackage[english]{babel}
\usepackage[utf8]{inputenc}
\usepackage{amsmath}
\usepackage[table]{xcolor}
\usepackage{amsfonts}
\usepackage{graphicx}
\usepackage{dsfont}
\usepackage[colorlinks=true, linkcolor=blue, citecolor=dred]{hyperref}
\usepackage{algorithm}
\usepackage[noend]{algpseudocode}
\usepackage{bbold}
\usepackage{mathtools}
\usepackage{amssymb}
\usepackage{bm}
\usepackage{bbm}
\usepackage{tabularx, booktabs}
\usepackage{xspace}
\usepackage{listings}
\usepackage{mleftright} 
\newcolumntype{Y}{>{\centering\arraybackslash}X}

\DeclareMathOperator*{\argmax}{arg\,max}
\newcommand{\ket}[1]{\ensuremath{|#1\rangle}\xspace}
\newcommand{\bra}[1]{\ensuremath{\langle #1|}\xspace}
\newcommand{\psh}[2]{\ensuremath{\langle #1|#2\rangle}\xspace}

\newcommand{\ii}{{\rm i}}
\newcommand{\lin}{\mathrm{span}}
\newcommand{\nonorthsuperscript}{\wr}

\newcommand{\Nloc}{{N_\mathrm{o}}}
\newcommand{\lmi}[1]{\phi_{#1}}

\newcommand{\lki}[1]{\ket{\lmi{#1}}}

\newcommand{\UTwo}{\hat U_{\{\theta\tau\}}}
\renewcommand{\epsilon}{\varepsilon}
\newcommand{\efrac}[2]{\emb{\frac{#1}{#2}}}

\newcommand{\abs}[1]{\left\vert#1\right\vert}

\newcommand{\zc}{\tilde c} 
\newcommand{\zs}{\tilde s} 
\definecolor{cdcolor}{rgb}{.6,.6,.6}

\definecolor{mypurple}{rgb}{.8,.0,.4}

\definecolor{myotherhl}{rgb}{0.,.4,.8}

\newcommand{\fn}[1]{\mathop{{}#1}}  

\newcommand{\fnargb}[2]{\fn{#1}\bigl(#2\bigr)}

\newcommand{\emb}[1]{\mbox{$#1$}} 
\renewcommand{\d}{\mathrm{d}}

\newcommand{\mat}[1]{\mathbf{#1}}

\definecolor{dgreen}{rgb}{0,.5,0}
\definecolor{dblue}{rgb}{0,0,.5}
\definecolor{dred}{rgb}{0.5,0,.5}

\title{Generalization of 
intrinsic orbitals to 
Kramers-paired quaternion spinors, molecular fragments and valence virtual spinors}

\author{Bruno Senjean}
\email{bsenjean@gmail.com}
\affiliation{Instituut-Lorentz, Universiteit Leiden, P.O. Box 9506, 2300 RA Leiden, The Netherlands}
\affiliation{Theoretical Chemistry, Vrije Universiteit, De Boelelaan 1083, NL-1081 HV, Amsterdam, The Netherlands}
\author{Souloke Sen}
\affiliation{Theoretical Chemistry, Vrije Universiteit, De Boelelaan 1083, NL-1081 HV, Amsterdam, The Netherlands}
\author{Michal Repisky}
\affiliation{Hylleraas Centre for Quantum Molecular Sciences, Department of Chemistry, UiT The Arctic University of Norway, N-9037 Troms{\o}, Norway}
\author{Gerald Knizia}
\affiliation{Department of Chemistry, The Pennsylvania State University, University Park, Pennsylvania 16802, United States}
\author{Lucas Visscher}
\affiliation{Theoretical Chemistry, Vrije Universiteit, De Boelelaan 1083, NL-1081 HV, Amsterdam, The Netherlands}

\begin{document}


\begin{abstract}
Localization of molecular orbitals finds its importance
in the representation of chemical bonding (and anti-bonding) and in
the local correlation treatments beyond mean-field approximation.
In this paper, we generalize the intrinsic atomic and bonding orbitals 
[G. Knizia, J. Chem. Theory Comput. 2013, 9, 11, 4834-4843]
to relativistic applications using complex and quaternion spinors, as 
well as to molecular fragments instead of atomic fragments only.
By performing a singular value decomposition, we show how 
localized valence virtual orbitals can be expressed in this intrinsic 
minimal basis.
We demonstrate our method on systems of increasing complexity, 
starting from simple cases such as benzene, acrylic-acid and ferrocene molecules, and then demonstrating
the use of molecular fragments and inclusion of relativistic effects for
complexes containing heavy elements such as tellurium, iridium and astatine.
The aforementioned scheme is implemented into a 
standalone program interfaced with several different quantum chemistry packages.
\end{abstract}

\maketitle

\section{Introduction}

Although chemical bonding models are shrouded 
in mystery~\cite{frenking2007unicorns} 
and even the concept of orbitals is sometimes met with scepticism~\cite{ogilvie1990nature}, 
one cannot deny the usefulness of orbitals in a
qualitative understanding of chemical concepts.
While delocalized canonical molecular orbitals (MOs)
resulting from a standard mean-field calculation 
such as Hartree--Fock (HF) are useful to understand electronic excitations and spectroscopy, it is of interest to consider
localized molecular orbitals (LMOs) when relating first principles calculations to simple intuitive models of chemical bonding.
Several schemes have been developed to obtain such localized orbitals
using different concepts, such as 
Foster--Boys~\cite{foster1960canonical}, 
Edminston--Ruedenberg~\cite{edmiston1963localized},
von Niessen~\cite{von1972density} and
Pipek-Mezey (PM)~\cite{pipek1989fast} methods,
among many others~\cite{magnasco1967uniform,kleier1974localized,reed1985natural,cioslowski1991partitioning,boughton1993comparison,raffenetti1993efficient,aquilante2006fast,jansik2011local,hoyvik2012trust,hoyvik2012orbital,hoyvik2013local,hoyvik2013localized,li2014localization,li2017localization,berghold2000general,alcoba2006orbital,lehtola2014pipek,hesselmann2016local,jonsson2017theory}. The various schemes were recently reviewed by H{\o}yvik  and  J{\o}rgensen~\cite{hoyvik2016characterization}.
Often only the set of occupied MOs is localized
as this is sufficient to analyse the self-consistent field (SCF) 
wavefunction, i.e. the single Slater determinant used in Hartree-Fock and Kohn-Sham~\cite{Kohn:1965p27} Density Functional Theory (DFT). For a more complete understanding of interacting molecules in terms of frontier orbitals it is, however, of interest to localize the virtual molecular orbitals as well.  Virtual LMOs can be determined using for
instance the PM method~\cite{hoyvik2013pipek} but they are much more
difficult to localize with standard schemes due to their more diffuse nature.
Specific approaches for virtuals
have been proposed to solve this issue, 
like the protohard-virtual MOs~\cite{subotnik2005fast},
the least-change algorithm~\cite{ziolkowski2009maximum}
or
the use of external quasi-atomic orbitals~\cite{west2013comprehensive}.
The use of powers of the second central
moment and powers of the fourth central moment localizations might also be a better alternative~\cite{hoyvik2013localized} for basis sets augmented by diffuse functions.
LMOs are not only useful to have a better representation of chemical
bonding and anti-bonding. They also have a significant importance in
local correlation treatments in
post-HF methods like second order
M{\o}ller Plesset~\cite{saebo1987fourth,saebo1993local,schutz1999low,weijo2007general},
coupled cluster~\cite{hampel1996local,schutz2001low,tatiana2003local,christiansen2006coupled},
embedding approaches~\cite{gomes2012quantum,bennie2015accelerating,hegely2016exact,wouters2016practical,libisch2017embedding,chulhai2017improved,sayfutyarova2017automated,lee2019projection,wen2019absolutely,claudino2019automatic,claudino2019simple,hermes2019multiconfigurational,hermes2020variational}, and
multireference methods~\cite{maynau2002direct,angeli2003use,ben2011direct,chang2012multi}.

In this work, a generalization of the intrinsic atomic and bonding 
orbitals introduced by Knizia~\cite{knizia2013intrinsic} (so-called IAOs and IBOs, respectively)
to molecular fragments and relativistic spinors is discussed.
The idea of using an intrinsic minimal basis able to exactly span the
occupied space is not
new~\cite{ruedenberg1982atoms,lee2000extracting,lu2004molecule,laikov2011intrinsic,knizia2013intrinsic,west2013comprehensive}, 
and IAOs have been shown to be related to other intrinsic minimal basis sets like the
quasi-atomic orbitals~\cite{janowski2014near}.
We like to stress that other works have 
generalized the PM localization to 
complex-valued orbitals~\cite{lehtola2013unitary} 
and one-, two- or four-component Kramers-restricted and unrestricted 
spinors~\cite{dubillard2006bonding,ciupka2011localization}. 
Our method
differs mainly by performing our localization
in a minimal reference 
basis of intrinsic fragment orbitals (IFOs), where
the term ``fragment'' denotes the ability to use
either intrinsic molecular or atomic orbitals.
We also show how localized valence 
virtual orbitals
can be obtained and expressed in this intrinsic basis.
The procedure described in this paper
has been implemented in a standalone program
called Reduction Of Orbital Extent (ROSE)
which can easily be interfaced with several quantum chemistry packages.
As expected from the use of an intrinsic orbital basis,
our intrinsic LMOs (ILMOs)
are basis set insensitive, and the localization procedure 
is cheaper and better behaved than PM 
localization~\cite{knizia2013intrinsic}.

The paper is organized as follows. 
After a brief review
of the construction of IFOs
(Sec.~\ref{sec:IFO}) and the localization
procedure (Sec.~\ref{sec:IBO}), we discuss 
how valence virtual orbitals can be expressed in this new minimal
basis (Sec.~\ref{sec:virtuals}).
We end this theory section by the generalization to 
Kramers spinors (Sec.~\ref{sec:quaternion}).
Computational details about the code and its interfaces are 
provided in Sec.~\ref{sec:compdetails}.
The attractiveness of the IFOs and the resulting ILMOs
is presented in Sec.~\ref{sec:results},
where we show well known results like the basis set insensitivity of the
partial charges, the chemically sound representation of orbital 
bonding and antibonding, as well as applications to relativistic
cases.
Conclusions and perspectives are finally
given in Sec.~\ref{sec:conclusions}.

\section{Theory}

In this section, we describe how the minimal
set of IFOs
is constructed and how the occupied
and valence 
virtual orbitals can be localized.
The equations derived in this work follow
closely the original scheme of Knizia~\cite{knizia2013intrinsic},
with as main difference that we express reference orbitals in terms of a fragment MO basis instead of directly in a Gaussian Type Orbital (GTO) basis.
This allows treatment of 4-component relativistic orbitals, which can not be straightforwardly expressed in terms of a single GTO function, in a simple 
manner. Furthermore, the difference 
between atomic and molecular fragments disappears, so
that both atomic and molecular fragments can be considered in the same manner.

\subsection{Intrinsic fragment orbitals}\label{sec:IFO}

Consider a molecule composed of $N_{\rm A}$ atoms
and partitioned into $N_{\rm F}$ different molecular (or atomic) fragments
labelled by $k$ ($k=1,\ldots,N_{\rm F}$).
Each fragment $k$ contains $N_{\rm A}^{k}$ atoms such that 
$N_{\rm A}=\sum_{k=1}^{N_{\rm F}}N_{\rm A}^{k}$.

Let us start with a basis $\mathcal{B}_1$ of 
orthonormal MOs $\lbrace \ket{\phi_p} \rbrace$ for the
full molecule, obtained from a linear combination of basis functions
$\lbrace \ket{\chi_\mu} \rbrace$ which form a (generally non-orthogonal) basis B$_1$,
\begin{eqnarray}\label{eq:MOcoeffB1}
\ket{\phi_p} = \sum_{\mu=1}^{\dim({\rm B}_1)} C_{\mu p} \ket{\chi_\mu},
\end{eqnarray}
where dim(B$_1$) $\geq$ dim($\mathcal{B}_1$). 
In the rest of the paper, the indexes $p, q, \hdots$ denote any MO of $\mathcal{B}_1$.
As the MOs are orthonormal, the overlap matrix of $\mathcal{B}_1$, 
$\mathbf{S}_{11}$, is the identity matrix and the closure relation can be written as
\begin{eqnarray}
1  = \sum_{p=1}^{\dim(\mathcal{B}_1)} \ket{\phi_p}\bra{\phi_p}.
\end{eqnarray}
This relation holds for all functions lying in the space spanned by $\mathcal{B}_1$.  We then introduce a projector on
the subspace spanned by the occupied MOs (indexed by $i,j,\hdots$) as
\begin{equation}
O = \sum_{i=1}^{N_{\rm occ}} \ket{\phi_i}\bra{\phi_i}.
\end{equation}
The complementary projector $1-O$ spans the subspace of 
the virtual MOs in 
$\mathcal{B}_1$ (indexed by $a, b, \hdots$).

Then, consider $N_{\rm F}$ other bases $\mathcal{B}^k$
(one per fragment $k$)
of orthonormal MOs $\lbrace \ket{\varphi_t^k}\rbrace$, 
obtained from a linear combination of (generally non-orthogonal) basis functions
$\lbrace \ket{\zeta_\nu^k} \rbrace$ forming the basis B$^k$, 
\begin{eqnarray}\label{eq:MOcoeffBk}
\ket{\varphi_t^k} = \sum_{\nu=1}^{\dim({\rm B}^k)} C_{\nu t}^k \ket{\zeta_\nu^k},
\end{eqnarray}
where dim(B$^k$) $\geq$ dim($\mathcal{B}^k$).
Note that the nature of the underlying 
basis functions $\lbrace \ket{\zeta_\nu^k} \rbrace$
can be different for each fragment. One could for instance choose a large GTO set for one fragment and a Slater type orbital (STO) set for another.
As the fragments represent
only a small part of the full molecule, obtaining the fragment orbitals $\mathcal{B}^k$ from an SCF calculation
 will typically be much cheaper
than the calculation that yielded $\mathcal{B}_1$.

We select from these sets
$\mathcal{B}^k$ a minimal set of MOs for the next step in which the sets of fragment MOs
are combined to form basis  $\mathcal{B}_2$ of
reference fragment orbitals 
(RFOs, indexed by $t, u, \hdots$). In analogy with the original scheme~\cite{knizia2013intrinsic}, the
RFOs constitute a minimal set of depolarized fragment orbitals which are only polarized inside the fragments 
but not in between different fragments.
In most cases, the size of the minimal basis of each fragment 
will simply be the sum of the minimal bases of each atom (i.e. its core and valence orbitals) composing the given fragment.
In that case, the total dimension of $\mathcal{B}_2$ is 
\begin{equation}
\dim(\mathcal{B}_2) = \sum_{k=1}^{N_{\rm F}} \sum_{a=1}^{N_{\rm A}^{k}} n_{a}^{\rm min}
\end{equation}
where $n_{a}^{\rm min}$ is the number 
of occupied and valence virtual orbitals of atom $a$.
One may, however, also use more advanced schemes in which the fragment orbitals themselves are the results of a previous localization procedure. This then allows selection of a smaller set of frontier orbitals that are localized in a region of interest. 
In the same spirit, one can also truncate the set of (occupied and virtual) canonical orbitals of the fragments, for instance by setting a cut off based on energy criteria as done in DIRAC~\cite{DIRAC19,DIRAC_article}. With these definitions we may proceed to construct IFOs which
form a minimal basis of polarized fragment orbitals 
that exactly span the occupied space of the full molecule in basis $\mathcal{B}_1$.

Let us now describe the steps to construct the IFOs, 
using matrix notation. The first step is the calculation of
the overlap matrices between $\mathcal{B}_1$ and $\mathcal{B}_2$.
The overlap matrix $\mathbf{S}_{11}$ in $\mathcal{B}_1$
is identity by construction,
while $\mathbf{S}_{22}$ in $\mathcal{B}_2$ is in 
general not diagonal due to non-orthogonality between
MOs belonging to different fragments.
For a partitioning of the molecule in two fragments $k$ and $k'$,
$\mathbf{S}_{22}$ is given by,
\begin{equation}
\mathbf{S}_{22} = 
\begin{pmatrix}
\mathbf{1} &  \mathbf{S}^{kk'}  \\
\mathbf{S}^{k'k} & \mathbf{1}
\end{pmatrix},
\end{equation}
which can of course be generalized to as many fragments as desired.
The diagonal blocks associated with each fragment 
are identity matrices
with dimension $(N_{\rm occ}^k + N_{\rm vir}^k)$, 
where $N_{\rm occ}^k$ and $N_{\rm vir}^k$
are the number of occupied and valence virtual MOs of fragment $k$,
respectively.
The off-diagonal blocks $\mathbf{S}^{kk'}$ 
have a rectangular form of dimension 
$(N_{\rm occ}^k + N_{\rm vir}^k) 
\times (N_{\rm occ}^{k'} + N_{\rm vir}^{k'})$, and are in general
filled by non-zero elements.
The overlap matrix between $\mathcal{B}_1$ and $\mathcal{B}_2$ is
denoted by $\mathbf{S}_{12} = \mathbf{S}_{21}^\dagger$.

A new set of orthonormal depolarized occupied MOs
$\{\ket{\tilde{\phi}_i}\}_{i=1}^{N_{\rm occ}}$
is constructed by projecting the original occupied MOs
of $\mathcal{B}_1$ into $\mathcal{B}_2$ and back 
into $\mathcal{B}_1$ as follows,
\begin{eqnarray}\label{eq:depolarized_MO_nonorth}
\mathbf{\tilde{c}} = \mathbf{P}_{1 \leftarrow 2} 
\mathbf{P}_{2 \leftarrow 1} \mathbf{C},
\end{eqnarray}
where
$\mathbf{C}$ is the coefficient matrix of the occupied MOs in 
$\mathcal{B}_1$. 
In the above [Eq.~(\ref{eq:depolarized_MO_nonorth})] we have introduced the convention of indicating coefficient matrices corresponding
to orthonormal sets of MOs by an uppercase symbol, while otherwise
using lowercase.
As $\mathcal{B}_1$ was defined as the MO basis itself, this
coefficient matrix is 0 everywhere except for the $N_{\rm occ}$ 
diagonal elements which are equal to 1.
$\mathbf{P}_{2\leftarrow 1}$  and
$\mathbf{P}_{1\leftarrow 2}$ are projector matrices
between the original MOs of $\mathcal{B}_1$ and
the RFOs of $\mathcal{B}_2$, and are 
implicitly defined by the following linear equations,
\begin{eqnarray}
\mathbf{S}_{22} \mathbf{P}_{2\leftarrow 1} = \mathbf{S}_{21},\label{eq:P21} \\
\mathbf{S}_{11} \mathbf{P}_{1\leftarrow 2} = \mathbf{S}_{12}.\label{eq:P12}
\end{eqnarray}
Note that because $\mathbf{S}_{11}$ is the identity matrix, 
$\mathbf{P}_{1\leftarrow 2}$
is simply equal to $\mathbf{S}_{12}$.
In Eq.~(\ref{eq:depolarized_MO_nonorth}), the depolarized
occupied MOs are non-orthonormal because space $\mathcal{B}_2$
does not fully span $\mathcal{B}_1$.
Before constructing the minimal polarized IFO basis, 
we symmetrically orthonormalize these depolarized MOs as follows,
\begin{eqnarray}\label{eq:symorth}
\mathbf{\tilde{C}} = \mathbf{\tilde{c}} 
\left[ \mathbf{\tilde{c}}^\dagger \mathbf{\tilde{c}} \right]^{-1/2}.
\end{eqnarray}
The coefficient matrix encoding the expression 
of the IFOs in basis $\mathcal{B}_1$ then reads~\cite{knizia2013intrinsic}
\begin{eqnarray}
\label{eq:IFO}
\mathbf{c}^{\rm IFO}=\Big[ \mathbf{C}\mathbf{C}^\dagger
\tilde{\mathbf{C}}
\tilde{\mathbf{C}}^\dagger +\left(\mathbf{1}-\mathbf{C} \mathbf{C}^\dagger \right) 
\left(\mathbf{1}-\tilde{\mathbf{C}} \tilde{\mathbf{C}}^\dagger \right)\Big] 
\mathbf{P}_{1 \leftarrow 2}, \nonumber \\
\end{eqnarray}
where the first term in the right hand side
of Eq.~(\ref{eq:IFO}) acts on the occupied MO subspace
and the second term on the virtual subspace. The result is
a number of IFOs corresponding to $\dim(\mathcal{B}_2)$.
The final step is to also symmetrically orthonormalize these IFOs
[see Eq.~(\ref{eq:symorth})] thus leading to $\mathbf{C}^{\rm IFO}$.
As a final note on the construction of the IFO basis,
another formulation is possible by replacing
Eq.~(\ref{eq:depolarized_MO_nonorth}) by
\begin{eqnarray}\label{eq:depolarized_MO_nonorth_v2}
\mathbf{\tilde{c}} =
\mathbf{P}_{2 \leftarrow 1} \mathbf{C},
\end{eqnarray}
which is then symmetrically orthogonalized in $\mathcal{B}_2$ as
\begin{eqnarray}\label{eq:symorth_v2}
\mathbf{\tilde{C}} = \mathbf{\tilde{c}} 
\left[ \mathbf{\tilde{c}}^\dagger \mathbf{S}_{22} \mathbf{\tilde{c}} \right]^{-1/2}.
\end{eqnarray}
The two definitions are identical if the space $\mathcal{B}_2$ is completely spanned by $\mathcal{B}_1$, but
will lead to slightly different results if this is not the case. The second definition, may then lead
to somewhat simpler final equations as is discussed in detail in the supplementary material. In our
implementation either definition can be used, with the first one, originally proposed in Ref.~\citenum{knizia2013intrinsic},
serving as the default option.

\subsection{Localized orbitals}\label{sec:IBO}

The IFOs are interesting in their own right, but are in this work primarily used
to define localized MOs that provide an alternative but fully equivalent representation 
of a single determinant wave function. The equivalence is guaranteed by defining the localization
as a sequence of unitary rotations among occupied orbitals that leave the density matrix and therefore
all observables unchanged. The localization procedure considered 
in this paper is inspired by the PM scheme~\cite{pipek1989fast}
where only orbital overlaps need to be computed. 
It consists of 2 by 2 rotations of the occupied MOs $\lbrace \ket{i} \rbrace$ until
the following function is maximized,
\begin{eqnarray}
L = \sum_{k=1}^{N_{\rm F}} \sum_{i'} [n^k_{i'}]^4,
\label{eq:L}
\end{eqnarray}
where $\ket{i'} = \sum_i U_{i'i}\ket{i}$ are the 
rotated occupied orbitals
and $n^k_{i'} = \sum_{t \in k} \psh{i'}{t}\psh{t}{i'}$
is the
number of $\ket{i'}$’s electrons located on all the IFOs $t$ of (atomic or 
molecular) fragment $k$. 
However, in contrast to PM~\cite{pipek1989fast}
the exponent is equal to 4 which leads to 
effectively identical results as the square exponent used in PM but avoids discrete localizations in 
aromatic systems~\cite{knizia2013intrinsic}. Increasing the exponent
in other localization procedures 
has also been shown to 
penalize delocalized orbitals~\cite{jansik2011local,hoyvik2012trust}.
Also, the numerical advantage of Eq.~\eqref{eq:L} compared to PM is that the 2 by 2 
rotations are performed in the minimal and orthogonal IFO
basis. As a result, the ILMOs are insensitive to the choice of AO basis set in 
contrast to standard PM orbitals.
The localization procedure
requires only few iterations, and is applied to the occupied MOs as
well as to the valence virtual MOs described in the next section.

\subsection{Valence virtual orbitals}\label{sec:virtuals}

The IFOs are designed to exactly span the occupied space and only provide 
a minimal description of the virtual space. 
While this space is too small to capture dynamic electron correlation,
these virtual orbitals are of interest for the purpose of analysis
(e.g. in frontier orbital theory), to connect to semi-empirical or
tight-binding approaches, and to provide starting orbitals for complete
active space SCF procedures.
We therefore consider localizing this limited set of valence virtual
orbitals (of dimension $N_{\rm vir}^{\rm val} = \dim(\mathcal{B}_2) -
N_{\rm occ}$) in addition to the occupied ones.
While valence virtual IBOs were previously defined in
Ref.~\citenum{steen2019SigmaNoninnocenceNiIV}, neither explicit
formulas nor the relativistic or fragment-orbital generalizations
discussed in this text were provided.

In order to construct the valence-virtual IFOs, we first project all
$N_{\rm vir} = \dim(\mathcal{B}_1) - N_{\rm occ}$ virtual orbitals
to the IFOs. The resulting unnormalized orbital 
coefficient matrix $\mathbf{C}^{\rm vir}$ is then subjected to a 
singular value decomposition (SVD)
\begin{eqnarray}\label{eq:SVD}
\mathbf{C}^{\rm vir} = \mathbf{U} \mathbf{\Sigma} \mathbf{V}^\dagger.
\end{eqnarray}
In this expression $\mathbf{C}^{\rm vir}$ has dimension 
$(\dim(\mathcal{B}_2) \times N_{\rm vir})$,
$\mathbf{U}$ has dimension $(\dim(\mathcal{B}_2) \times \dim(\mathcal{B}_2))$,
$\mathbf{\Sigma}$ has dimension $(\dim(\mathcal{B}_2) 
\times N_{\rm vir})$
and 
$\mathbf{V}^\dagger$ has dimension $(N_{\rm vir}
\times N_{\rm vir})$.
Because the 
IFOs already span the space
of the occupied orbitals, only $N_{\rm vir}^{\rm val}$ 
singular values (diagonal elements of $\mathbf{\Sigma}$)
will be nonzero. The eigenvectors  $\mathbf{U}$ 
corresponding to these nonzero singular values 
form a new set of valence virtuals which, together with the occupied,
form the minimal set of orthonormal 
MOs that can be expressed exactly in terms of IFOs.
Those valence virtual MOs can then be localized
with the same procedure as
for the occupied, described in the previous section.

Note that many correlating, or ``hard'', virtuals which cannot be expressed 
in terms of IFOs are lost in this process. These hard virtuals are anyhow known
to be more difficult to localize compared to the valence
virtual orbitals\cite{subotnik2005fast}.
The inclusion of additional hard virtuals
can be done by increasing the size of $\dim(\mathcal{B}_2)$, i.e. by
defining a lighter truncation in the number of RFOs, 
or by adding additional 
correlation functions to the IFO basis on which hard 
virtuals can be expressed on, for instance using quasi-atomic external 
orbitals~\cite{west2013comprehensive} or 
protohard-virtuals~\cite{subotnik2005fast}.
As shown in the following, the standard valence virtual MOs
nicely depict $\sigma$- and $\pi$-antibonding orbitals 
and are analogous to the split-localized 
orbitals~\cite{bytautas2003split,west2013comprehensive}.
Together with the occupied localized orbitals,
the localized valence virtuals
provide a very good approximation to the
weakly occupied correlating multiconfigurational SCF orbitals 
in the full valence
space~\cite{west2013comprehensive}, and are thus a very good effective
configurational basis for the treatment of valence-internal correlation~\cite{bytautas2003split}.
Inclusion of 
localized hard virtuals for correlated post-SCF calculations 
is out of scope for the present paper but an implementation of such
a scheme is planned for the near future.

\subsection{Approximate energy ordering}\label{sec:approxenergy}
The ILMOs are in general not eigenfunctions
of the original Fock operator. This invalidates concepts such as the
highest occupied molecular orbital (HOMO) or lowest unoccupied molecular
orbital (LUMO) that are based on energy ordering.
Depending on the degree of localization one may devise an alternative ordering
of orbitals by grouping together orbitals that are primarily localized on one fragment, 
but also then it is useful to have a secondary, approximately energy-based, ordering
criterion. For this we compute the diagonal matrix elements of the Fock operator expressed
in the ILMO basis. This quantity is meaningful as long as primarily MOs
with similar energies are mixed upon the generation of the ILMOs. 
In our experience this
is often the case and this additional label is then useful for instance in quickly
separating the energetically lower lying sigma orbitals from the pi-orbitals in aromatic
systems.

\subsection{Treatment of complex and quaternion orbitals}\label{sec:quaternion}

Up to this point we did not specify the algebra of the basis functions and MO-coefficients. In this section we will discuss the generalizations needed to
work with orbitals resulting from relativistic calculations in which the coefficient
and overlap matrices are in general complex. 

We will start by the easier
discussion of restricted versus unrestricted SCF calculations in non-relativistic theory for which the spin and spatial part of the orbitals can be considered separately. The simplest case (i) is found when both the calculation giving rise to
$\mathcal{B}_1$ as well as the calculations for $\mathcal{B}_2$ are carried out
with spin-restricted SCF. In this case the coefficient matrices for $\alpha$- and $\beta$-spinorbitals are identical. It then suffices to work with matrices that
follow the dimension of the number of spatial orbitals, a reduction of a factor of 2 as compared to the use of spinorbitals. These matrices are also typically restricted to be real as there is little advantage in defining complex orbitals.
The second case (ii) occurs if an unrestricted SCF calculation is used for $\mathcal{B}_1$. While it is still possible to use restricted SCF calculations for the fragments, it is then necessary to apply two separate localization procedures for $\alpha$- and $\beta$-spinorbitals. As overlap between these sets of orbitals is strictly zero, both the original and localized sets consist of spinorbitals written as a product of a spin and a spatial part. Also in this case, the algebra can be kept real.

For relativistic calculations, that include the effect of spin-orbit coupling (SOC) in the generation of the MOs, the situation is different. The spin and spatial parts
of the wave function can not be factored in terms of a simple product, and overlap and coefficient matrices are in general complex. The simplest treatment is to consider
this as a generalization (iii) of the unrestricted case in which the procedures
described above are carried out for complex hermitian matrices. This is what we anticipated in our notation in the previous sections where we indicated hermitian conjugation for matrices rather than a simple transpose as would be applicable for real algebra. No special implementations are needed for this relativistic unrestricted case, as complex versions of matrix diagonalization, singular value decompositions and linear equation solvers are readily available in standard linear algebra libraries like  LAPACK~\cite{LAPACK} (\texttt{zheevd}, \texttt{zgesvd}, and \texttt{zposv}). While this
approach is useful for truly unrestricted calculations, the procedure leads to
the undesirable loss of Kramers symmetry for orbitals that were generated with a Kramers-restricted SCF algorithm~\cite{Saue:1999ur}. In this generalized form of restricted SCF, orbitals can be made to adhere to a strict pairing in which the coefficients of one orbital can be obtained by operating with the anti-unitary Kramers operator on the coefficients of its ``Kramers partner''. Such a pairing is automatically guaranteed
in quaternion algebra, in which the Fock matrix and MO coefficients are
block diagonalized by a quaternion matrix transformation. Like in the nonrelativistic
restricted case (i) it is possible to work with matrices that have the dimension
of the number of spatial orbitals, albeit now with matrix elements that are quaternion. Compared to the non-relativistic (or scalar relativistic) case (i) we have 4 times more unique real numbers in the overlap or coefficient matrices. Compared to the relativistic unrestricted case (iii) there are, however, 2 times less unique real numbers (as this corresponds to complex matrices of twice the dimension). 
Carrying out the projections and localizations in quaternion algebra is furthermore advantageous as this guarantees keeping proper pairing of orbitals, as needs to be safeguarded explicitly in complex algebra~\cite{ciupka2011localization}. We will below discuss how the above-mentioned linear algebra techniques were adapted for use with quaternion
orbitals. For this purpose we define the quaternion MO-coefficients as:

\begin{eqnarray}\label{eq:quaternion_coeff_matrix}
\mathbf{C} = \,^0\mathbf{C} + \breve{i}\,\, ^{1}\mathbf{C} +
\breve{j}\,\, ^2\mathbf{C} + \breve{k}\,\, ^3\mathbf{C}, ~~~ \mathbf{C}^\dagger = \left( \mathbf{C}^* \right) ^T,
\end{eqnarray}
where $^x\mathbf{C}$ ($x=0,\hdots,3$)
are real coefficient matrices,
and $\breve{i}^2 = \breve{j}^2 = \breve{k}^2 = \breve{i}\breve{j}\breve{k} = - 1$ are a basis for quaternion algebra.

We note that, since we work in the MO basis, the overlap matrix $\mathbf{S}_{11}$ is
the identity matrix and therefore real. 

\subsubsection{Gauss--Seidel method to solve linear equation}

One of the first steps in the IFO's construction is the determination
of the projection matrices. To this end the linear equation in Eq.~(\ref{eq:P21}) has to be solved, for which we chose the Gauss--Seidel algorithm.
As $\mathbf{S}_{22}$ is identity in the diagonal blocks
and should have relatively small values in the off-diagonal blocks,
it is safe to say that $\mathbf{S}_{22}$ is close to diagonally dominant,
which is a sufficient but not necessary condition for convergence of the Gauss--Seidel
algorithm. In addition the $\mathbf{S}_{22}$ matrix is (except for ill-defined exactly 
overlapping fragments) symmetric and 
positive definite which is another sufficient condition for convergence.
This algorithm is straightforward to implement with quaternion
matrices as described in the appendix (Algorithm~\ref{algo:solvelineq}).

\subsubsection{Jacobi rotations for diagonalization and singular value decomposition}

Symmetrical orthogonalization [Eq.~(\ref{eq:symorth})] 
and singular value decompositions can be carried out with the aid of
a diagonalization. For the latter purpose it us convenient to employ
the Jacobi eigenvalue algorithm as this can be adapted to the use of quaternion
matrices. Our implementation (Algorithm~\ref{algo:diag} in the appendix) follows closely the one in Ref.~\citenum{le2007jacobi},
except that our Jacobi rotation matrices are not quaternion matrices
but real matrices. This is possible by
rotating the matrix into the real plane by scaling the 
$j$-th basis vector by the phase $P_{ij}$ 
of the pivot $M_{ij}$ 
(where the pivot denotes the largest off-diagonal matrix element),
\begin{eqnarray}
P_{ij} = M_{ij} / ||M_{ij}||
\end{eqnarray}
such that (for all $n=1,\hdots,\dim(\mathbf{M})$),
\begin{eqnarray}
M_{nj} &\longleftarrow & M_{nj}/P_{ij}, \nonumber \\
M_{jn} &\longleftarrow & P_{ij}\times M_{jn}.
\end{eqnarray}
Then, Jacobi rotations in the real plane can be applied straightforwardly
as
\begin{eqnarray}
\begin{bmatrix}
M_{ik} \\
M_{jk}
\end{bmatrix}
\longleftarrow
\begin{bmatrix}
\cos (\theta) & -\sin(\theta) \\
\sin(\theta) & \cos(\theta)
\end{bmatrix}
\begin{bmatrix}
M_{ik} \\
M_{jk}
\end{bmatrix}
\end{eqnarray}
after scaling (dividing) the eigenvectors by the phase factor.
The rotation matrices are defined by $\cos(\theta)$ and $\sin(\theta)$ and
are determined as follows,
\begin{eqnarray}
w &=& \dfrac{M_{jj} - M_{ii}}{2||M_{ij}||}\\
\tan(\theta) &=& \left\lbrace \begin{array}{ll}
-w + \sqrt{w^2 + 1}, {\text{ if $w \leq 0$}}\nonumber \\
-w - \sqrt{w^2 + 1}, {\text{ otherwise}}\nonumber \\
\end{array}\right.\nonumber\\
\cos(\theta) &=& \dfrac{1}{\sqrt{1 + \tan(\theta)^2}},\nonumber \\
\sin(\theta) &=& \tan(\theta)\cos(\theta).
\end{eqnarray}

Turning to the SVD algorithm (Algorithm~\ref{algo:SVD} in the appendix) used in Eq.~(\ref{eq:SVD}),
one can again apply the above Jacobi eigenvalue algorithm
on the quaternion Hermitian matrix 
$\mathbf{H} = \mathbf{C}^{\rm vir}\left(\mathbf{C}^{\rm vir}\right)^\dagger$ 
if $\dim(\mathcal{B}_2) \leq N_{\rm vir}$,
or on $\mathbf{H} = \left(\mathbf{C}^{\rm vir}\right)^\dagger \mathbf{C}^{\rm vir}$
otherwise.
One can then relate the eigenvalues and 
eigenvectors of $\mathbf{H}$ to the singular values (square roots of the
eigenvalues) and the
unitary matrices $\mathbf{U}$ and $\mathbf{V}$, as also discussed in
Ref.~\citenum{le2007jacobi}.

\subsubsection{Jacobi rotations in the localization procedure}\label{sec:jac_loc}

In order to maximise $L$ in Eq.~(\ref{eq:L}), one has to rotate
each pair of occupied orbitals $\ket{i}$ and $\ket{j}$ ($j < i$).
In real algebra~\cite{knizia2013intrinsic} the angle $\theta$ of the rotation
\begin{eqnarray}\label{eq:rotation}
\ket{i'} &=& \cos(\theta)\ket{i} + \sin(\theta)\ket{j},\nonumber\\
\ket{j'} &=& -\sin(\theta)\ket{i} + \cos(\theta)\ket{j},
\end{eqnarray}
is defined as
\begin{equation}\label{eq:rotation_angle}
\theta = (1/4)\,\text{atan2}\,(B_{ij},-A_{ij})
\end{equation}
where atan2($x$,$y$) = arctan($x/y$) with the resulting value being
restricted to lie in the interval $[-\pi,\pi]$.
In this expression $B_{ij}$ is the actual gradient and $A_{ij}$
is (an approximation to) the second derivative at $\theta=0$.
They both are functions of
the charge matrix elements of fragment $k$,
\begin{eqnarray}\label{eq:charge_matrix}
Q^k_{ij} = \sum_{t\in k} C_{ti}^{\rm occ*} C_{tj}^{\rm occ},
\end{eqnarray}
as follows (in real algebra):
\begin{eqnarray}
A_{ij} &=& \sum_F\biggl[3 \left(\bigl(Q_{ii}^F\bigr)^2+\bigl(Q_{jj}^F\bigr)^2\right) \bigl(Q_{ij}^F\bigr)^2-\frac{1}{2} \bigl(Q_{ii}^F-Q_{jj}^F\bigr)\left(\bigl(Q_{ii}^F\bigr)^3 - \bigl(Q_{jj}^F\bigr)^3\right) \biggr], \label{eq:A} \\
B_{ij} &=& \sum_F\Bigl[2 \left(\bigl(Q_{ii}^F\bigr){}^3-\bigl(Q_{jj}^F\bigr){}^3\right) Q_{ij}^F \Bigr] \label{eq:B}
\end{eqnarray}
When localizing valence virtual orbitals, $\mathbf{C}^{\rm occ}$ needs 
to be replaced by the eigenvectors $\mathbf{U}$ with non-zero 
singular values (see Sec.~\ref{sec:virtuals}).

Let us now switch to cases of complex or quaternion algebra.
To compute the rotation angle $\theta$ in Eq.~(\ref{eq:rotation_angle}), both $A_{ij}$ and $B_{ij}$ needs to be real.
As the matrix $\mathbf{Q}^k$ is Hermitian, $Q^k_{ii}$ elements are always real, but the $Q^k_{ij}$ elements are not.
For complex and quaternion algebra, the rotation matrix  
should actually be replaced by (see Appendix~\ref{app:OrbitalLocalization} for more details)
\begin{align}\label{eq:AppRotationComplex_maintext}
\ket{i'} &=  e^{-\ii \tau}\cos(\theta)\ket{i} + e^{\ii \tau}\sin(\theta)\ket{j},\nonumber\\
   \ket{j'} &= -e^{-\ii \tau}\sin(\theta)\ket{i} + e^{\ii \tau}\cos(\theta)\ket{j},
\end{align}
where $\tau$ is an additional scalar rotation parameter. This choice leads to
\begin{eqnarray}
A_{ij}(\tau) &=& \sum_F\biggl[3 \left(\bigl(Q_{ii}^F\bigr)^2+\bigl(Q_{jj}^F\bigr)^2\right) \fnargb{\Re}{e^{2\ii \tau}Q_{ij}^F}^2-\frac{1}{2} \bigl(Q_{ii}^F-Q_{jj}^F\bigr)\left(\bigl(Q_{ii}^F\bigr)^3 - \bigl(Q_{jj}^F\bigr)^3\right) \biggr], \label{eq:A} \\
B_{ij}(\tau) &=& \sum_F\Bigl[2 \left(\bigl(Q_{ii}^F\bigr){}^3-\bigl(Q_{jj}^F\bigr){}^3\right) \fnargb{\Re}{e^{2\ii \tau}Q_{ij}^F} \Bigr] \label{eq:B}
\end{eqnarray}
We deal with this additional degree of freedom by choosing it to maximize the real part of $B_{ij}(\tau)$, i.e.
$e^{2\ii \tau}=\frac{\tilde{B}_{ij}^*}{|\tilde{B}_{ij}|}$ with
\begin{eqnarray}
\tilde{B}_{ij} &=& \sum_F\Bigl[2 \left(\bigl(Q_{ii}^F\bigr){}^3-\bigl(Q_{jj}^F\bigr){}^3\right) Q_{ij}^F \Bigr], \label{eq:B_complex}
\end{eqnarray}
where $Q_{ij}^F$ (and thus $\tilde{B}_{ij}$) follows the algebra of the MO coefficients.
These phase-adjusted orbitals are used in the computation of the second derivative $A_{ij}$ and the rotation angle $\theta$.
After the rotation, the phase of the orbitals is restored.
More details on this procedure and on the derivations of $A_{ij}$ and $B_{ij}$ are provided in Appendix~\ref{app:OrbitalLocalization}, and a pseudo-code can be found in Algorithm~\ref{algo:loc} in Appendix~\ref{app:localization}.
The exact same procedure is applied to the valence virtual MOs expressed
in terms of IFOs in Eq.~(\ref{eq:SVD}). In both cases the result is a set
of orbitals that correspond to half the space. The remaining orbitals
can be generated via the Kramers' operator if needed.

\section{Computational details}\label{sec:compdetails}

The IFOs and ILMOs are generated in a standalone program called
Reduction of Orbital Space Extent (ROSE)~\cite{rose}.
ROSE has been interfaced with DIRAC~\cite{DIRAC19,DIRAC_article} using
quaternion spinors, Psi4~\cite{psi4} and PySCF~\cite{PYSCF} using real spinors,
as well as the ADF program~\cite{ADF2001} which is
based on Slater type orbitals (STO).
As complex spinors can also be used in ROSE,
an interface with the relativistic DFT code Respect~\cite{respect}
is currently under development.
Note that formatted checkpoint files are also read and generated in ROSE,
thus allowing for an interface with GAUSSIAN~\cite{g16}.
For Gaussian Type Orbitals (GTO) codes, only
the information on the basis set and on the MO coefficients
are required (for both the full molecule in basis $\mathcal{B}_1$ and
fragments in basis $\mathcal{B}^k$). The overlap matrices are then computed within ROSE by a standalone routine.
For the STO basis of ADF, they are extracted from the ADF data files with the
Python Library for Automating Molecular Simulation 
(PLAMS)~\cite{PLAMS} of the Amsterdam Modeling Suite AMS~\cite{AMS19}.
Our current version only works with 
uncontracted basis sets and only the non-relativistic, the exact two-component
X2C with or without (i.e., scalar-X2C) spin-orbit coupling, and
the molecular mean-field X2C (X2Cmmf) Hamiltonians
can be used.
Both restrictions are only of technical nature and we plan to lift these in the next release of the ROSE software.
The use of ROSE has been detailed in a manual 
accessible online~\cite{rose} together with several
examples including the interfaces with DIRAC, Psi4, PySCF and ADF.

\section{Results and discussions}\label{sec:results}

In this section, we investigate the IFOs and ILMOs obtained by ROSE
for different systems of increasing complexity. 
First and in order to compare with
existing results in the literature, the benzene and the acrylic acid
molecules with atomic fragments are considered. 
Then, our generalizations to quaternion spinors and to
molecular fragments are tested on the ferrocene molecule,
tellurazol oxide complexes (monomer and dimer), a linear chain of tellurium-substituted poly-ethylene
glycol oligomers (Te-PEG-4), an iridium complex, and finally a system composed of an astatine anion surrounded by ten water molecules.
To visualize the complex-valued spinors, 
we will plot the orbital densities. 
When the inclusion of spin-orbit coupling
does not affect the shape of the orbitals, we plot
the scalar-X2C (real-valued) orbitals to also display information on the phase.

\subsection{Benzene and acrylic acid}

\begin{figure}
\resizebox{\columnwidth}{!}{
\includegraphics[scale=1]{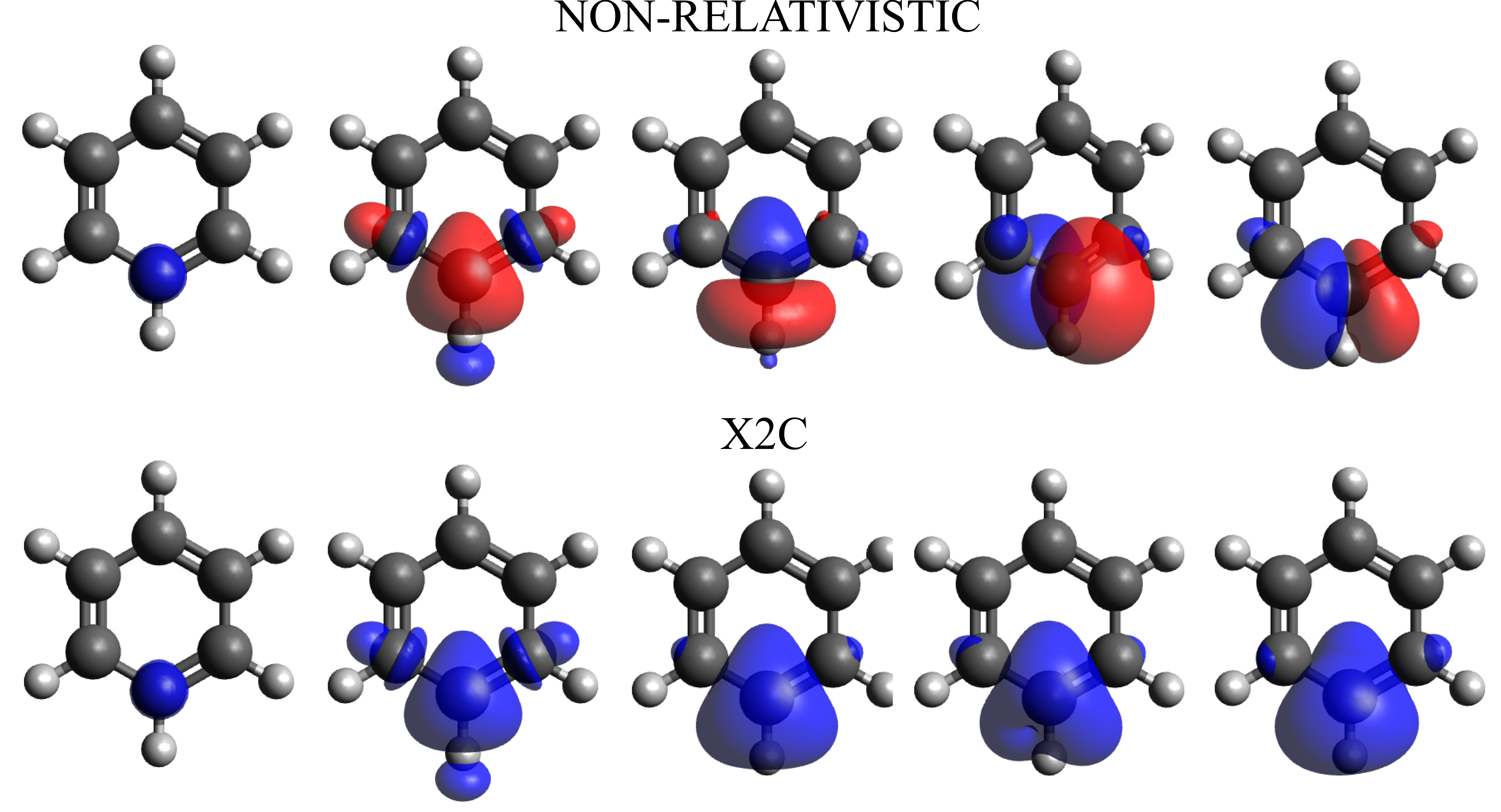}
}
\caption{
First row: Non-relativistic IFOs of a carbon atom in the benzene molecule. 
Second row: density of the IFOs from a X2C calculation.
}
\label{fig:benzene_IAO}
\end{figure}

As a first test case, we consider the benzene molecule using
atomic fragments, for which IBOs have been reported in Ref.~\citenum{knizia2013intrinsic}.
The cc-pVTZ and cc-pV5Z basis are used to construct the MO basis $\mathcal{B}_1$ and the RFO bases $\mathcal{B}^k$, respectively.
To see the difference between a non-relativistic and a relativistic
calculation with spin-orbit coupling, 
we plot the IFOs (which reduces to
IAOs in this case) of one carbon atom. 
There are 5 of them, one for the 1s and the 2s and 
three for the p$_x$, p$_y$ and p$_z$ atomic orbitals.
As readily seen in the first row of Fig.~\ref{fig:benzene_IAO}, the
IFOs are polarized but
remain atom-centered as expected. 
One can clearly distinguish between the p$_x$, p$_y$ and p$_z$ orbitals
in the non-relativistic calculation, while in the relativistic
(second row in Fig.~\ref{fig:benzene_IAO}) IFOs of the carbon atom,
one can recognize a spherical $p_{1/2}$ orbital as well as 
two $p_{3/2}$ orbitals. All these orbitals are much more spherical than the non-relativistic ones\footnote{
Note that also in the non-relativistic case, use of complex or quaternion coefficients can yield
spherical orbitals as this allows for recombination of the real orbitals to form complex spherical
harmonics. This can be observed in DIRAC by setting the speed of light to a very large value within the 
relativistic X2C framework: the orbital densities are then spherical but this calculation is otherwise non-relativistic.}.
We then localize separately the
sets of (canonical) occupied and 
valence virtual MOs [constructed with
the SVD in Eq.~(\ref{eq:SVD})]
by maximising Eq.~(\ref{eq:L}) in the IFO basis.

\begin{figure}
\resizebox{\columnwidth}{!}{
\includegraphics[scale=1]{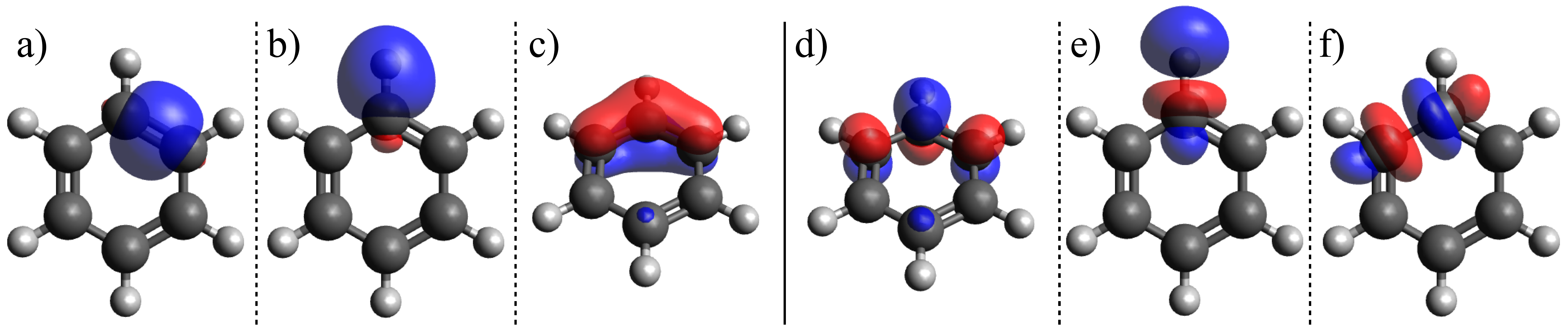}
}
\caption{Scalar-X2C ILMOs of benzene (1s-type orbitals not shown). The full vertical line separates occupied (left) from virtual (right) ILMOs.}
\label{fig:benzene}
\end{figure}

\begin{figure*}
\resizebox{\textwidth}{!}{
\includegraphics[scale=1]{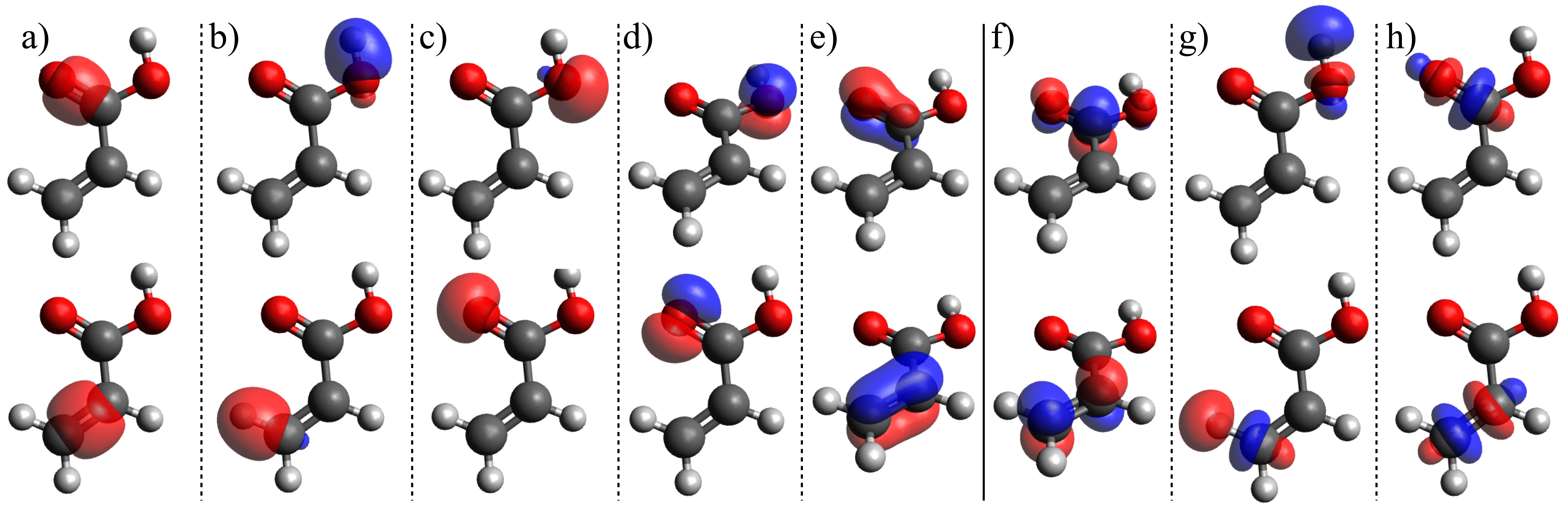}
}
\caption{Scalar-X2C ILMOs of the acrylic acid molecule (1s-type orbitals not shown). The full vertical line separates occupied (left) from virtual (right) ILMOs.
}
\label{fig:acrylicacid}
\end{figure*}

Turning to the shape of the resulting ILMOs in Figs.~\ref{fig:benzene} 
and \ref{fig:acrylicacid} for benzene and the acrylic acid, respectively,
one recognize the IBOs obtained in Ref.~\citenum{knizia2013intrinsic}, as well as the $\sigma$-antibonding ones in Fig.~6 of Ref.~\citenum{west2013comprehensive}.
Interestingly, even while the IFOs were not the same
between the non-relativistic and the relativistic case (see Fig.~\ref{fig:benzene_IAO}), we do not see any
difference for the ILMOs. This is indicative of the quenching of the
atomic spin-orbit coupling in the formation of molecular bonds~\cite{Powell:lQb96T_h}.
Starting with benzene,
we can attribute the following to the occupied ILMOs
of Fig.~\ref{fig:benzene}:
{\bf a)} C--C $\sigma$-bonding orbital (6 times degenerate),
{\bf b)} C--H $\sigma$-bonding orbital (6 times degenerate),
and {\bf c)} delocalized (aromatic) $\pi$-bonds (3 times degenerate).
Let us now turn to the valence virtual ILMOs, which could be called
intrinsic antibonding orbitals. Indeed,
as readily seen in the right part of Fig.~\ref{fig:benzene},
the virtual ILMOs are:
{\bf d)} delocalized (aromatic) $\pi$-antibonding orbital (3 times degenerate),
{\bf e)} C--H $\sigma$-antibonding orbital (6 times degenerate),
and {\bf f)} C--C $\sigma$-antibonding orbital (6 times degenerate).
Only the $\pi$-system is delocalized on more than two centers, i.e. four centers. The $\sigma$-system shows bonds localized on two centers only. For both the bonding and antibonding $\pi$ and $\pi^*$ orbitals, the square of the orbital coefficients in terms of IFOs shows compositions of
50\% on the central carbon atom, 22.2\% on the neighboring two carbon atoms and 5.6\% on the carbon atom of the opposite side.

The exact same analysis can be done on the acrylic acid
shown in Fig.~\ref{fig:acrylicacid}, i.e.,
all C--H, C--C, O--C and O--H
$\sigma$-bonding orbitals are clearly identified, as well as 
the $sp$-hybrid orbitals, oxygen lone pairs and the 
delocalized $\pi$-systems.
The orbital's labels in Fig.~\ref{fig:acrylicacid} therefore corresponds
to the following occupied ILMOs:
{\bf a)} O--C and C--C $\sigma$-bonding orbitals,
{\bf b)} O--H and C--H $\sigma$-bonding orbitals,
{\bf c)} $sp$-hybrid orbitals of the oxygen atoms,
{\bf d)} lone pairs of the oxygen atoms,
and {\bf e)} O=C and C=C $\pi$-bonding orbitals.
Turning to the virtual ILMOs, we have:
{\bf f)} O=C and C=C $\pi$-antibonding orbitals,
{\bf g)} O--H and C--H $\sigma$-antibonding orbitals,
and {\bf h)} O=C and C=C $\sigma$-antibonding orbitals.
As for benzene, 24 of the 29 ILMOs are more than 99\% localized on two atomic fragments.
The slightly more spread orbitals are the lone pairs of the oxygen atom of the C=O, 
the C=C $\pi$-bonding orbital, the O=C $\pi$-antibonding orbital, and the 
C--O and C--C $\sigma$-antibonding orbitals.

The localization procedure for benzene converged in 16 and 24 iterations
for the occupied and the valence virtual space,
respectively, when no spin-orbit coupling is considered
(i.e. for both the non-relativistic case as well as the scalar-X2C one). 
However, adding spin-orbit coupling increased the number
of iterations to 
24 for the occupied space. 
Importantly, we observed that the 
quartic exponent in Eq.~(\ref{eq:L}) is necessary
to achieve convergence for this aromatic system, as 
a quadratic exponent would lead to a
continuum of maximal localizations, as discussed
in Ref.~\citenum{knizia2013intrinsic}.
For the acrylic acid, only 11 and 13 iterations achieved convergence
without spin-orbit coupling, against only a small increase
to 12 for X2C in the occupied space.

\subsection{Ferrocene with molecular fragments}

Let us now look at the ILMOs of the ferrocene molecule, where
molecular fragments are composed of each cyclopentadienyl ring C$_5$H$_5^{-}$
and the isolated Fe$^{2+}$ cation. 
The cc-pVDZ and the cc-pVTZ basis were used to construct the MO basis $\mathcal{B}_1$ and the RFO bases $\mathcal{B}^k$, respectively.
Compared to the
IAOs which are centred on the atoms, the IFOs based on molecular 
fragments are now centred on the different fragments and can be 
delocalized on several atoms belonging to a given fragment.
This example is a perfect illustration of the usefulness of 
using molecular fragments
instead of atomic ones. We furthermore note that the localization procedure
(using Jacobi rotations)
is extremely hard to converge when atomic fragments are considered. 
For instance, in the X2C case convergence is reached in the occupied space after 336 iterations, while the localization in the virtual space
reaches a plateau with a final gradient of $10^{-8}$
(instead of $10^{-15}$ like for
the other systems studied in this paper).
For cases like this, in which simple Jacobi optimization is problematic, it is possible to consider more robust
optimizations based on an exponential parametrization of the unitary transformation, 
where the gradient and Hessian of the localization function with respect to orbital rotations are estimated~\cite{berghold2000general,dubillard2006bonding,hoyvik2012orbital}.

\begin{figure}
\resizebox{\columnwidth}{!}{
\includegraphics[scale=1]{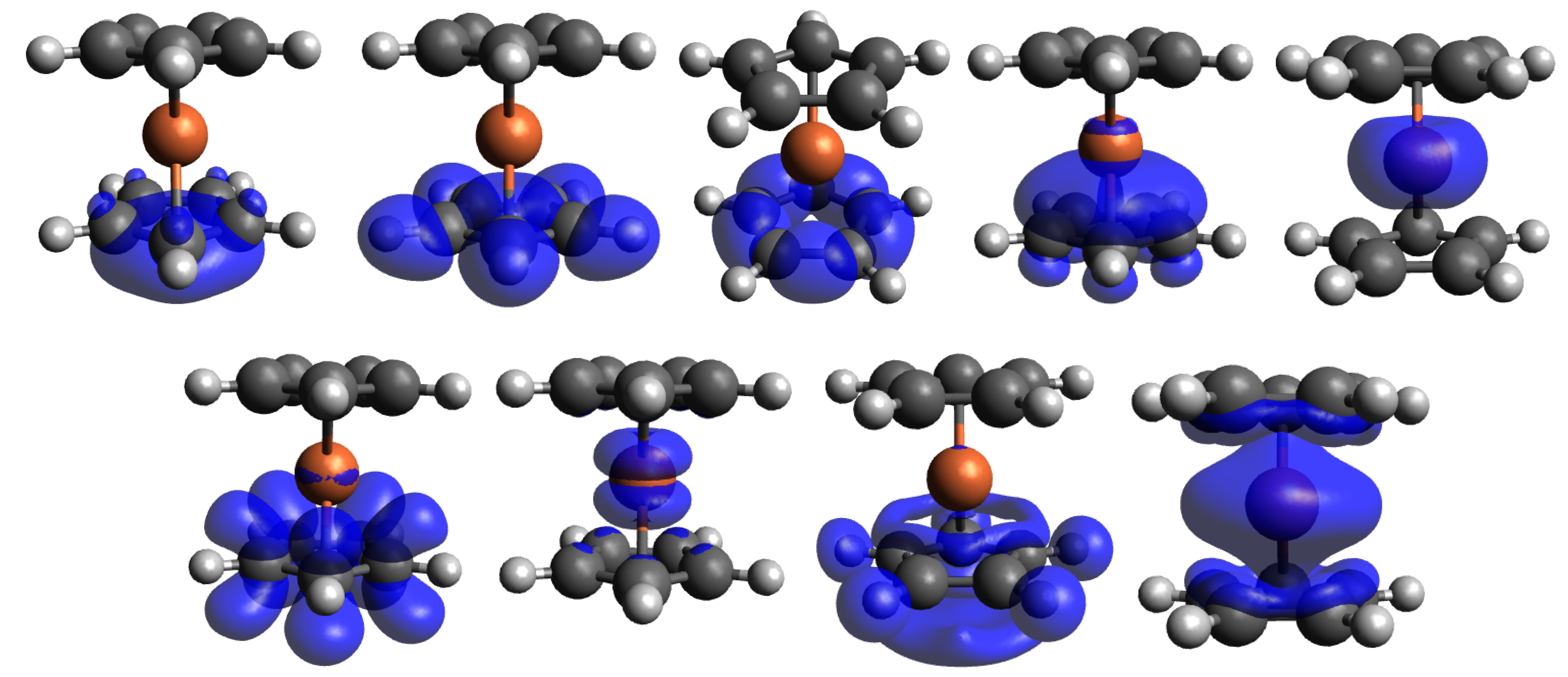}
}
\caption{Density of the ILMOs of ferrocene from a X2C calculation. First row: occupied orbitals, second row: virtual orbitals.}
\label{fig:ferrocene}
\end{figure}

In contrast, if the above molecular fragments are considered instead of
atomic ones to build the IFOs, then
convergence is reached in less than 12 iterations 
for both occupied and virtual orbitals.
As their are too many ILMOs to display, we selected a few 
illustrative ones in Fig.~\ref{fig:ferrocene}.
Before localization, all orbitals not fully localized on the Fe atom were
equally distributed on both cyclopentadienyl rings due to the symmetry of the system.
After localization they are localized on a single cyclopentadienyl ring (first three
orbitals) with sometimes some donation onto the Fe ion visible (fourth orbital), or on the metal (fifth orbital). 
Of the total of 48 occupied ILMOs, 39 are more than 99.7\% localized on a single fragment 
(9 on the Fe atom and 15 on each cyclopentadienyl ring), while 6 are localized on two fragments: one
cyclopentadienyl ring and the Fe atom (two and four of them more than 85.3\% and 90.1\% localized on a cyclopentadienyl ring, respectively, and the rest on the Fe atom).
The last three occupied ILMOs come
from the $d$ orbitals of the Fe atom. 
They are more than 93.2\% localized on the atom, with a small
delocalization equally spread on the two cyclopentadienyl rings.
A similar analysis holds for the virtual ILMOs: 24 out of 27 are largely localized on a single cyclopentadienyl
ring (more than 99.8\% for 18 of them, and more than 96.5\% for the 6 others) 
with sometimes a small contribution on the Fe atom. The three remaining virtual ILMOs are 72.2\%, 
80.9\% and 81.3\% localized on the Fe atom, the rest being equally spread on 
the cyclopentadienyl rings.

\subsection{Tellurazol oxide complexes}

To probe the influence of spin-orbit coupling on localization requires inclusion of elements  of the fourth or lower rows of the periodic table. For this purpose
we selected the Te-substituted polyethylene glycol oligomers that comprise a suitable benchmark system for relativistic algorithms~\cite{helmich2019relativistic}.
In this section, we consider two different calculations. First,
we consider one tellurazol oxide complex for which atomic fragments
are used. Second, we consider a dimer of tellurazol oxide complexes, 
each monomer defining one molecular fragment used to construct the IFOs.
The uncontracted dyall.2zp~\cite{dyall-dzp} basis was used to construct $\mathcal{B}_1$ 
and $\mathcal{B}^k$.
Geometries are extracted from Ref~\citenum{helmich2019relativistic}.

\subsubsection{Atomic fragments}

\begin{figure*}
\resizebox{\textwidth}{!}{
\includegraphics[scale=1]{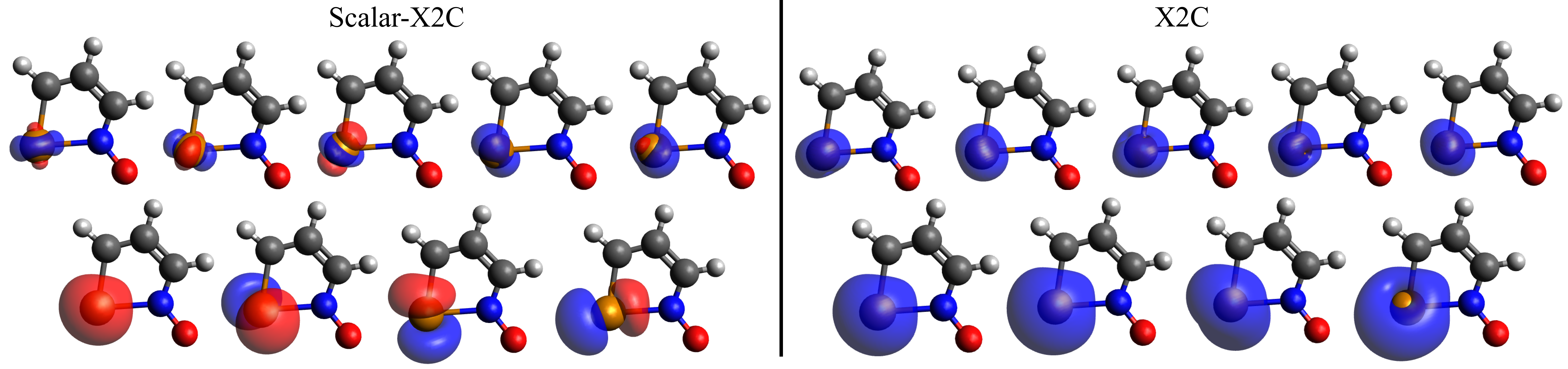}
}
\caption{IFOs of the Tellurium atom in the tellurazol oxide complex. Left panel: scalar-X2C, right panel: X2C.
}
\label{fig:Te_IAO}
\end{figure*}

Starting with atomic fragments,
we first take a look at the difference in the IFOs
with and without spin-orbit coupling.
As readily seen in Fig~\ref{fig:Te_IAO},
the IFOs (which reduces to IAOs here) are very close to the 
well-known $d$ (first row), $s$ and $p$ (second row) orbitals 
in the scalar-relativistic case (left panel).
However, adding spin-orbit coupling (X2C, right panel) leads
to a mixture of the above orbitals, like seen in Fig.~\ref{fig:benzene_IAO} for benzene. For the $d$-orbitals we now have two degenerate $4d_{3/2}$ orbitals and
three degenerate $4d_{5/2}$ orbitals with all a rather spherical density.

\begin{figure}
\resizebox{\columnwidth}{!}{
\includegraphics[scale=1]{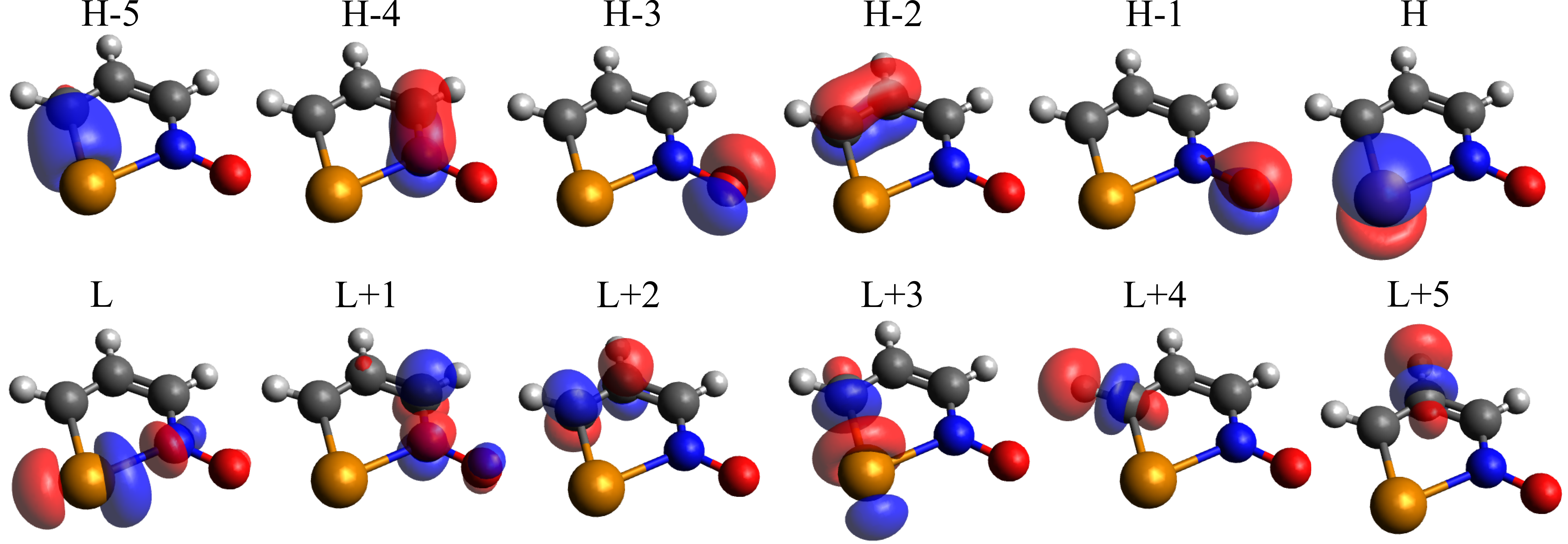}
}
\caption{Scalar-X2C ILMOs of the tellurazol oxide complex.}
\label{fig:Te-ox_monomer}
\end{figure}

Somewhat surprisingly we do not see the increased effect of spin-orbit coupling
in the shape of the ILMOs that look very similar in the non-relativistic, scalar
relativistic and full relativistic calculations.
The scalar-X2C spinors
are plotted in Fig.~\ref{fig:Te-ox_monomer}. Again, the occupied ILMOs
(first row of Fig.~\ref{fig:Te-ox_monomer}) give
a very good representation of chemical
bonding. One can easily recognize the 
Te--C $\sigma$-bonding orbital (H-5 in Fig.~\ref{fig:Te-ox_monomer})
more than 99\% localized on two atomic centers, 
the $\pi$-bonding orbitals more than 95\% localized on two atomic centers (H-4 and H-2), 
as well as the lone pairs of the
oxygen (H-1 and H-3) and the tellurium (H) atoms.
Turning to the valence virtual ILMOs 
(second row of Fig.~\ref{fig:Te-ox_monomer}),
they also efficiently represent Te--N and Te--C 
$\sigma$-antibonding orbitals
(L and L+3, respectively), $\pi$-antibonding orbitals (L+1 and L+2)
and C--H $\sigma$-antibonding orbitals (L+4 and L+5).

One of the advantages of the intrinsic orbitals is that they are not
particularly tied to the basis set compared to 
PM orbitals~\cite{dubillard2006bonding,knizia2013intrinsic,janowski2014near}, leading
to almost basis set independent charges.
This is illustrated in Table~\ref{tab:charges_X2C}.
This table features Mulliken and IFO charges of fragments $k$, the latter
given by
\begin{eqnarray}
q_k = Z_k - \sum_{i}\sum_{t \in k}
C_{ti}^{{\rm occ}*} C_{ti}^{\rm occ},
\end{eqnarray}
where $Z_k$ is the fragment's nuclear charge (i.e. the sum of the nuclear charge of all atoms constituting the molecular fragment
minus the charge of the fragment) and $\mathbf{C}^{\rm occ}$ is
the coefficient matrix of dimension
$(\dim(\mathcal{B}_2) \times N_{\rm occ})$ representing 
the canonical occupied orbitals (labelled $i$)
expressed in terms of the IFOs (labelled $t$).
\begin{table}
\begin{center}
\begin{tabularx}{\columnwidth}{llrrrrrrrrr}
\hline
Meth. & basis & Te & O & N & C$_1$ & C$_2$ &  C$_3$ & H$_1$ & H$_2$ & H$_3$\\
\hline
IFO & v2z&    0.63& -0.62&  0.03& -0.38& -0.15& -0.02&  0.17&  0.16&  0.19\\
IFO & av2z&   0.64& -0.63&  0.03& -0.39& -0.15& -0.02&  0.17&  0.16&  0.18\\
IFO & v3z&    0.64& -0.63&  0.03& -0.39& -0.15& -0.02&  0.17&  0.16&  0.18\\
IFO & av3z&   0.64& -0.63&  0.03& -0.39& -0.15& -0.02&  0.17&  0.16&  0.18\\
\hline
Mu. & v2z & 0.53& -0.47& -0.19& -0.31& -0.06& 0.20& 0.14& 0.07& 0.10\\
Mu. & av2z & 0.83& -0.68& 0.23& -0.32& 0.35& -0.40& -0.07& -0.01& 0.07 \\
Mu. & v3z& 0.42& -0.55& 0.01& -0.15& -0.10& 0.12& 0.07& 0.07& 0.11\\
Mu. & av3z& 0.61& -0.81& 0.05& -0.18& 0.26& 0.30& -0.11& -0.10& -0.02 \\
\hline
\end{tabularx}
\caption{X2C partial charges of the tellurazol oxide complex as a function of the GTO basis set used to form both $\mathcal{B}_1$
and $\mathcal{B}^k$. First four rows: IFO charges; last four rows: Mulliken charges.}
\label{tab:charges_X2C}
\end{center}
\end{table}

\subsubsection{Molecular fragments}

\begin{figure}
\resizebox{\columnwidth}{!}{
\includegraphics[scale=1]{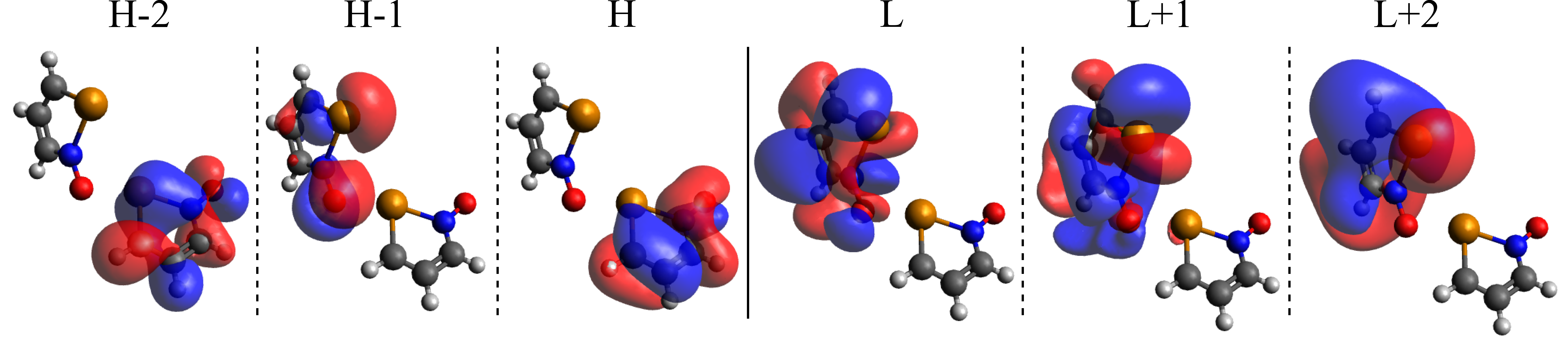}
}
\caption{Scalar-X2C ILMOs of the tellurazol oxide complex dimer.}
\label{fig:Te-ox_dimer}
\end{figure}

Let us now look at the dimer of tellurazol oxide complexes, 
where each monomer is used as a molecular fragment to 
construct the IFO basis.
Six ILMOs around the HOMO and LUMO are plotted in 
Fig.~\ref{fig:Te-ox_dimer}. As expected from the localization procedure,
all orbitals are indeed very localized on a given monomer but not on
a particular atom inside this fragment. This is well-suited for embedding 
approaches in which the density of one or more fragments is to be kept frozen
in an electron correlation calculation.
After localization, all orbitals are localized by more than 89\% on a single monomer. 100 out of 110 ILMOs
are localized by more than 99\% and 8 by more than 97\%.

\subsection{Te-PEG-4}

\begin{figure}
\resizebox{\columnwidth}{!}{
\includegraphics[scale=1]{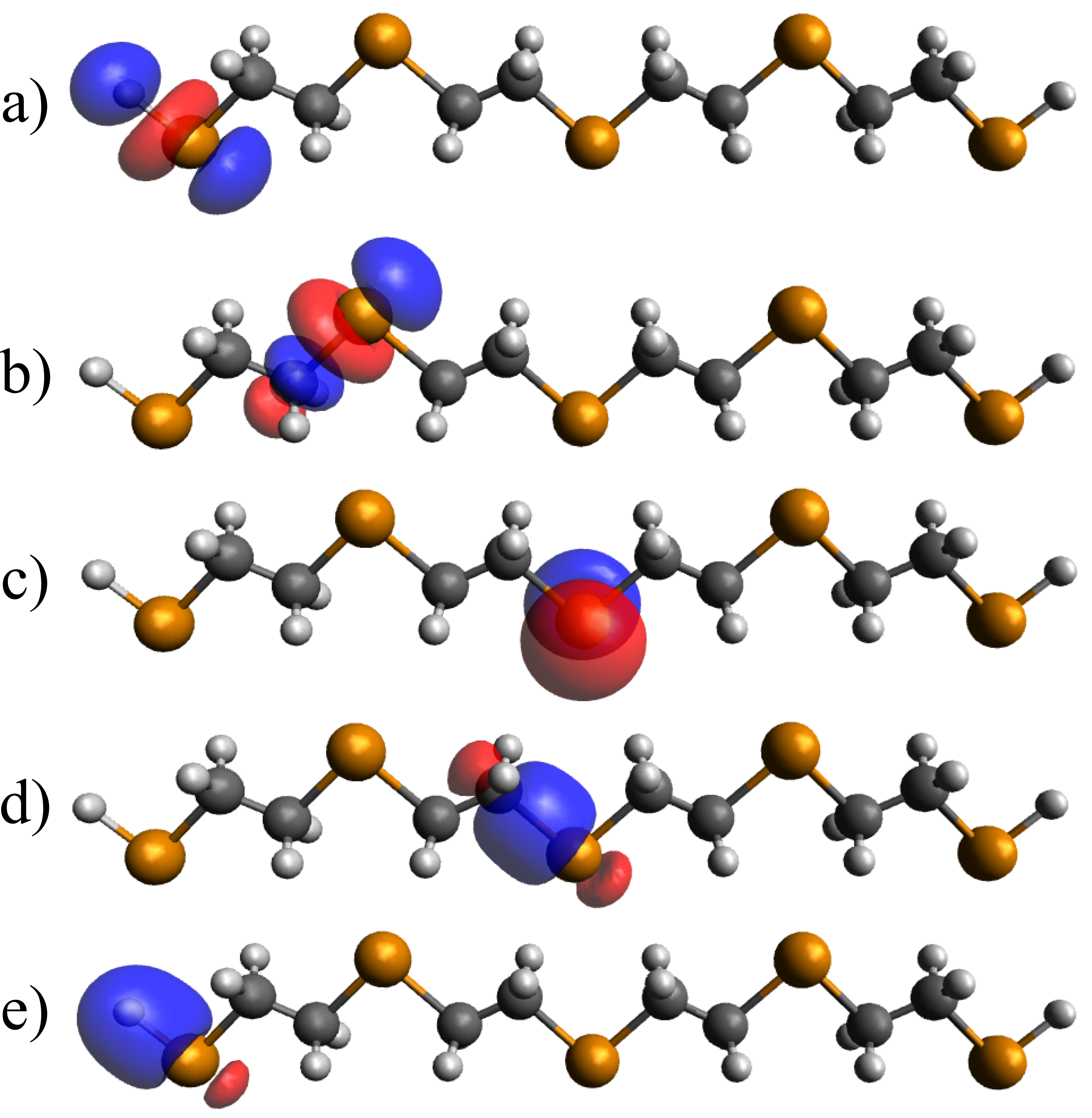}
}
\caption{Virtual (a-b) and occupied (c-e) scalar-X2C ILMOs of Te-PEG-4.}
\label{fig:Te-PEG-4}
\end{figure}

We then investigated a simpler (without double
bonds) but larger system, i.e. a
linear chain of tellurium-substituted poly-ethylene
glycol oligomers (Te-PEG-4)~\cite{helmich2019relativistic}.
The uncontracted dyall.2zp~\cite{dyall-dzp} basis were used for both $\mathcal{B}_1$ and $\mathcal{B}^k$, together with atomic fragments.
While the original canonical MOs can be delocalized over all the atoms of the molecule,
the ILMOs, both occupied and virtual, are localized on at most two atomic-centers. In Fig.~\ref{fig:Te-PEG-4} we show some of 
the highest occupied
and lowest virtual ILMOs.
Just like in the other systems, one can easily identify the
bonding nature of the orbitals, that is
{\bf a)} a Te--H $\sigma$-antibonding orbital,
{\bf b)} a Te--C $\sigma$-antibonding orbital,
{\bf c)} lone pair orbital of the Tellurium atom,
{\bf d)} Te--C $\sigma$-bonding orbital and
{\bf e)} Te--H $\sigma$-bonding orbital.
Note that due to the symmetry of the system, {\bf a)}, {\bf b)}, {\bf d)} and {\bf e)} are
exactly two times degenerate, but also very close in energy to other
related orbitals (for instance, orbital in {\bf a)} with other Te--H $\sigma$-antibonding orbitals involving the other Tellurium atoms). We remind that the term `energy' should be regarded following the discussion in Sec.~\ref{sec:approxenergy}.
Lower in energy, we find C--H and C--C 
$\sigma$-bonding orbitals followed by the one-center-localized
4s and 5d orbitals of the Tellurium atoms, etc.
Higher in energy, the remaining valence virtual MOs are 
C--H and C--C $\sigma$-antibonding orbitals.
All ILMOs are more than 99\% localized on maximally two atomic centers.

\subsection{The \textit{fac}-Irppy$_3$ complex}

\begin{figure}
\resizebox{1.1\columnwidth}{!}{
\includegraphics[scale=1]{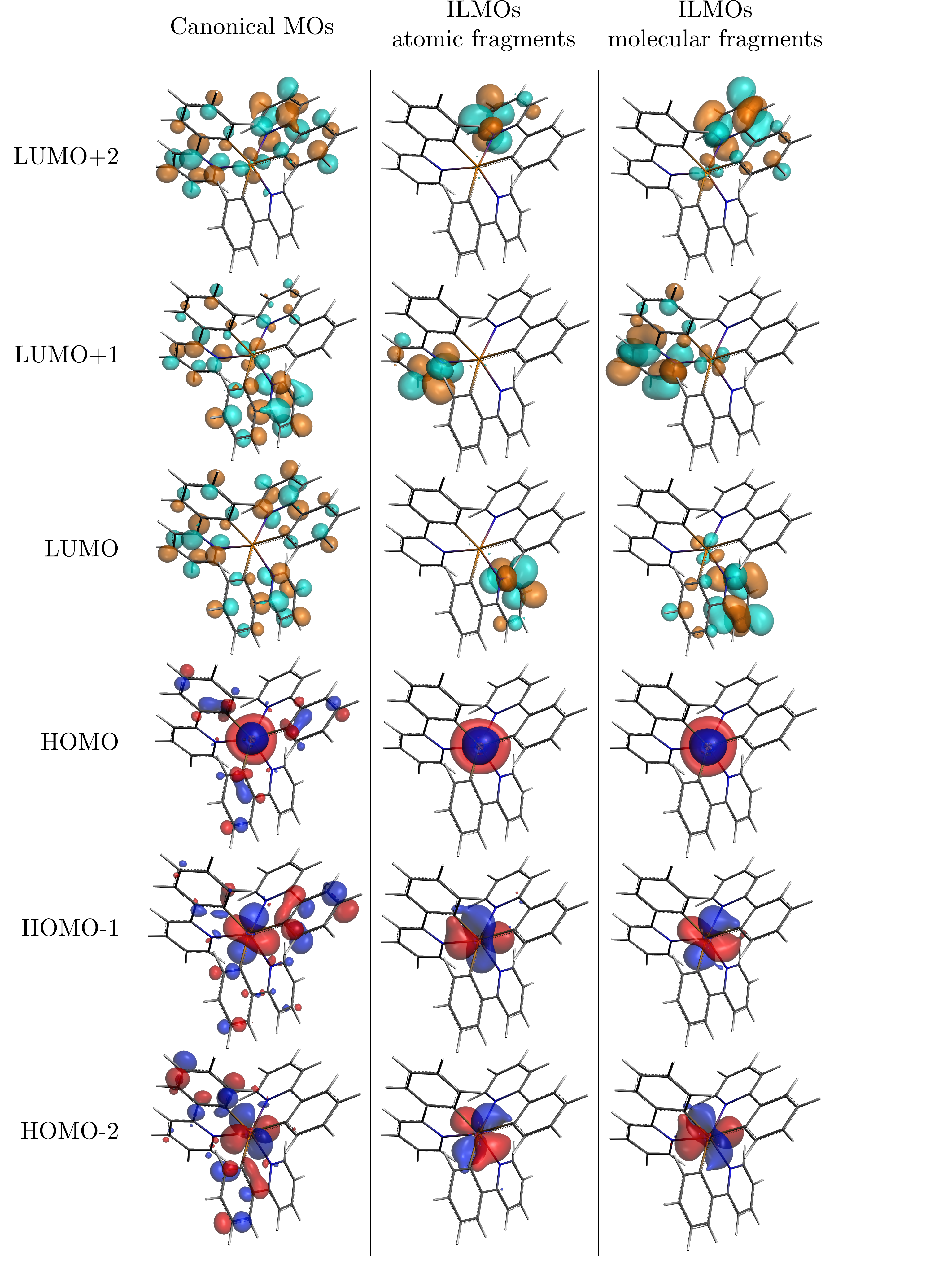}
}
\caption{Scalar-ZORA orbitals of the Irppy$_3$ complex.}
\label{fig:irppy3}
\end{figure}

To demonstrate the ADF interface we choose a well-known complex of iridium with three phenylpyridinate (ppy) ligands. This complex is one of the first phosphors used to emit light from triplet excitons in organic light emitting diode (OLED) 
devices~\cite{BaryshnikovAgren:2017} and similar compounds with different heavy metals and different ligands are being investigated to further enhance OLED 
performance~\cite{Kim.Kim.2014}. For analysis of the anisotropy of the emission and coupling of the phosphor to the host material it is convenient to work in a local picture. This is easily possible using ILMOs.
The uncontracted DZP Slater type basis~\cite{van2003optimized} that is supplied with the ADF program were used for both $\mathcal{B}_1$ and $\mathcal{B}^k$.
We consider here two possible ways to obtain such localized orbitals. The simplest procedure is to define atomic fragments and this approach does largely remove the ligand delocalization of the canonical 5d orbitals 
(HOMO-2 to HOMO) and produces almost pure 5d orbitals (11\% delocalization over the ppy fragments for HOMO-2 and HOMO-1 and 5\% for HOMO, compared to 48\% and 36\% for canonical orbitals, respectively). One also obtains a total of 84 virtual orbitals that are mostly localized on the ppy fragments. A more compact description of the valence virtual space is possible by defining the ppy units as molecular fragments and keeping just the lowest canonical virtuals of each ppy unit in the IFO basis. Such a reduction of the basis set by removing virtual orbital from fragments is already an option when ADF is started in fragment mode, but this procedure does also affect the resulting occupied orbital space and 
SCF energy as effects of polarization and charge-transfer are only accounted for in an approximate way. This problem does obviously not occur in post-SCF localization. We chose to limit basis $\mathcal{B}_2$ to include just the lowest 2 
virtuals of each ppy fragment and kept the valence space of the Ir identical to the atomic fragment run. In the occupied space we then obtain three almost atomic 5d orbitals (99\% for HOMO, 93 \% for HOMO-1 and HOMO-2) and in the valence virtual space we have 9 orbitals consisting of three sets of three-fold degenerate orbitals. The lowest three orbitals are displayed in Fig.~\ref{fig:irppy3} and are 95 \% localized on a single ppy ligand with 5\% admixture of Ir orbitals. 
The fact that considering molecular fragments led to an even better localization of the occupied 5d orbitals than was obtained with atomic fragments is
an indirect effect of the truncation of basis $\mathcal{B}_2$ for the former. Keeping all the 27 valence virtual orbitals of each ppy fragment does lead
to a localization of 91\% for HOMO-2, and 85\% for HOMO-1 and HOMO (not shown).
We believe that the compact and transparent description of the valence space offered by the use of molecular fragments will
prove useful for analysis. The possibilities for truncation can be beneficial in post-SCF applications as a smaller and more localized virtual space can reduce the computational cost of the algorithms significantly.

\subsection{A microsolvated astatine anion}

\begin{figure*}
\resizebox{\textwidth}{!}{
\includegraphics[scale=1]{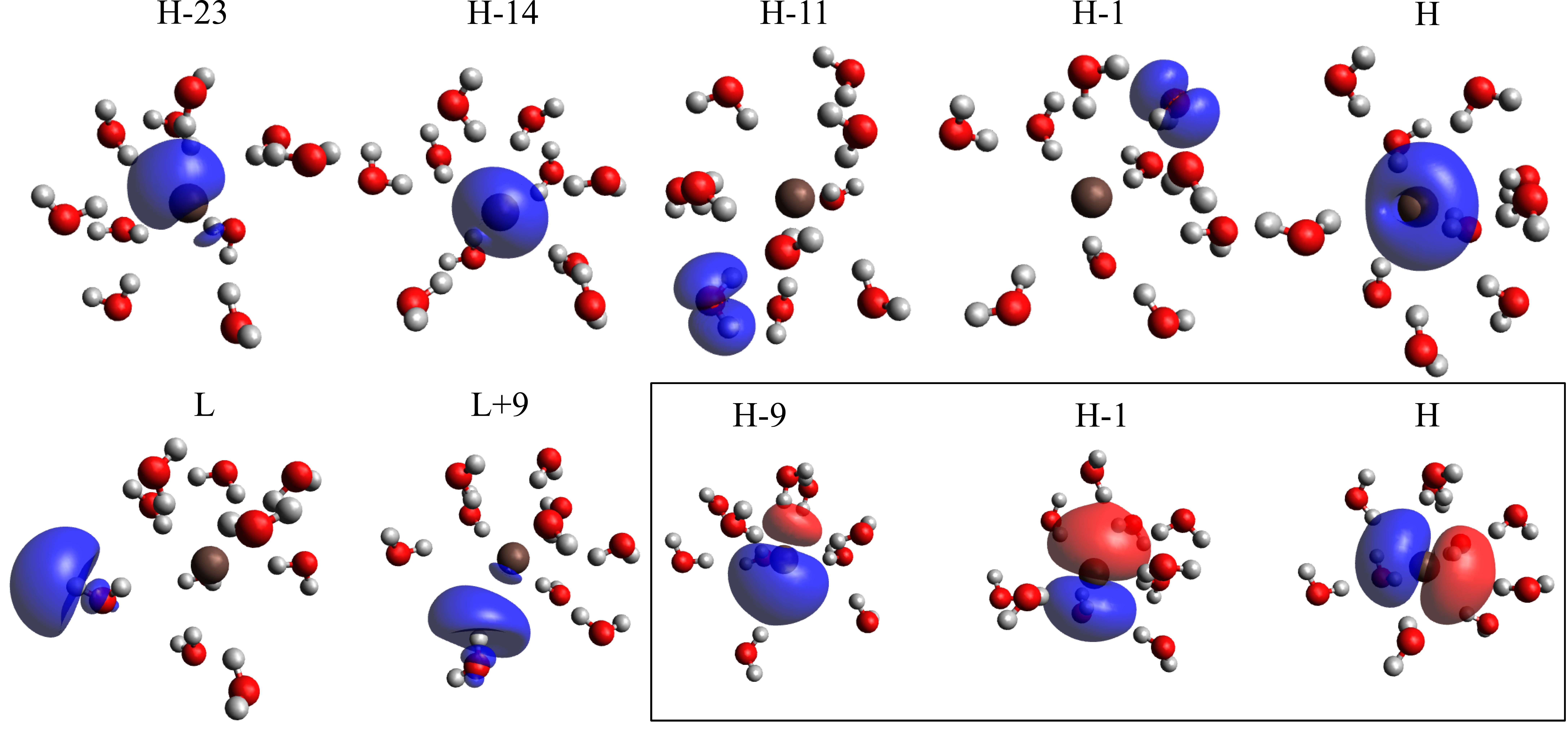}
}
\caption{Density of the ILMOs of the astatine anion surrounded
by ten water molecules.
The last three framed orbitals are non-relativistic spinors (thus the possibility to show phases).}
\label{fig:astatine}
\end{figure*}

The last system studied in this paper is the hydrated astatine anion. The radioactive element astatine is the next-to-last member of the halogens in the periodic table of elements (it was the last until
the synthesis of Tennessine). It is much studied by theoreticians interested by the
effects of spin-orbit coupling on chemical bonding. Another motivation to study this element is the potential use of $^{210}$At and $^{211}$At in radiotherapy~\cite{at-review}. For this purpose, the behavior of its anion in water is of relevance, spurring interest in embedding techniques for an efficient treatment of its electronic structure. One possibility
is to \textit{a priori} partition the electron density in terms of astatine and surrounding water molecules. We will here consider whether this goal can also be achieved
by \textit{a posteriori} localization of the supermolecular orbitals.
We chose an environment of ten water molecules,
where the geometry is extracted from Ref.~\citenum{bouchafra2018predictive}.
The uncontracted dyall.2zp~\cite{dyall-dzp} basis was used to construct both 
$\mathcal{B}_1$ and $\mathcal{B}^k$. 
Molecular fragments comprise the individual water molecules and the 
isolated astatine anion.
ILMOs are plotted in Fig.~\ref{fig:astatine}.
In contrast to the canonical MOs which
are delocalized on several water molecules, most of the ILMOs
are 90\%  to more than 99\% localized on a single fragment (see for instance H-11 and  H-1 in Fig.~\ref{fig:astatine}).
H-11 corresponds to a bonding orbital 
between the oxygen and the two hydrogens of the water molecule, while 
H-1 features the lone pair of the oxygen.
The ILMOs with a strong contribution from the valence $p$ orbitals 
of the astatine anion (see H-23, H-14 and H in 
Fig.~\ref{fig:astatine}) are for more
than 80\% localized
on the astatine atom with the remainder spread on the surrounding water molecules.

In this case the ILMOs do not look like the $p$-type orbitals of 
the astatine anion, as shown in the non relativistic case (see H-9, H-1 
and H in the frame of Fig~\ref{fig:astatine}), as the atomic
spin-orbit coupling is not quenched in the molecular environment. This is the first system studied here
for which there is a clear difference 
in the ILMOs upon adding spin-orbit coupling.
This \textit{a posteriori} localization lends support to embedding approaches in which the density is partitioned \textit{a priori}, the $p$-orbitals of the astatine anion can be considered fully occupied and are not delocalized much over the water 
molecules. Thus, those ILMOs are practically the IFOs for which
the presence of spin-orbit coupling leads to rather spherical
$p$ orbital densities.
Finally, all our twenty virtual LMOs are anti-bonding orbitals between
the oxygen and each of the hydrogen of each water molecules.
Some of them are
oriented towards the astatine anion and thus partially delocalized (see 
L+9 in Fig.~\ref{fig:astatine}), while others are fully localized on the 
water fragments (see L in Fig.~\ref{fig:astatine}).
Again, as
the orbitals of the astatine anion are all filled, there is no
virtual orbital left in the minimal basis $\mathcal{B}_2$
for the astatine. Thus, no virtual IFO and ILMO are found for the astatine anion.

\section{Conclusions}\label{sec:conclusions}

A generalization of the intrinsic atomic (and bonding) orbitals of
Knizia~\cite{knizia2013intrinsic} to complex and quaternion
spinors, as
well as an extension to molecular fragments instead of atomic ones are 
described in this paper. The calculations are performed with
the Reduction of Orbital Space Extent (ROSE).
ROSE is a flexible standalone code
which can be interfaced with several quantum chemistry codes.
For now, interfaces with DIRAC~\cite{DIRAC19,DIRAC_article}, Psi4~\cite{psi4},
PySCF~\cite{PYSCF} and
ADF~\cite{ADF2001} have been implemented.
As expected, our orbitals have the same advantages of 
the original ones, i.e. 
they are not tied to the basis set and they are a very
good representation of chemical bonding.
As the IFOs form a minimal polarized basis, only a limited set of
virtual orbitals can be localized in this basis.
By performing a singular value decomposition, 
we first express a reduced set of valence virtual orbitals in terms
of IFOs.
The latter are then as easy to localize as the occupied ones and
provide a good representation of anti-bonding
orbitals.
In order to test our method, different systems with increasing complexity 
are investigated. We started with simple molecules such as benzene and
the acrylic acid with atomic fragments. Then, we showed how considering
molecular fragments instead of atomic ones can improve the convergence of
the localization procedure, as seen for the ferrocene, or provide a more
compact representation of the ligand virtual space for Irppy$_3$. Systems
with significant relativistic effects such as tellurazol oxide complexes, a
linear chain of tellurium-substituted poly-ethylene
glycol oligomers Te-PEG-4, and the astatine anion surrounded by water 
molecules were studied, and orbitals were successfully localized.
We think this implementation will be useful for embedding techniques,
like in the context of the automated valence active 
space~\cite{sayfutyarova2017automated},
density matrix embedding theory~\cite{wouters2016practical} or the 
localized
active space self-consistent field 
method~\cite{hermes2019multiconfigurational},
to cite a few. The investigation of
our localized occupied and virtual orbitals
in the context of local correlated methods is left for future work,
but we expect that they can provide a very good approximation to the
weakly occupied correlating multiconfigurational SCF orbitals 
in the full valence
space~\cite{west2013comprehensive}, and are thus a very good effective
configurational basis for the treatment of valence-internal 
correlation~\cite{bytautas2003split}.

\section{Acknowledgments}

BS thanks Saad Yalouz, André Severo Pereira Gomes and Johann Pototschnig for fruitful discussions.
BS acknowledges support from the Netherlands Organization for Scientific Research (NWO) and Shell Global Solutions BV.
SS and LV acknowledge support from NWO via the CSER programme.
MR acknowledges support from the Research Council of Norway through a Center of Excellence Grant 
(Grant No. 262695).
GK acknowledges support from the NSF CAREER program under award number CHE-1945276.

\section{Supporting Information}
A clarification of some mathematical aspects of the intrinsic atomic orbital construction.

\newcommand{\Aa}[0]{Aa}
\providecommand{\latin}[1]{#1}
\makeatletter
\providecommand{\doi}
  {\begingroup\let\do\@makeother\dospecials
  \catcode`\{=1 \catcode`\}=2 \doi@aux}
\providecommand{\doi@aux}[1]{\endgroup\texttt{#1}}
\makeatother
\providecommand*\mcitethebibliography{\thebibliography}
\csname @ifundefined\endcsname{endmcitethebibliography}
  {\let\endmcitethebibliography\endthebibliography}{}

\appendix

\section{Pseudo-codes}

\subsection{Gauss--Seidel algorithm}

The pseudocode to solve Eq.~(\ref{eq:P21}) with the Gauss--Seidel method
applied on quaternion spinors is given in Algorithm~\ref{algo:solvelineq}.
This algorithm is easily applied in a complex 
or quaternion form. The only non-standard operations are the
quaternion multiplication in the calculation of $X_{ji}$ and the
definition of an Euclidean norm for
quaternion algebra in the ``do until" loop.

\begin{algorithm}[H]
\caption{Gauss--Seidel algorithm. Solve linear equation $\mathbf{A}\mathbf{X} = \mathbf{B}$ where $\mathbf{A}$ is a diagonally-dominant or symmetric and positive matrix (sufficient but not necessary condition).}
\label{algo:solvelineq}
\begin{algorithmic}[1]
\State Inputs: $\mathbf{A} \equiv \mathbf{S}_{22}$, $\mathbf{B} \equiv \mathbf{S}_{21}$. Outputs: $\mathbf{X} \equiv \mathbf{P}_{21}$
\vspace{0.2cm}
\State Initialization: $\mathbf{X} = \mathbf{S}_{21}$
\vspace{0.2cm}
\State {\bf for} $i=1$ {\bf to} $\dim(\mathcal{B}_2)$ (columns)
\vspace{0.2cm}
    \State \hspace{0.4cm} {\bf do until} $|| \mathbf{X}^{n}_i - \mathbf{X}^{n+1}_i || < 10^{-15}$
    \vspace{0.2cm}
        \State \hspace{0.8cm} $\mathbf{X}^{n}_{i} = \mathbf{X}_{i}$
        \vspace{0.2cm}
        \State \hspace{0.8cm} {\bf for} $j=1$ {\bf to} $\dim(\mathcal{B}_1)$ (rows)
        \vspace{0.2cm}
            \State \hspace{1.2cm} $X_{ji} = \dfrac{B_{ji} - {\displaystyle \sum_{k\neq j}^{\dim(\mathcal{B}_1)}} A_{jk} \times X_{ki}}{A_{ii}}$
            \vspace{0.2cm}
        \State \hspace{0.8cm} {\bf end}
        \vspace{0.2cm}
        \State \hspace{0.8cm} $\mathbf{X}^{n+1}_{i} = \mathbf{X}_{i}$
        \vspace{0.2cm}
    \State \hspace{0.4cm} {\bf end}
    \vspace{0.2cm}
\State {\bf end}
\end{algorithmic}
\end{algorithm}

\subsection{Diagonalization algorithm}

\begin{algorithm}[H]
\caption{Jacobi eigenvalue algorithm of a Hermitian square matrix $\mathbf{M}$ with dimension $n$, such that $\mathbf{E}^\dagger \mathbf{M} \mathbf{E} = \mathbf{D}$}
\label{algo:diag}
\begin{algorithmic}[1]
\State Input: $\mathbf{M}$. Outputs: $\mathbf{E}$, $\mathbf{D}$
\vspace{0.2cm}
\State Initialization: $|| p || = 1$ ({\it pivot} norm), $\mathbf{E} = \mathbf{1}$
\vspace{0.2cm}
\State {\bf do until} $||p|| < 10^{-15}$
\vspace{0.2cm}
    \State \hspace{0.3cm} Find the {\it pivot} $p$ (maximal off-diagonal element of the upper triangular matrix $\mathbf{M}$)
    \vspace{0.2cm}
    \State \hspace{0.3cm} $p = M_{ij}$
    \vspace{0.2cm}
    \State \hspace{0.3cm} Compute the phase $P = p/||p||$
    \vspace{0.2cm}
    \State \hspace{0.3cm} {\bf for} $k=1$ {\bf to} $n$
    \vspace{0.2cm}
        \State \hspace{0.6cm} $E_{kj} = E_{kj}/P$
        \vspace{0.2cm}
        \State \hspace{0.6cm} {\bf if} $k = j$ {\bf then} cycle 
        \vspace{0.2cm}
        \State \hspace{0.6cm} $M_{kj} = M_{kj}/P$ ; $M_{jk} = M_{jk}\times P$
        \vspace{0.2cm}
    \State \hspace{0.3cm} {\bf end}
    \vspace{0.2cm}
    \State \hspace{0.3cm} $w = (M_{jj} - M_{ii})/||2p||$
    \vspace{0.2cm}
    \State \hspace{0.3cm} {\bf if} $w \leq 0$ {\bf then} 
    \vspace{0.2cm}
        \State \hspace{0.6cm} $\tan(\theta) = - w + \sqrt{w^2 + 1}$
        \vspace{0.2cm}
    \State \hspace{0.3cm} {\bf else} 
    \vspace{0.2cm}
        \State \hspace{0.6cm} $\tan(\theta) = - w - \sqrt{w^2 + 1}$
        \vspace{0.2cm}
    \State \hspace{0.3cm} {\bf end}
    \vspace{0.2cm}
    \State \hspace{0.3cm} $\cos(\theta) = \dfrac{1}{\sqrt{\tan(\theta)^2 + 1}}$ ; $\sin(\theta) = \dfrac{\tan(\theta)}{\sqrt{\tan(\theta)^2 + 1}}$
    \vspace{0.2cm}
    \State \hspace{0.3cm} $M_{ii} = M_{ii} - \tan(\theta)||p||$ ; $M_{jj} = M_{jj} + \tan(\theta)||p||$
    \vspace{0.2cm}    
    \State \hspace{0.3cm} $M_{ij} = M_{ji} = 0$
    \vspace{0.2cm}
    \State \hspace{0.3cm} {\bf for} $k = 1$ {\bf to} $n$
    \vspace{0.2cm}
        \State \hspace{0.6cm} $\begin{bmatrix}
        E_{ki}\\
        E_{kj}
        \end{bmatrix} = \begin{bmatrix}
        \cos(\theta) & -\sin(\theta)\\
        \sin(\theta) & \cos(\theta)
        \end{bmatrix}\begin{bmatrix}
        E_{ki}\\
        E_{kj}
        \end{bmatrix}$
        \vspace{0.2cm}
        \State \hspace{0.6cm} {\bf if} $(k=i)$ or $(k=j)$ {\bf then} cycle
        \vspace{0.2cm}
        \State \hspace{0.6cm} $\begin{bmatrix}
        M_{ik}\\
        M_{jk}
        \end{bmatrix} = \begin{bmatrix}
        \cos(\theta) & -\sin(\theta)\\
        \sin(\theta) & \cos(\theta)
        \end{bmatrix}\begin{bmatrix}
        M_{ik}\\
        M_{jk}
        \end{bmatrix}$
        \vspace{0.2cm}
        \State \hspace{0.6cm} $M_{ki} = M_{ik}^*$ ; $M_{kj} = M_{jk}^*$
        \vspace{0.2cm}
    \State \hspace{0.3cm} {\bf end}
    \vspace{0.2cm}
\State {\bf end}
\vspace{0.2cm}
\State $\mathbf{D} \equiv \mathbf{M}$
\vspace{0.2cm}
\State Sort $\mathbf{E}$ and $\mathbf{D}$ by ascending order
\end{algorithmic}
\end{algorithm}

The pseudocode of the Jacobi eigenvalue algorithm for quaternion spinors
is given in Algorithm~\ref{algo:diag}. The crucial step is to
rotate the matrix into the real plane by scaling the $j$-th 
basis vector by the phase factor, before performing the rotation in the
real plane. Similarly, the eigenvectors are also to be
rotated only after scaling them with the phase factor.

\subsection{SVD algorithm}

The pseudocode of the Jacobi singular value decomposition algorithm 
for quaternion spinors is given in Algorithm~\ref{algo:SVD}.

\begin{algorithm}[H]
\caption{Jacobi singular value decomposition algorithm, solve $\mathbf{A} = \mathbf{U}\mathbf{\Sigma}\mathbf{V}^\dagger$. $\mathbf{A}$
and $\mathbf{\Sigma}$ are rectangular matrices of dimension $m \times n$, $\mathbf{U}$ and $\mathbf{V}$ are unitary matrices of dimension $m\times m$ and $n\times n$, respectively.}
\label{algo:SVD}
\begin{algorithmic}[1]
\State Input: $\mathbf{A}$. Outputs: $\mathbf{U}$, $\lbrace \Sigma_{ii}\rbrace_{i=1,\hdots,n_{\rm val}}$, $\mathbf{V}$
\vspace{0.2cm}
\State Initialization: $l = \min(m,n)$
\vspace{0.2cm}
\State {\bf if} $l = m$ {\bf then}
\vspace{0.2cm}
    \State \hspace{0.3cm} $\mathbf{M} = \mathbf{A} \mathbf{A}^\dagger$
    \vspace{0.2cm}
    \State \hspace{0.3cm} call Algorithm~\ref{algo:diag}(input: $\mathbf{M}$, output: $\mathbf{E}$, $\mathbf{D}$)
    \vspace{0.2cm}
    \State \hspace{0.3cm} Sort $\mathbf{E}$ and $\mathbf{D}$ by descending order
    \vspace{0.2cm}
    \State \hspace{0.3cm} $\mathbf{U} = \mathbf{E}$ ; $\Sigma_{ii} = D_{ii}$, $i=1,\hdots,l$
    \vspace{0.2cm}
    \State \hspace{0.3cm} {\bf for} $i = 1$ {\bf to} $l$
    \vspace{0.2cm}
        \State \hspace{0.6cm} {\bf for} $j = 1$ {\bf to} $n$
        \vspace{0.2cm}
            \State \hspace{0.9cm} $V_{ji} = \left(\mathbf{A}^\dagger\mathbf{U}\right)_{ji} / \Sigma_{ii}$
            \vspace{0.2cm}
        \State \hspace{0.6cm} {\bf end}
        \vspace{0.2cm}
    \State \hspace{0.3cm} {\bf end}
    \vspace{0.2cm}
\State {\bf else}
\vspace{0.2cm}
    \State \hspace{0.3cm} $\mathbf{M} = \mathbf{A}^\dagger \mathbf{A}$
    \vspace{0.2cm}
    \State \hspace{0.3cm} call Algorithm~\ref{algo:diag}(input: $\mathbf{M}$, output: $\mathbf{E}$, $\mathbf{D}$)
    \vspace{0.2cm}
    \State \hspace{0.3cm} Sort $\mathbf{E}$ and $\mathbf{D}$ by descending order
    \vspace{0.2cm}
    \State \hspace{0.3cm} $\mathbf{V} = \mathbf{E}$ ; $\Sigma_{ii} = D_{ii}$, $i=1,\hdots,l$
    \vspace{0.2cm}
    \State \hspace{0.3cm} {\bf for} $i = 1$ {\bf to} $l$
    \vspace{0.2cm}
        \State \hspace{0.6cm} {\bf for} $j = 1$ {\bf to} $m$
        \vspace{0.2cm}
            \State \hspace{0.9cm} $U_{ji} = \left(\mathbf{A}\mathbf{V}\right)_{ji} / \Sigma_{ii}$
            \vspace{0.2cm}
        \State \hspace{0.6cm} {\bf end}
        \vspace{0.2cm}
    \State \hspace{0.3cm} {\bf end}
    \vspace{0.2cm}
\State {\bf end}
\end{algorithmic}
\end{algorithm}

Note that we do not get the whole $\mathbf{U}$ or $\mathbf{V}$ matrix
depending on the value of $\min(m,n)$. This is not
an issue as only $N^{\rm val}_{\rm vir}$ will have non-zero eigenvalues
and only the corresponding eigenvectors are of interest to us.
Furthermore, one has to be careful when dividing by a singular value close or
equal to 0. But again, for the purpose of this paper, we did need not 
consider such singular values and their associated vectors.

\subsection{Localization procedure}\label{app:localization}

The pseudocode to maximise Eq.~(\ref{eq:L}) by performing $2 \times 2$ Jacobi rotations on quaternion spinors is given in Algorithm~\ref{algo:loc}.
Just as in Sec.~\ref{sec:jac_loc}, we omit the superscript $k$ of $Q^k_{ij}$ [Eq.~(\ref{eq:charge_matrix})]
for ease of notation.

\begin{algorithm}[H]
\caption{Jacobi $2\times 2$ rotations to maximize Eq.~(\ref{eq:L}).}
\label{algo:loc}
\begin{algorithmic}[1]
\State Input \& Output: $\mathbf{C} \equiv \mathbf{C}^{\rm occ}_{\rm IFO}$
\vspace{0.2cm}
\State Initialization: Grad = 1
\vspace{0.2cm}
\State {\bf do until} Grad $< 10^{-15}$
\vspace{0.2cm}
    \State \hspace{0.3cm} {\bf for} $i=2$ {\bf to} $N_{\rm occ}$
    \vspace{0.2cm}
        \State \hspace{0.6cm} {\bf for} $j=1$ {\bf to} $i-1$
        \vspace{0.2cm}
            \State \hspace{0.9cm} {\bf for} $k=1$ {\bf to} $N_{\rm F}$ (number of fragments)
            \vspace{0.2cm}
                \State \hspace{1.2cm} Compute $Q_{ii}$, $Q_{jj}$, $Q_{ij}$
                \vspace{0.2cm}
                \State \hspace{1.2cm} $\tilde{B}_{ij} = \tilde{B}_{ij} + 4Q_{ij}(Q_{ii}^3 - Q_{jj}^3)$
            \State \hspace{0.9cm} {\bf end}
                \vspace{0.2cm}
                \State \hspace{0.9cm} Compute $|| \tilde{B}_{ij} ||$, $P_{ij} = \tilde{B}_{ij} / ||\tilde{B}_{ij} ||$
                \vspace{0.2cm}
            \State \hspace{0.9cm} {\bf for} $k=1$ {\bf to} $N_{\rm F}$ (number of fragments)
                \State \hspace{1.2cm} $A_{ij} = A_{ij} - Q_{ii}^4 - Q_{jj}^4 + 6(Q_{ii}^2 + Q_{jj}^2)  \fnargb{\Re}{Q_{ij}\times P_{ij}^*}^2 + Q_{ii}^3Q_{jj} + Q_{ii}Q_{jj}^3$
                \vspace{0.2cm}
            \State \hspace{0.9cm} {\bf end}
            \vspace{0.2cm}
                \State \hspace{0.9cm} Compute $\theta = (1/4)\,\text{atan2}\,(|| \tilde{B}_{ij} ||,-A_{ij})$
            \vspace{0.2cm}
            \State \hspace{0.9cm} {\bf for} $t=1$ {\bf to} $\dim(\mathcal{B}_2)$
            \vspace{0.2cm}
                \State \hspace{1.2cm} $C'_{ti} = \cos (\theta)C_{ti} + \sin(\theta) (C_{tj}\times P_{ij}^*)$
                \vspace{0.2cm}
                \State \hspace{1.2cm} $C'_{tj} = - \sin(\theta)(C_{ti}\times P_{ij}) + \cos(\theta) C_{tj}$
                \vspace{0.2cm}
                \State \hspace{1.2cm} $C_{ti} = C'_{ti}$ and $C_{tj} = C'_{tj}$
                \vspace{0.2cm}
            \State \hspace{0.9cm} {\bf end}
            \vspace{0.2cm}
            \State \hspace{0.9cm} Compute Grad = Grad + $||B_{ij}||^2$
            \vspace{0.2cm}
        \State \hspace{0.6cm} {\bf end}
        \vspace{0.2cm}
    \State \hspace{0.3cm} {\bf end}
    \vspace{0.2cm}
    \State \hspace{0.3cm} Grad = $\sqrt{\text{Grad}}/(N_{\rm occ}(N_{\rm occ}-1)/2)$
    \vspace{0.2cm}
\State {\bf end}
\end{algorithmic}
\end{algorithm}

Just as in the diagonalization procedure described in Algorithm~\ref{algo:diag}, the crucial step of the algorithm is to multiply the $i$-th vector
by the phase $P_{ij}$ and the $j$-th vector by the phase $P_{ij}^*$. Once this is done, real rotation matrices can be
applied to the quaternion spinors until Eq.~(\ref{eq:L}) is maximized.

\section{Orbital localization: Technical details}\label{app:OrbitalLocalization}
\subsection{Formal goal and notation}

For a given set of orbitals $\{\lmi{i};\,i\in\{1,\ldots,\Nloc\}\}$ (typically either the set of occupied molecular orbitals, valence-virtual orbitals, or some subset of either), one can define a multitude of scalar functionals $L(\{\lmi{i}\})$ which characterize the degree of locality of the orbital set.
Here we concentrate on a class of generalized PM-like\cite{pipek1989fast} functionals, 
\begin{align}
   L\bigl(\{\lmi{i}\}\bigr) = \sum_{i=1}^\Nloc \sum_{F}  h\bigl(n_F(\lmi{i})\bigr), \label{eq:LpAppx}
\end{align}
where $h(n)$ is a scalar function mapping the real interval $[0,1]$ to $\mathbb{R}$ (we will consider $h(n)=n^2$ and $h(n)=n^4$ as special cases), and $n_F(\lmi{i})$ is a function which quantifies which fraction of orbital $\lmi{i}$ is attributed to system fragment $F$.
We consider $n_F(\lmi{i})$ which fulfill $\forall F:\,n_F(\lmi{i}) \geq 0$ and $\sum_F n_F(\lmi{i})=1$ (at least approximately) and which are concretely given by an expression
\begin{align}
   n_F(\lmi{i}) = \bra{\lmi{i}}\hat P_F\ket{\lmi{i}}  \label{eq:nFviaPF}
\end{align}
for some definition of fragment-dependent operators $\{\hat P_F\}$.\footnote{The operators $\hat P_F$ can be most naturally defined if the system fragments $F$ originate from a partitioning the full-system Hilbert space; alternatively, they can be defined by one of  several atomic partitionings of the electron charge distribution.\cite{pipek1989fast,lehtola2014pipek}.}


Let $\{\lmi{i};\,i\in\{1,\ldots,\Nloc\}\}$ denote the set of input orbitals.
The goal of the localization procedure is to find an unitary transformation $\hat U$ within $\lin\{\lmi{i}\}$
for which the transformed orbitals $\{\hat U \lmi{i}\}$ maximize the localization functional Eq.~\eqref{eq:LpAppx}:
\begin{align}
   \hat U = \argmax_{\text{unitary } \hat U} L\bigl(\{\hat U \lmi{i}\}\bigr).
   \label{eq:UnitaryArgmax}
\end{align}

\subsection{Localization by incremental 2$\times$2-updates in $\mathbb{C}^2$}\label{app:UpdateBy2by2Rotations}
In Eq.~\eqref{eq:rotation} of the main text and appendix \ref{app:localization} we describe a simple, but generally well-working, algorithm for orbital localization in the sense of Eq.~\eqref{eq:UnitaryArgmax}.
The algorithm is a revision of the method described by PM\cite{pipek1989fast}, and, like PM's, works by incrementally applying 
$2\times2$ Jacobi rotations
\begin{align}\label{eq:AppRotation}
   \hat U_\theta \lki{i} &:= \cos(\theta)\lki{i} + \sin(\theta)\lki{j}\nonumber\\
   \hat U_\theta \lki{j} &:= -\sin(\theta)\lki{i} + \cos(\theta)\lki{j},
\end{align}
to all unique pairs of orbitals $i,j\in\{1,\ldots,\Nloc\}$ (with $i\neq j$) in the localization set (we also formally define $\hat U_\theta$ to act as identity operator outside $\lin\{\lmi{i},\lmi{j}\}$).
In this, each such $2\times 2$ rotation individually either exactly or approximately solves the localization problem described by Eq.~\eqref{eq:UnitaryArgmax} in the one-dimensional subspace of rotations (parameterized by the rotation angle $\theta$).
In this section, we explain the construction of the $2\times 2$ update formula in Eqs.~\eqref{eq:rotation} and \eqref{eq:rotation_angle}.
The real version of the angle parameterization in Eq.~\eqref{eq:rotation_angle} (which differs from PM's~\cite{pipek1989fast}) and the concrete formulas for $A_{ij}$ and $B_{ij}$ were already provided in Ref.~\citenum{knizia2013intrinsic}, but no technical details of their derivation.

In the complex case, Eq.~\eqref{eq:AppRotation} is not the most general $2\times 2$ unitary transformation which covers all degrees of freedom which physically could potentially become relevant.
Rather, as explained in Sec.~\ref{app:General22UnitaryTrafo}, formally it should be replaced by
\begin{align}\label{eq:AppRotationComplex}
   \UTwo \lki{i} &:=  e^{-\ii \tau}\cos(\theta)\lki{i} + e^{\ii \tau}\sin(\theta)\lki{j},\nonumber\\
   \UTwo \lki{j} &:= -e^{-\ii \tau}\sin(\theta)\lki{i} + e^{\ii \tau}\cos(\theta)\lki{j},
\end{align}
where $\theta,\tau\in\mathbb{R}$ are the \emph{two} scalar rotation parameters.

\subsection{Solution to the $2\times2$ localization problem}\label{app:LocalizationUpdateFormula}
To proceed, let $i$ and $j$ denote two fixed orbital indices,
and consider the maximization of localization functional $L$ Eq.~\eqref{eq:LpAppx} with respect to \emph{only} the unitary transformation $\UTwo$ between $\phi_i$ and $\phi_j$ described by Eq.~\eqref{eq:AppRotationComplex}.
Eq.~\eqref{eq:LpAppx} makes clear that $\UTwo$ cannot affect any term of $L$'s orbital sum except the two involving $\phi_i$ or $\phi_j$;
therefore, maximizing the full $L$ of Eq.~\eqref{eq:LpAppx} with respect to $\UTwo$ is equivalent to maximizing the simpler two-term target function
\begin{equation}
L(\theta,\tau) := \sum_{F}\left[ h\mleft(n_F\!\left(\UTwo\lmi{i}\right)\mright) + h\mleft(n_F\!\left(\UTwo\lmi{j}\right)\mright) \right]. \label{eq:LpTwoOrbs}
\end{equation}
To proceed, we first evaluate $n_F\!\left(\UTwo\lmi{i}\right)$ and $n_F\!\left(\UTwo\lmi{j}\right)$.
Defining the hermitian-matrix elements (\emph{cf.} Eq.~\eqref{eq:nFviaPF})
\begin{align}
   Q^F_{kl} := \bra{\lmi{k}}\hat P_F\ket{\lmi{l}}
\end{align}
and abbreviating $c:=\cos(\theta)$ ($\in\mathbb{R}$), $s:=\sin(\theta)$ ($\in\mathbb{R}$), $z:=e^{\ii \tau}$ ($\in\mathbb{C}$, $\abs{z}=1$), $\zc=z\,c$, and $\zs:=z\,s$, we get
\begin{align}
   n_F\!\left(\UTwo\lmi{i}\right) &= \bra{\UTwo\lmi{i}}\hat P_F\ket{\UTwo\lmi{i}}
\notag\\ &= \bra{\zc^* \lmi{i} + \zs \lmi{j}}\hat P_F\ket{\zc^* \lmi{i} + \zs \lmi{j}}
\notag\\ &= \abs{\zc^*}^2 \bra{\lmi{i}}\hat P_F\ket{\lmi{i}}
          + \abs{\zs}^2 \bra{\lmi{j}}\hat P_F\ket{\lmi{j}} + 
          \underbrace{\zc\zs \bra{\lmi{i}}\hat P_F\ket{\lmi{j}}
          + \zs^*\zc^* \bra{\lmi{j}}\hat P_F\ket{\lmi{i}}}_{2cs \Re\left( z^2 \bra{\lmi{i}}\hat P_F\ket{\lmi{j}} \right)}
\notag\\ &= c^2 Q^F_{ii} + s^2 Q^F_{jj} + 2 cs \,\Re(z^2 Q^F_{ij}). \label{eq:nUTwoI}
\end{align}
In the last step, we used $c,s\in\mathbb{R}$.
Proceeding in the same way for $n_F$ of $\UTwo\lki{j}$, we find
\begin{align}
   n_F\!\left(\UTwo\lmi{j}\right) &= \bra{\UTwo\lmi{j}}\hat P_F\ket{\UTwo\lmi{j}}
\notag\\ &= \bra{-\zs^* \lmi{i} + \zc \lmi{j}}\hat P_F\ket{-\zs^* \lmi{i} + \zc \lmi{j}}
\notag\\ &= s^2 Q^F_{ii} + c^2 Q^F_{jj}
          \underbrace{-\zs \zc Q^F_{ij} - \zc^* \zs^* Q^F_{ji}}_{-2cs \Re\left(z^2 Q^F_{ij}\right)}
\notag\\ &= s^2 Q^F_{ii} + c^2 Q^F_{jj} - 2 cs \,\Re(z^2 Q^F_{ij}).\label{eq:nUTwoJ}
\end{align}
Inserting these back into the functional, we get
\begin{align}
   L(\theta,z) &= \sum_{F} h\mleft(c^2 Q^F_{ii} + s^2 Q^F_{jj} + 2 cs \,\Re(z^2 Q^F_{ij})\mright) + \sum_{F} h\mleft(s^2 Q^F_{ii} + c^2 Q^F_{jj} - 2 cs \,\Re(z^2 Q^F_{ij})\mright). \label{eq:LttOriginal}
\end{align}

For some functions $h(n)$, this expression can be evaluated into a simple closed-form result,
which allows immediately reading off the resulting conditions for stationarity.
For example, for the special case $h(n)=n^2$, Eq.~\eqref{eq:LttOriginal} yields
\begin{align}
   L_2(\theta,z) = \sum_{F}\bigg[&
   \frac{1}{4}\sin(4\theta)\left(4(Q^F_{ii}-Q^F_{jj}) \Re\bigl(z^2 Q^F_{ij}\bigr)\right)+\frac{1}{4} \cos(4\theta)\left((Q^F_{ii}-Q^F_{jj})^2-4\Re\bigl(z^2 Q^F_{ij}\bigr)^2\right)
\notag\\ &+\frac{1}{4} \left(3 (Q^F_{ii})^2+2 Q^F_{ii} Q^F_{jj}+4 \Re\bigl(z^2 Q^F_{ij}\bigr)^2+3 (Q^F_{jj})^2\right)
\bigg],\label{eq:L2explicitCosSin}
\end{align}
and the exact maximum of this expression upon variation of $\theta$ can be straightforwardly written down (see below).

However, not all $h$ can be treated just like that (e.g., $h(n)=n^4$ cannot, which was the original default for the IBO construction\cite{knizia2013intrinsic}).
For the general case, we therefore first construct an approximation to $L(\theta,z)$, and then use it to determine an approximate extremum of $L$.
To this end, we substitute $\theta=\frac{1}{4}\arctan(x)$ into $L$ of Eq.~\eqref{eq:LttOriginal}, and then expand $L$ in a power series around $x=0$:
\begin{align}
   L\bigl(\efrac{1}{4}\operatorname{arctan}(x),z\bigr) = \sum_{k=0}^{\infty} \frac{x^k}{k!} \bigg[\frac{\d^k L\bigl(\efrac{1}{4}\operatorname{arctan}(x),z\bigr)}{\d x^k}\bigg\vert_{x=0}\bigg].
\end{align}
Retaining only terms up to second order, this yields
\begin{align}
   L\bigl(\theta(x),z\bigr) = \frac{1}{2} A_{ij}(z)\,x^2 + B_{ij}(z)\,x
   + C_{ij} + \mathcal{O}(x^3). \label{eq:LxExpansion2ndOrder}
\end{align}
where $A_{ij}(z)$, $B_{ij}(z)$ and $C_{ij}$ denote the 2nd, 1st, and 0th order coefficients, respectively.
For a general smooth function $h(n):\,[0,1]\mapsto\mathbb{R}$ with first derivative $h'$ and second derivative $h''$, these coefficients can evaluated as
\begin{align}
   A_{ij}(z) &= \sum_F\bigg[
   \frac{1}{4} \Re\mleft(z^2 Q_{ij}^F\mright){}^2
   \left(h''\mleft(Q_{ii}^F\mright)+h''\mleft(Q_{jj}^F\mright)\right)
   -\frac{1}{8}\left(Q_{ii}^F-Q_{jj}^F\right)
   \left(h'\mleft(Q_{ii}^F\mright)-h'\mleft(Q_{jj}^F\mright)\right) \bigg]
\notag\\
   B_{ij}(z) &= \sum_F\bigg[ \frac{1}{2} \Re\mleft(z^2 Q_{ij}^F\mright) \left(h'\mleft(Q_{ii}^F\mright)-h'\mleft(Q_{jj}^F\mright)\right)\bigg]
\notag\\
   C_{ij} &= \sum_F \left(h\mleft(Q_{ii}^F\mright)+\mathop h\mleft(Q_{jj}^F\mright)\right).
   \label{eq:ABCSeriesCoefficients}
\end{align}
Concrete expressions for the special cases of $h(n)=n^2$ and $h(n)=n^4$, yielding the two formulas presented with the original IBO method, are provided in Tab.~\ref{tab:ABCexpressions}. 
However, a large variety of other convex functions $h(n)$ yield stable localization algorithms; the properties of more general $h(n)$ will be considered elsewhere.
\begin{table*}
\caption{Coefficients $A_{ij}(z)$, $B_{ij}(z)$, and $C_{ij}$ of the expansion
$L(\theta(x),z)=\sum_F \fnargb{h}{\fnargb{n_F}{\UTwo \lmi{i}}} + \sum_F \fnargb{h}{\fnargb{n_F}{\UTwo \lmi{j}}}=\frac{1}{2}A_{ij}(z)\,x^2+B_{ij}(z)\,x+C_{ij}+\mathcal{O}(x^3)$ [Eq.~\eqref{eq:LxExpansion2ndOrder}],
where $\theta(x)=\frac{1}{4}\arctan(x)$,
obtained for various functions viable as $h(n)$ in $L$ [Eq.~\eqref{eq:LpAppx}].}
\label{tab:ABCexpressions}
\begin{tabular}{l@{\hspace{2em}}l}
\toprule
General $h(n)$
&
{$\begin{aligned}
   A_{ij}(z)&=\sum_F\bigg[\frac{1}{4} \fnargb{\Re}{z^2 Q_{ij}^F}^2 \left(\fnargb{h''}{Q_{ii}^F}+\fnargb{h''}{Q_{jj}^F}\right)-\frac{1}{8}\bigl (Q_{ii}^F-Q_{jj}^F\bigr) \left(\fnargb{h'}{Q_{ii}^F}-\fnargb{h'}{Q_{jj}^F}\right)\bigg]
\\ B_{ij}(z)&=\sum_F\bigg[\frac{1}{2} \fnargb{\Re}{z^2 Q_{ij}^F} \left(\fnargb{h'}{Q_{ii}^F}-\fnargb{h'}{Q_{jj}^F}\right)\bigg]
\\ C_{ij}&=\sum_F\Bigl[\fnargb{h}{Q_{ii}^F}+\fnargb{h}{Q_{jj}^F}\Bigr]
\end{aligned}$}
\\[1ex]  \midrule
$h(n)=n^2$
&
{$\begin{aligned}
    A_{ij} (z)&=\sum_F\bigg[\fnargb{\Re}{z^2 Q_{ij}^F}^2-\frac{1}{4} \bigl(Q_{ii}^F-Q_{jj}^F\bigr)^2 \bigg]
\\  B_{ij}(z)&=\sum_F\Bigl[\bigl(Q_{ii}^F-Q_{jj}^F\bigr) \fnargb{\Re}{z^2 Q_{ij}^F} \Bigr]
\\  C_{ij}&=\sum_F\Bigl[\bigl(Q_{ii}^F\bigr)^2+\bigl(Q_{jj}^F\bigr)^2\Bigr]
\end{aligned}$}
\\[1ex] \midrule
$h(n)=n^4$
&
{$\begin{aligned}
A_{ij}(z)&=\sum_F\biggl[3 \left(\bigl(Q_{ii}^F\bigr)^2+\bigl(Q_{jj}^F\bigr)^2\right) \fnargb{\Re}{z^2 Q_{ij}^F}^2-\frac{1}{2} \bigl(Q_{ii}^F-Q_{jj}^F\bigr)\left(\bigl(Q_{ii}^F\bigr)^3 - \bigl(Q_{jj}^F\bigr)^3\right) \biggr]
\\ B_{ij}(z)&=\sum_F\Bigl[2 \left(\bigl(Q_{ii}^F\bigr){}^3-\bigl(Q_{jj}^F\bigr){}^3\right) \fnargb{\Re}{z^2 Q_{ij}^F} \Bigr]
\\ C_{ij}&=\sum_F\Bigl[\bigl(Q_{ii}^F\bigr)^4+\bigl(Q_{jj}^F\bigr)^4 \Bigr]
\end{aligned}$}
\\\bottomrule
\end{tabular}
\end{table*}

For $L(\theta,\tau)$ to be extremal with respect to variations in $\theta$, $\partial L/\partial \theta$ must vanish.
Translated to $L$'s second order approximation Eq.~\eqref{eq:LxExpansion2ndOrder}, this yields the stationarity condition
\begin{align}
   \frac{\partial L(x,z)}{\partial x} = A_{ij}(z)\,x + B_{ij}(z) = 0 \label{eq:StationaryX}
\end{align}
for $x$, and therefore as approximate extremum of $L(\theta,z)$ any rotation angle
\begin{align}
   \theta = \frac{1}{4}\left(\arctan\mleft(-\frac{B_{ij}(z)}{A_{ij}(z)}\mright) + \pi n\right)\quad(n\in\mathbb{Z})\label{eq:MinimumArctan1Arg}
\end{align}
However, not all of those solutions are minima, and some lead to spurious rotations by $180{}^{\circ{}}$ (i.e., swaps of the output orbitals $\UTwo\lmi{i}$ and $\UTwo\lmi{j}$, which are mathematically admissible but physically meaningless).

In order to establish the correct choice of solutions in Eq.~\eqref{eq:MinimumArctan1Arg}, and also rationalize our original choice of the parametrization $\theta=\frac{1}{4}\arctan(x)$, we consider the special case of $h(n)=n^2$.
For this special case $h(n)=n^2$, one can show by direct calculation that $L(\theta,z)$ can be \emph{exactly} expressed in terms of $\cos(4\theta)$ and $\sin(4\theta)$ and the series coefficients $A_{ij}(z)$, $B_{ij}(z)$, and $C_{ij}$ defined in Eq.~\eqref{eq:ABCSeriesCoefficients}:
\begin{align}
   L_2(\theta,z) = \bigl(C_{ij}+A_{ij}(z)\bigr) + B_{ij}(z)\,\sin(4\theta)  - A_{ij}(z)\,\cos(4\theta).
\end{align}
For $h(n)=n^2$, this coincides with $L_2$ from Eq.~\eqref{eq:L2explicitCosSin}.
The stationary condition $\partial L_2(\theta,z)/\partial \theta=0$ then becomes
\begin{align}
   B_{ij}(z)\,\cos(4\theta) = -A_{ij}(z)\,\sin(4\theta),
\end{align}
which has the solution set
\begin{align}
   \theta \in &\mleft\{ \tfrac{1}{4}\bigl(2\pi n + \operatorname{arctan2}\mleft(B_{ij}(z),\,-A_{ij}(z)\mright)\bigr);\, n\in\mathbb{Z}\mright\}\cup 
\notag\\&
   \mleft\{ \tfrac{1}{4}\bigl(2\pi n + \operatorname{arctan2}\mleft(-B_{ij}(z),\,A_{ij}(z)\mright)\bigr);\, n\in\mathbb{Z}\mright\}.\label{eq:L2StationaryPointsTheta}
\end{align}
The set described by Eq.~\eqref{eq:L2StationaryPointsTheta} is equivalent to the one described by Eq.~\eqref{eq:MinimumArctan1Arg}, but here rewritten in terms of the two-argument $\operatorname{arctan2}(y,x)$ function\footnote{The two-argument $\arctan(y,x)$ function is available in many programming languages. For $x>0$, it evaluates to $\arctan(y,x)=\arctan(y/x)$, and otherwise takes account of the signs of both the $x$ and the $y$ arguments to yield a unique arc angle to point $(x,y)$ in the full range $[-\pi,\pi]$; note that $\arctan(y/x)$ itself cannot distinguish the cases of $(x,y)$ and $(-x,-y)$, and therefore is only uniquely defined for a half-circle arc.}.
So which of those $\theta$ should we choose as rotation parameter?
To clarify that, we evaluate $L''_2(\theta,z):=\partial^2 L_2(\theta,z)/\partial \theta^2$ for the stationary points.
After simplifying and abbreviating $A_{ij} := A_{ij}(z)$ and $B_{ij} := B_{ij}(z)$, one finds
\begin{align}
   L_2''\mleft(\tfrac{n\pi}{2} + \tfrac{1}{4}\operatorname{arctan2}\mleft(-B_{ij},\,A_{ij}\mright),\tau\mright) &= +16\sqrt{A_{ij}^2+B_{ij}^2}
\notag\\
   L_2''\mleft(\tfrac{n\pi}{2} + \tfrac{1}{4}\operatorname{arctan2}\mleft(B_{ij},\,-A_{ij}\mright),\tau\mright) &= -16\sqrt{A_{ij}^2+B_{ij}^2}.
\end{align}
This implies that the upper set of solutions characterizes the minima of $L_2$
(positive second derivative at stationary points),
while the lower set characterizes the maxima.
Since our localization functionals are meant to be maximized, we need to select a solution from the lower set.
Additionally, we can just use the $n=0$ solution without loss of generality:
while the solutions for all $n\in\mathbb{Z}$ are technically admissible, choosing a $n\neq 0$ will change $\theta\mapsto \theta+n\cdot 90{}^{\circ{}}$, and therefore only cause physically inconsequential swaps and/or sign-flips the output orbitals in Eq.~\eqref{eq:AppRotationComplex}.
Therefore, for $\theta$ we can indiscriminately employ the update formula described in the main text:
\begin{align}
   \theta = \frac{1}{4}\operatorname{arctan2}\mleft(B_{ij}(z),\;-\,A_{ij}(z)\mright).\label{eq:MinimumArctan2Arg}
\end{align}

\subsection{Determining the relative phase factor $\tau$}\label{app:General22UnitaryTrafo}
A general (2,2)-shape complex unitary transformation\footnote{In linear algebra, the $\mathbb{C}^n$ unitary group is called $U(n)$; while the group name would simplify notation here, we have many other $U$ symbols, and using it may cause ambiguities. We therefore describe $U(n)$ as ``(n,n)-shape unitary matrices''.} $\mat U$
can have four real scalar degrees of freedom.
However, in our application, there are two redundant degrees of freedom as the resulting two new orbitals can be multiplied with arbitrary phase factors without
influencing the value of the functional $n$ 
(and $L$). 
The only relevant additional degree of freedom, compared to the case of real orbitals, is therefore a larger freedom to change their 
relative phase $z^2=e^{2 \ii \tau}$, as this phase may take a complex value rather than just reducing to a sign. 
The gradient $B_{ij}(\tau)$ depends on this value as we have 
\begin{align}
   B_{ij}(\tau)&=\sum_F b_{ij}^F \fnargb{\Re}{e^{2 \ii \tau} Q_{ij}^F}\\ \nonumber
   &=\sum_F b_{ij}^F \bigl[ \cos( 2 \tau) \fnargb{\Re}{Q_{ij}^F} - \sin( 2 \tau) \fnargb{\Im}{Q_{ij}^F} \bigr]\\ \nonumber
   &= \cos( 2 \tau) \fnargb{\Re}{\tilde{B}_{ij}} - \sin( 2 \tau) \fnargb{\Im}{\tilde{B}_{ij}}
      \label{eq:Bij_phasedependence}
\end{align}
where $b_{ij}^F$ are real coefficients that depend on the definition of $h(n)$ and $\tilde{B}_{ij}=\sum_F b_{ij}^F Q_{ij}^F$
is the (complex) gradient before we apply the phase transformation. 
In principle, one could consider evaluating the functional Eq.~\eqref{eq:LxExpansion2ndOrder} at its extremum $x=-B_{ij}(\tau)/A_{ij}(\tau)$ (Eq.~\eqref{eq:StationaryX}) as a function of $\tau$,
\begin{align}
   L_\mathrm{max}(\tau) = C_{ij} - \frac{1}{2} \frac{B_{ij}(\tau)^2}{A_{ij}(\tau)}
\end{align}
and determining which phase factor would make this stationary.
Unfortunately, this is not straight-forward due to the $\Re(e^{2\ii \tau} Q_{ij}^F){}^2$ term in $A_{ij}(\tau)$, which offers no obvious way of deferring the real component extraction in the computation of the sum over the fragments $F$.
Instead, we choose the value of  $\tau$ such that $B_{ij}$ is maximized, so that we locate an extremum of $B_{ij}$ with respect to a variation in $\tau$.
Since $C_{ij}$ is real and independent of $\tau$, and since close to a solution $x\approx 0$ the $B_{ij}\,x$ term will also dominate the $\frac{1}{2}A_{ij}\,x^2$ term in the second order expansion $L(x)$ in Eq.~\eqref{eq:LxExpansion2ndOrder}, this procedure should be able to reach a stationary point of $L$ with respect to both $\theta$ and $\tau$ at convergence.
This choice of phase factor may be viewed as writing the unitary
transformation of Eq. (\ref{eq:AppRotation}) as 
\begin{align}
   \mat U(\theta,\tau) & =
   \begin{bmatrix}
     e^{-\ii \tau} \cos(\theta) & e^{\ii \tau} \sin(\theta)
   \\ -e^{-\ii \tau} \sin(\theta) & e^{\ii \tau}\cos(\theta)
   \end{bmatrix}\\
   & =  
    \begin{bmatrix}
      \cos(\theta) & \sin(\theta)
   \\ -\sin(\theta) &   \cos(\theta) \end{bmatrix}
       \begin{bmatrix}
      e^{-\ii \tau} & 0
   \\ 0 & e^{\ii \tau}
   \end{bmatrix}
 \nonumber,
\end{align}
comprising a phase transformation of the two orbitals to make the gradient real, followed by rotation in the real plane. After the rotation, one may 
transform back to the original plane by choosing
\begin{align}
   \mat U^\prime(\theta,\tau) & =
   \begin{bmatrix}
      \cos(\theta) & e^{2 \ii \tau} \sin(\theta)
   \\ -e^{-2 \ii \tau} \sin(\theta) & \cos(\theta)
   \end{bmatrix}\\
   & =  
    \begin{bmatrix}
      e^{\ii \tau} & 0
   \\ 0 & e^{-\ii \tau}
   \end{bmatrix}
    \begin{bmatrix}
      \cos(\theta) & \sin(\theta)
   \\ -\sin(\theta) &   \cos(\theta) \end{bmatrix}
    \begin{bmatrix}
      e^{-\ii \tau} & 0
   \\ 0 & e^{\ii \tau}
   \end{bmatrix} \nonumber,
\end{align}
We implemented the latter transformation by first accumulating $\tilde B_{ij}$ as a complex number to obtain $e^{-2\ii \tau}=\frac{\tilde{B}_{ij}^*}{|\tilde{B}_{ij}|}$. Multiplying orbital $j$ with this phase factor yields:
\begin{align}
\tilde B^\prime_{ij}=\tilde{B}_{ij}\frac{\tilde{B}_{ij}^*}{|\tilde{B}_{ij}|}=\frac{|\tilde{B}_{ij}|^2}{|\tilde{B}_{ij}|}=|\tilde{B}_{ij}|
\end{align}
so that the gradient is maximized, as desired. The phase-adjusted orbitals are used in the computation of the second derivative $A_{ij}$ and the rotation angle $\theta$ as discussed in section \ref{app:UpdateBy2by2Rotations}. After the rotation, the phase of the orbitals is restored.

\clearpage

\noindent\rule{\columnwidth}{2pt}
\\[-8pt]\noindent\rule{\columnwidth}{2pt}

\begin{flushleft}
{\large Supporting Information (SI) for:}
\begin{flushleft}
\Large\textbf{Generalization of intrinsic orbitals to Kramers-paired quaternion spinors, molecular fragments, and valence virtual spinors}
\end{flushleft}
\textsf{\rule{0pt}{14pt}\large Bruno Senjean, Souloke Sen, Michal Repisky, Gerald Knizia, Lucas Visscher}
\end{flushleft}

\noindent\rule{\columnwidth}{2pt}
\setcounter{page}{1}
\setcounter{figure}{0}
\setcounter{table}{0}
\setcounter{equation}{0}

\def\thesection{S\arabic{section}}
\def\thesubsection{S\arabic{section}.\arabic{subsection}}
\def\thesubsubsection{S\arabic{section}.\arabic{subsection}.\arabic{subsubsection}}
\def\thefigure{S\arabic{figure}}
\def\thetable{S\arabic{table}}
\def\theequation{S\arabic{equation}}
\makeatletter
\def\p@subsection{}
\def\p@subsubsection{}
\makeatother




\section{Review of the intrinsic atomic orbital construction}

\newcommand{\oo}[1]{\phi_{#1}} 
\newcommand{\oto}[1]{\tilde\phi_{#1}} 
\newcommand{\ov}[1]{\phi_{#1}} 
\newcommand{\otv}[1]{\tilde\phi_{#1}} 
\newcommand{\noto}[1]{\tilde\phi_{#1}^\nonorthsuperscript} 
\newcommand{\SnotoName}{\mat {\tilde s}_{\rm pdo}}
\newcommand{\Snoto}[2]{\bigl[\SnotoName\bigr]_{#1#2}} 
\newcommand{\InvSnoto}[2]{\bigl[\SnotoName^{-1}\bigr]^{#1#2}} 
\newcommand{\PmhSnoto}[2]{\bigl[\SnotoName^{-1/2}\bigr]_{#1#2}} 
\newcommand{\mainbf}[1]{\chi_{#1}} 
\newcommand{\minbf}[1]{\zeta_{#1}} 
\newcommand{\iao}[1]{\smash{\phi_{#1}^\nonorthsuperscript}} 

\newcommand{\tOprj}{\smash{\hat {\tilde O}}} 
\newcommand{\tVprj}{\smash{\hat {\tilde V}}} 

The main text provides a new generalization of intrinsic atomic orbitals to intrinsic fragment orbitals and four-component relativistic orbitals.
In order to provide some background information, and make this work self-contained, in this appendix we review and clarify mathematical aspects of the IAO construction in the original non-relativistic AO context.
Apart from providing background information on the current work, this also provides a complete derivation of the previously rather terse theoretical details of the IAO method in Ref.~\citenum{knizia2013intrinsic}, and discusses a slight revision with improves formal properties.
Concretely:
\begin{itemize}
   \item The argument of why the IAOs span the occupied space is explicitly formulated in terms of equations (instead of text as in the original article), and full derivations of all involved equations are given.
   \item The IAO definition is slightly revised; the revised IAOs are simpler to construct, and provide near-indistinguishable results to the original IAOs.
   \item The matrix formulation of the IAO construction is derived, and an optimal set of working equations is given.
\end{itemize}

\subsection{Computational, free-atom AO, and intrinsic AO bases}

The original IAO method aims to construct a set of one-particle states---the IAOs---which have the same dimension and conceptual meaning as a minimal basis, but at the same time is capable of exactly spanning the occupied orbitals of a previously computed SCF (Hartree-Fock or Kohn-Sham) wave function $\ket{\Phi}$ (and thereby is also capable of exactly representing the corresponding many-electron SCF determinant itself).

In the previous version of the method,\cite{knizia2013intrinsic} this is achieved by applying a combination of subspace projections (described next) to an input set of tabulated free-atom atomic orbitals; these tabulated functions are expected to closely represent the ``chemical'' AOs of free atoms, and by placing them on the positions of the atoms in a molecule they can be used as a minimal molecular AO basis.
This minimal basis then \emph{could} be used for interpretative purposes, because each of its AO basis functions uniquely corresponds to an actual atomic orbital in the chemical sense.
However, due the complete lack of polarization functions in the basis, which are required to model the shifts in electronic structure of atoms as they come in contact with each other in a molecule, in general a minimal basis of free-atom AOs---no matter how accurate these represent the actual free atoms---does not have the capability of representing molecular wave functions with any degree of quantitative accuracy.
The IAO method aims to rectify this by retaining a minimal basis, but changing the original free-atom AOs it contained into molecule-intrinsic AOs by incorporating information obtained from an already computed molecular wave function.

\subsection{Rationalization of the IAO subspace projection formula}\label{sec:RationalizationOfTheIaoFormula}

In order to rationalize the construction achieving this, imagine the following scenario.
Assume we have been given a molecule and computed for it an accurate SCF wave function $\ket{\Phi}$ in a large computational basis set\footnote{In Ref.~\citenum{knizia2013intrinsic}, $B_1$ denoted a raw basis consisting of a set of non-orthogonal functions. In the main text we distinguish for clarity between this basis and the basis $\mathcal{B}_1$ resulting from orthonormalization of these functions. In this note we assume for simplicity that no near-linear dependencies need to be removed so that $\dim(\mathcal{B}_1)=\dim(B_1)$} $\mathcal{B}_1$.
As a result of the SCF computation, we first obtain a set of $N_{\rm occ}$ occupied molecular orbitals
\begin{eqnarray}\label{eq:MOcoeffoccB1}
   \ket{\oo{i}} = \sum_{\mu=1}^{\dim(\mathcal{B}_1)} \ket{\mainbf{\mu}} C_{\mu i}\qquad\text{$(i\in\{1,\ldots,N_{\rm occ}\})$},
\end{eqnarray}
which are needed for representing the many-electron wave function $\ket{\Phi}$.
We furthermore obtain a set of $N_{\rm vir}=\dim(\mathcal{B}_1)-N_{\rm occ}$ virtual molecular orbitals
\begin{eqnarray}\label{eq:MOcoeffvirB1}
   \ket{\ov{a}} = \sum_{\mu=1}^{\dim(\mathcal{B}_1)} \ket{\mainbf{\mu}} C_{\mu a} \qquad\text{$(a\in\{1,\ldots,N_{\rm vir}\})$},
\end{eqnarray}
which are \emph{not} needed for representing $\ket{\Phi}$, but rather describe the orthogonal complement of the occupied subspace $\lin\{\ket{\oo{i}}\}$ in the full one-particle space $\lin(\mathcal{B}_1)$.

Now let us consider the hypothetical situation in which we were to compute (e.g. by a separate SCF procedure) a \emph{second} single determinant wave function $\ket{\tilde\Phi}$ for the same molecule in the same electronic state---but this time employing only the minimal basis set of free-atom atomic orbitals, for now called $\mathcal{B}_2$, to expand its molecular orbitals.
[Note that $\mathcal{B}_2$ takes a more general meaning in the main text; also, in practice $\ket{\tilde\Phi}$ is not computed separately, but will be extracted from the original $\ket{\Phi}$ (see Appx.~\ref{eq:DepolarizationProjector})].
This calculation would yield another set of occupied and virtual orbitals, expanded over the minimal basis functions $\{\ket{\minbf{\mu}};\,\minbf{\mu} \in\mathcal{B}_2\}$:
\begin{align}\label{eq:MOcoeffoccB2}
   \ket{\oto{i}} &= \sum_{\mu=1}^{\dim(\mathcal{B}_2)} \ket{\minbf{\mu}} \tilde C_{\mu i}\qquad\text{$(i\in\{1,\ldots,N_{\rm occ}\})$},
\\ \ket{\otv{a}} &= \sum_{\mu=1}^{\dim(\mathcal{B}_2)} \ket{\minbf{\mu}} \tilde C_{\mu a} \qquad\text{$(a\in\{1,\ldots,\tilde N_{\rm vir}\})$}.
\end{align}
The number of occupied molecular orbitals remains at $N_{\rm occ}$ as in Eq.~\eqref{eq:MOcoeffoccB1}, because the number of electrons did not change.
The number of virtual orbitals $\{\otv{a}\}$ reduces to $\tilde N_{\rm vir}=\dim(\mathcal{B}_2)-N_{\rm occ}$, because, as a minimal basis set, $\mathcal{B}_2$ \emph{only} spans the core and valence orbitals of the molecule---in particular, $\mathcal{B}_2$ lacks the polarization functions and diffuse functions which's span would normally make up the largest part of the virtual one-particle space.

If combined, $\{\ket{\oto{i}}\}$ and $\{\ket{\oto{a}}\}$ obviously allow representing the entirety of $\lin(\mathcal{B}_2)$; consequently, for any of the minimal-basis functions $\ket{\minbf{\mu}}\in\mathcal{B}_2$, the expression
\begin{align}
   \ket{\minbf{\mu}} = \bigg(\sum_{i=1}^{N_{\rm occ}} \ket{\oto{i}}\bra{\oto{i}} + \sum_{a=1}^{\tilde N_{\rm vir}}  \ket{\oto{a}}\bra{\oto{a}}\bigg) \ket{\minbf{\mu}}
   \label{eq:B2RIofMinimalBasisFn}
\end{align}
describes a resolution of the identity (RI) over the approximate occupied space $\lin\{\ket{\oto{i}}\}\subset \mathcal{B}_2$ and its orthogonal complement $\lin\{\ket{\otv{a}}\}\subset \mathcal{B}_2$.
Eq.~\eqref{eq:B2RIofMinimalBasisFn} can be rephrased as
\begin{align}
   \ket{\minbf{\mu}} &= \bigl(\tOprj + \tVprj\bigr) \ket{\minbf{\mu}},
   \label{eq:B2RIofMinimalBasisFnOV}
\end{align}
where we defined the following projectors onto the occupied and virtual subspaces within $\lin(\mathcal{B}_2)$:
\begin{align}
\tOprj := \sum_{i=1}^{N_{\rm occ}} \ket{\oto{i}}\bra{\oto{i}}
\qquad \tVprj := \sum_{a=1}^{\tilde N_{\rm vir}} \ket{\otv{a}}\bra{\otv{a}}.
   \label{eq:tildeProjectors}
\end{align}
Note that if applied to vectors inside the span of $\mathcal{B}_2$, the operators $\smash{\tVprj}$ and $\smash{1-\tOprj}$ have identical effects, because the $\{\ket{\otv{a}}\}$ form a basis of the orthogonal complement of $\lin\{\ket{\oto{i}}\}$ inside $\lin(\mathcal{B}_2)$; in particular, we have
\begin{align}
   \tVprj \ket{\minbf{\mu}} = \bigl(1 - \tOprj\bigr) \ket{\minbf{\mu}}
   \label{eq:tildeVisOneMinusTildeOForB2Fn}
\end{align}
for the $\ket{\minbf{\mu}}$ of Eq.~\eqref{eq:B2RIofMinimalBasisFnOV}, which will play a role later.

In the IAO construction, the idea is to retain the beneficial aspects of having a meaningful minimal basis of atomic orbitals $\{\ket{\minbf{\mu}}\}$, but \emph{changing} its basis functions $\{\ket{\minbf{\mu}}\}$ in such a way that afterwards their occupied subspace spans \emph{exactly} the occupied space of $\ket{\Phi}$, and their virtual subspace lies \emph{exactly} inside its orthogonal complement.
This is easily achieved by an adjustment to Eq.~\eqref{eq:B2RIofMinimalBasisFn}, which as written is a RI over the approximate occupied and virtual subspaces making up $\lin({\mathcal{B}_2})$.
We first define the projectors onto the accurate occupied and virtual subspaces obtained from $\ket{\Phi}$:
\begin{align}
\hat{O} := \sum_{i=1}^{N_{\rm occ}} \ket{\oo{i}}\bra{\oo{i}}
\qquad \hat{V} := \sum_{a=1}^{N_{\rm vir}} \ket{\ov{a}}\bra{\ov{a}}.
   \label{eq:nontildeProjectors}
\end{align}
(with $\{\ket{\oo{i}}\}$ from Eq.~\eqref{eq:MOcoeffoccB1} and $\{\ket{\ov{a}}\}$ from Eq.~\eqref{eq:MOcoeffvirB1}).
Next, we use these projectors to \emph{separately} project the approximate occupied and virtual subspaces of the RI in Eq.~\eqref{eq:B2RIofMinimalBasisFnOV} onto their accurate counterparts:
\begin{align}
   \ket{\iao{\mu}} &= \bigl(\hat O \tOprj + \hat V \tVprj\bigr) \ket{\minbf{\mu}}.
   \label{eq:IaoIdeaOV}
\end{align}
Provided that resulting functions $\{\ket{\iao{\mu}}\}$ are not linearly dependent (which, in particular, implies the weaker condition that the $(N_{\rm occ},N_{\rm occ})$-shape overlap matrix with elements
\begin{align}
   \left[\mat S^{{\rm o},\tilde{\rm o}}\right]_{ij} := \left[\psh{\oo{i}}{\oto{j}}\right]_{ij},\label{eq:ApproxAccurateOccOverlap}
\end{align}
has no vanishing singular values), these $\{\ket{\iao{\mu}}\}$ will \emph{exactly} span the accurate occupied space $\lin\{\ket{\oo{i}}\}$, as explained next.
The quantities defined in Eq.~\eqref{eq:IaoIdeaOV} are the proto-IAOs we aim to construct.
[The ``proto''-prefix only indicates that the orbitals are not yet orthogonal, and may be marked with a ``$\nonorthsuperscript$'' superscript on state vectors if necessary for disambiguation; coefficient matrices of such non-orthogonal quantities will be written in lower case, as in the main text].

To see that the proto-IAOs span the occupied space (i.e., that $\lin\{\ket{\oo{i}}\}\subset \lin\{\ket{\iao{\mu}}\}$), consider the following argument:
First, as a direct consequence of the definition of the IAOs $\{\ket{\iao{\mu}}\}$ in Eq.~\eqref{eq:IaoIdeaOV} combined with the fact that the mutually orthogonal $\{\ket{\oto{i}}\}$ and $\{\ket{\otv{a}}\}$ together form a basis of $\lin(\mathcal{B}_2)$, the space 
\begin{align}
   \lin\bigl\{\ket{\iao{\mu}};\,\mu\in\{1,\ldots,\dim(\mathcal{B}_2)\}\bigr\}
\end{align}
can be exactly split into the two orthogonal subspaces
\begin{align}
   \mathbb{A}_o &:= \lin\big\{ \hat O \ket{\oto{i}};\,i\in\{1,\ldots,N_{\rm occ}\} \big\} \subset \lin(\mathcal{B}_1)
\\
   \mathbb{A}_v &:= \lin\big\{ \hat V \ket{\otv{a}};\,a\in\{1,\ldots,\tilde N_{\rm vir}\} \big\} \subset \lin(\mathcal{B}_1)
\end{align}
($\mathbb{A}_o$ and $\mathbb{A}_v$ are orthogonal to each other because both the original $\{\ket{\oto{i}}\}$ and $\{\ket{\otv{a}}\}$ are mutually orthogonal, and the occupied and virtual subspaces they are projected to by $\hat O$ and $\hat V$ are orthogonal, too; so they can be considered entirely independently of each other).
Of these, the subspace $\lin\{ \hat O \ket{\oto{i}}\}$ obviously lies entirely \emph{inside} $\lin\{\ket{\oo{i}}\}$;
additionally, the vectors $\{\hat O \ket{\oto{i}}\}$ are all linearly independent, and therefore both $\lin\{ \hat O \ket{\oto{i}}\}$ and $\lin\{\ket{\oo{i}}\}$ are vector spaces of the same dimension ($N_{\rm occ}$).
It is now elementary linear algebra to recognize that if a vector space $\mathbb{A}$ is a subspace of another vectorspace $\mathbb{B}$, and both $\mathbb{A}$ and $\mathbb{B}$ have the same dimension, then $\mathbb{A}$ and $\mathbb{B}$ are in fact identical.
Under the given premise of $\mat S^{{\rm o},\tilde{\rm o}}$ (Eq.~\eqref{eq:ApproxAccurateOccOverlap}) not being singular, this is exactly the case here.
Consequently, $\mathbb{A}_o$, which is a subspace of the span of IAOs, is identical to $\ket{\Phi}$'s occupied space $\lin\{\ket{\oo{i}}\}$.
This means in particular that, if needed, all $\ket{\oo{i}}$ could be \emph{exactly} represented as linear combinations of the $\dim(\mathcal{B}_2)$ IAOs of Eq.~\eqref{eq:IaoIdeaOV}.

So why would we expect the premise of $\mat S^{{\rm o},\tilde{\rm o}}$ being non-singular to be the applicable?
As described above, the main difference between a free-atom minimal basis $\mathcal{B}_2$ and an accurate computational basis $\mathcal{B}_1$ is the former's lack of diffuse and polarization functions.
However, while important for quantitative accuracy, the diffuse and polarization functions which $\mathcal{B}_2$ lacks are generally \emph{not needed} to represent the qualitative molecular electronic structure of core or valence states; in fact, $\mathcal{B}_2$ \emph{does} contain all functions required to model those qualitatively.
For this reason, we would expect that apart from some exceptional cases (if $\ket{\tilde\Phi}$ and $\ket{\Phi}$ describe qualitatively different states; e.g., if $\ket{\Phi}$ describes a Rydberg state or non-valence anion), the space spanned by the $\mathcal{B}_2$-basis occupied orbitals $\{\ket{\oto{i}}\}$ of Eq.~\eqref{eq:MOcoeffoccB2} has a high overlap with, and closely resembles, the occupied space spanned by the accurate large-basis molecular orbitals $\{\ket{\oo{i}}\}$ of Eq.~\eqref{eq:MOcoeffoccB1}.
Similarly, $\lin\{\ket{\otv{a}}\}$, should be entirely sufficient to qualitatively represent all anti-bonding and unoccupied non-bonding orbitals in the valence space of the molecule; the rationale for this is even stronger, because those orbitals can with good reason be \emph{defined} as a localized representation of the orthogonal complement of the occupied space ($\approx\lin\{\ket{\oto{i}}\}$) inside the valence space ($\approx\lin(\mathcal{B}_2)$).

Let us define the projectors onto the entire linear span of $\mathcal{B}_1$ and $\mathcal{B}_2$ as $\hat P_1$ and $\hat P_2$:
\begin{align}
   \hat P_1 = \sum_{\mu,\nu=1}^{\dim(\mathcal{B}_1)} \ket{\mainbf{\mu}}[\mat S_{11}^{-1}]^{\mu\nu} \bra{\mainbf{\nu}}
\notag\\
   \hat P_2 = \sum_{\mu,\nu=1}^{\dim(\mathcal{B}_2)} \ket{\minbf{\mu}}[\mat S_{22}^{-1}]^{\mu\nu} \bra{\minbf{\nu}}
\label{eq:P1P2Abstract}
\end{align}
(if the functions of $\mathcal{B}_1$ already are orthonormal, as in the main text of this work, the inverse overlap matrix is an identity matrix, and the projector can be accordingly simplified; however, for the current section we retain the general form to simplify comparison to previous formulas).
For the occupied and virtual space projectors of Eq.~\eqref{eq:nontildeProjectors}, we then find
\begin{align}
   \hat P_1 \hat O = \hat O
\qquad
   \hat P_1 \hat V = \hat V
\qquad
   \hat O + \hat V = \hat P_1
\label{eq:P1oP1vOVP1}
\end{align}
because the $\{\ket{\oo{i}}\}$ and $\{\ket{\ov{a}}\}$ are expanded over $\mathcal{B}_1$ to begin with, and together form a basis of $\lin(\mathcal{B}_1)$.
Taken together with Eq.~\eqref{eq:tildeVisOneMinusTildeOForB2Fn}, these expression
allow rewriting Eq.~\eqref{eq:IaoIdeaOV} as
\begin{align}
   \ket{\iao{\rho}} &= \left(\hat O \tOprj + \hat V \tVprj\right) \ket{\minbf{\rho}}
\notag\\
   &= \left(\hat P_1\hat O \tOprj + \hat P_1\bigl(\hat P_1-\hat O\bigr) \bigl(1-\tOprj\bigr)\right) \ket{\minbf{\rho}}
\notag\\
   &= \hat P_1\left(\hat O \tOprj + \bigl(1-\hat O\bigr) \bigl(1-\tOprj\bigr)\right) \ket{\minbf{\rho}}.
   \label{eq:OOplus1mO1mOFormalP1left}
\end{align}
This formula can be compared to Eq.~(2) of Ref.~\citenum{knizia2013intrinsic}, which in the current notation would read as
\begin{align}
   \ket{\iao{\rho}} = \left(\hat O \tOprj  + \bigl(1 - \hat O\bigr)\bigl(1 - \tOprj\bigr)\right) \hat P_1 \ket{\minbf{\rho}}.\label{eq:OOplus1mO1mOFormalP1right}
\end{align}
Eqs.~\eqref{eq:OOplus1mO1mOFormalP1left} and \eqref{eq:OOplus1mO1mOFormalP1right} subtly differ
in that Eq.~\eqref{eq:OOplus1mO1mOFormalP1left} applies $\hat P_1$
as the last step in the construction (leftmost), rather than as first step (rightmost), as Eq.~\eqref{eq:OOplus1mO1mOFormalP1right} does.
This and other subtle aspects of choices in the IAO construction are discussed in detail in Appx.~\ref{appx:DetailsAndDifferencesOfIaoVariants}.
In the current case, it is easily established (Appx.~\ref{appx:DetailsAndDifferencesOfIaoVariants}) that Eqs.~\eqref{eq:OOplus1mO1mOFormalP1left} and \eqref{eq:OOplus1mO1mOFormalP1right} are mathematically equivalent if the $\{\ket{\oto{i}}\}$ are defined as in Ref.~\citenum{knizia2013intrinsic} (Eq.~\eqref{eq:noto13} below), so this apparent difference is spurious.
Eq.~\eqref{eq:OOplus1mO1mOFormalP1left} is the more general form, and will therefore be used in the following discussion.

\subsection{Construction of the depolarized occupied orbitals $\{\ket{\oto{i}}\}$ and their subspace projector $\tOprj=\sum_i\ket{\oto{i}}\bra{\oto{i}}$}\label{eq:DepolarizationProjector}
The IAO formula Eq.~\eqref{eq:OOplus1mO1mOFormalP1left} involves the projector $\smash{\tOprj}$,
for which we require the occupied orbitals $\{\ket{\oto{i}}\}$ of Eq.~\eqref{eq:MOcoeffoccB2}.
Despite the outline in Appx.~\ref{sec:RationalizationOfTheIaoFormula},
neither the original IAO method\cite{knizia2013intrinsic} nor this work requires an actual independent computation in the minimal basis $\mathcal{B}_2$ to obtain these $\{\ket{\oto{i}}\}$.
We only introduced this conceptual possibility because it affords a cleaner outline of the core of the IAO construction in Eq.~\eqref{eq:OOplus1mO1mOFormalP1left} (which affords the spanning property), before introducing the essentially unrelated technical details of viable $\{\ket{\oto{i}}\}$ constructions used in practice.

Instead of computing the $\{\ket{\oto{i}}\}$ a priori, we obtain them from a simplification of the full-basis occupied states $\{\ket{\oo{i}}\}$, which we already have computed---in the simplest case, by just a projection onto $\lin(\mathcal{B}_2)$, followed by a re-orthogonalization.
[Apart from being much simpler than performing an SCF to obtain a single determinant wave function, this also ensures that the full-basis wave function $\ket{\Phi}$ and the minimal-basis wave function $\ket{\tilde \Phi}$ are automatically generated in a consistent fashion: if the wave function $\ket{\tilde\Phi}$ \emph{were} computed with an actual SCF procedure, it could happen that both SCF procedures arrive at qualitatively inequivalent electronic states of the molecule (e.g., in molecules with multiple viable electronic states).]

Concretely, two variants of constructing the $\ket{\oto{i}}$ are used in practice.
Both start by computing what we here denote as ``proto-depolarized occupied'' (pdo) molecular orbitals;
these are obtained by ``depolarizing'' the large-basis occupied orbitals $\{\ket{\oo{i}};\,i\in\{1,\ldots,N_{\rm occ}\}\}$.
The ``proto''-prefix again indicates that the orbitals are not yet orthogonal, and indicated with a ``$\nonorthsuperscript$'' superscript if necessary for disambiguation.
The original 2013 IAO article suggested computing the dpo as
\begin{align}
   \ket{\noto{i}} &:= \hat P_1 \hat P_2 \ket{\oo{i}}.\label{eq:noto13}
\end{align}
In a (slight) revision of the IAO method (which since 2014 was outlined and recommended in GK's reference implementation, but has never been formally published academically), the dpo are instead computed as
\begin{align}
   \ket{\noto{i}} &:= \hat P_2 \ket{\oo{i}},\label{eq:noto14}
\end{align}
without the additional $\mathcal{B}_1$-span projection $\hat P_1$.
As argued before,\cite{knizia2013intrinsic} in practice $\hat P_1$ acts \emph{almost} as an identity operator if applied to functions in $\lin(\mathcal{B}_2)$, because just about any realistic computational basis can almost perfectly represent the minimal-basis functions; however, in most practical situations $\lin(\mathcal{B}_2)\subset\lin(\mathcal{B}_1)$ is not \emph{exactly} fulfilled, so Eqs.~\eqref{eq:noto13} and \eqref{eq:noto14} are not strictly equivalent, either.
As the construction resulting from the choice of Eq.~\eqref{eq:noto13} has already been described in the original IAO article, we will here focus on Eq.~\eqref{eq:noto14}; however, we will point out details regarding the formal and numerical differences in Appx.~\ref{sec:IaoMatrixFormulation} and ~\ref{appx:DetailsAndDifferencesOfIaoVariants}.

The original molecular orbitals $\{\ket{\oo{i}}\}$ are orthonormal, but the depolarization in Eq.~\eqref{eq:noto14} strips off all the polarization contributions to the $\ket{\oo{i}}$ (because these cannot be represented in the span of the minimal basis $\mathcal{B}_2$), and therefore the $\ket{\noto{i}}$ are generally not exactly orthogonal as written.
To obtain a set of $\{\ket{\oto{i}}\}$ which can serve as replacement of Eq.~\eqref{eq:MOcoeffoccB1}, we therefore orthogonalize the $\ket{\noto{i}}$.
To this end, we first compute the $(N_{\rm occ}, N_{\rm occ})$-shape pdo-MO overlap matrix
\begin{align}
   \Snoto{i}{j} &= \psh{\noto{i}}{\noto{j}}
   = \psh{\oo{i}}{\hat P_2 \hat P_2|\oo{j}}
   = \psh{\oo{i}}{\hat P_2 |\oo{j}}.
\label{eq:Snoto}
\end{align}
In the second step, we used $\hat P_2^2=\hat P_2$ (the idempotency property shared by all projection operators).
A set of orthogonalized $\{\ket{\oto{i}}\}$ can then be obtained as
\begin{align}
   \ket{\oto{i}} = \sum_{j=1}^{N_{\rm occ}} \ket{\noto{j}} \PmhSnoto{j}{i}.
   \label{eq:tildePhiOccFromSnoto}
\end{align}
While this formally describes a symmetric orthogonalization, the IAO construction is invariant to the type of orthogonalization used, and the final formulas need not actually involve any explicit orthogonalization of the $\{\ket{\oto{i}}\}$ states at all.
In fact, Eq.~\eqref{eq:OOplus1mO1mOFormalP1left} makes clear that as far as the $\{\ket{\oto{i}}\}$ are concerned, only a representation of the projector $\tOprj$ is needed for the construction of the IAOs.
And by inserting Eqs.~\eqref{eq:Snoto} and \eqref{eq:tildePhiOccFromSnoto}, we can obtain one as follows:
{\allowdisplaybreaks
\begin{align}
   \tOprj &= \sum_{i=1}^{N_{\rm occ}}\ket{\oto{i}}\bra{\oto{i}}
\notag\\
   &= \sum_{j,k=1}^{N_{\rm occ}}\sum_{i=1}^{N_{\rm occ}} \ket{\noto{j}} \PmhSnoto{j}{i} \PmhSnoto{i}{k} \bra{\noto{k}}
\notag\\
   &= \sum_{j,k=1}^{N_{\rm occ}} \ket{\noto{j}} \biggl(\sum_{i=1}^{N_{\rm occ}}\PmhSnoto{j}{i} \PmhSnoto{i}{k}\biggr) \bra{\noto{k}}
\notag\\
   &= \sum_{j,k=1}^{N_{\rm occ}} \ket{\noto{j}} \InvSnoto{j}{k} \bra{\noto{k}}
\notag\\
   &= \sum_{j,k=1}^{N_{\rm occ}} \hat P_2\ket{\oo{j}} \InvSnoto{j}{k} \bra{\oo{k}} \hat P_2.
\label{eq:TildeOv14}
\end{align}
}

\subsection{Formal simplifications of Eq.~\eqref{eq:OOplus1mO1mOFormalP1left}}
Appx.~\ref{sec:RationalizationOfTheIaoFormula} described the emergence of Eq.~\eqref{eq:OOplus1mO1mOFormalP1left}---this is the core formula of the IAO construction.
In principle, Eq.~\eqref{eq:OOplus1mO1mOFormalP1left} can be translated into an implementable matrix formulation directly as is (and, indeed, that is exactly how the ``Standard/2013'' variant of the IAO construction was described in Appendix C of Ref.~\citenum{knizia2013intrinsic}).
However, with the slight revision of replacing Eq.~\eqref{eq:noto13} by Eq.~\eqref{eq:noto14}, Eq.~\eqref{eq:OOplus1mO1mOFormalP1left} can be formally simplified before doing so.
To this end, first note that Eq.~\eqref{eq:OOplus1mO1mOFormalP1left} can be rearranged as follows:
\begin{align}
   \ket{\iao{\rho}} &= \hat P_1\left(\hat O \tOprj  + \bigl(1 - \hat O\bigr)\bigl(1 - \tOprj\bigr)\right) \ket{\minbf{\rho}}
\notag\\
   &= \hat P_1\left(\hat O \tOprj + 1 - \hat O - \tOprj + \hat O \tOprj\right) \ket{\minbf{\rho}}
\notag\\
   &= \hat P_1\left(1 - \hat O - \tOprj + 2 \hat O \tOprj\right) \ket{\minbf{\rho}}
\notag\\
   &= \hat P_1\left(1 + \hat O - \tOprj + 2 \hat O \tOprj - 2 \hat O\right) \ket{\minbf{\rho}}
\notag\\
   &= \hat P_1\left(1 + \hat O - \tOprj + 2 \bigl(\hat O \tOprj - \hat O\bigr)\right) \ket{\minbf{\rho}}
\notag\\
   &= \hat P_1\left(1 + \hat O - \tOprj - 2 \hat O \bigl(1 - \tOprj\bigr)\right) \ket{\minbf{\rho}}.\label{eq:Fsdfkj}
\end{align}
We will next establish that the $\tOprj$ of Eq.~\eqref{eq:TildeOv14} fulfills
\begin{align}
   \hat O \bigl(1 - \tOprj\bigr)\ket{\minbf{\rho}} = 0.\label{eq:Otimes1minusTildeOisZero}
\end{align}
As a consequence, Eq.~\eqref{eq:Fsdfkj} reduces to
\begin{align}
   \ket{\iao{\rho}} &= \hat P_1\bigl(1 + \hat O - \tOprj\bigr) \ket{\minbf{\rho}}. \label{eq:1plusOminusOtFormalP1left}
\end{align}
This is a significant formal simplification, and also affords a simpler and more efficient matrix formulation than obtained by translating Eq.~\eqref{eq:OOplus1mO1mOFormalP1left} directly (see Appx.~\ref{sec:IaoMatrixFormulation}).

The relation Eq.~\eqref{eq:Otimes1minusTildeOisZero} can be established by direct calculation after inserting Eq.~\eqref{eq:nontildeProjectors} for the occupied space projector $\hat O$
and Eq.~\eqref{eq:TildeOv14} for its depolarized counterpart $\tOprj$:
\begin{align}
   &\hat O \bigl(1 - \tOprj\bigr) \ket{\minbf{\rho}}
\notag\\
   &= \sum_{i=1}^{N_{\rm occ}} \ket{\oo{i}}\bra{\oo{i}} \bigg(1 - \sum_{j,k=1}^{N_{\rm occ}} \hat P_2\ket{\oo{j}} \InvSnoto{j}{k} \bra{\oo{k}} \hat P_2\bigg)\ket{\minbf{\rho}}
\notag\\
   &= \sum_{i=1}^{N_{\rm occ}} \ket{\oo{i}}\bigg(\bra{\oo{i}} - \sum_{j,k=1}^{N_{\rm occ}} \underbrace{\psh{\oo{i}}{\hat P_2|\oo{j}}}_{=\Snoto{i}{j}} \InvSnoto{j}{k} \bra{\oo{k}} \hat P_2\bigg)\ket{\minbf{\rho}}
\notag\\
   &= \sum_{i=1}^{N_{\rm occ}} \ket{\oo{i}}\bigg(\bra{\oo{i}} - \sum_{k=1}^{N_{\rm occ}} \sum_{j=1}^{N_{\rm occ}}\Snoto{i}{j} \InvSnoto{j}{k} \bra{\oo{k}} \hat P_2\bigg)\ket{\minbf{\rho}} \label{eq:DFsjlkjw}
\end{align}
To arrive here, we inserted $\Snoto{i}{j}$ from Eq.~\eqref{eq:Snoto}.
Note this yields a contraction of $\SnotoName$ to $\SnotoName^{-1}$, which can be evaluated as
\begin{align}
\sum_{j=1}^{N_{\rm occ}}\Snoto{i}{j} \InvSnoto{j}{k} = \delta^k_i.
\end{align}
Substituting this back into Eq.~\eqref{eq:DFsjlkjw}, we find
\begin{align}
   \hat O \bigl(1 - \tOprj\bigr) \ket{\minbf{\rho}} &= \sum_{i=1}^{N_{\rm occ}} \ket{\oo{i}}\bigg(\bra{\oo{i}} - \sum_{k=1}^{N_{\rm occ}} \delta_i^k\bra{\oo{k}} \hat P_2\bigg)\ket{\minbf{\rho}}
\notag\\
   &= \sum_{i=1}^{N_{\rm occ}} \ket{\oo{i}}\bigg(\bra{\oo{i}} - \bra{\oo{i}} \hat P_2\bigg)\ket{\minbf{\rho}}
\notag\\
   &= \sum_{i=1}^{N_{\rm occ}} \ket{\oo{i}}\bigg(\psh{\oo{i}}{\minbf{\rho}} - \underbrace{\psh{\oo{i}}{\hat P_2|\minbf{\rho}}}_{=\psh{\oo{i}}{\minbf{\rho}}}\bigg) = 0
\end{align}
In the last step we used
\begin{align}
   \psh{\oo{i}}{\hat P_2|\minbf{\rho}} = \psh{\oo{i}}{\minbf{\rho}}.
\end{align}
This holds because $\ket{\minbf{\rho}}\in\mathcal{B}_2$; that is, it already lies inside the subspace $\hat P_2$ projects to, so $\hat P_2$ does not affect it.

Eq.~\eqref{eq:1plusOminusOtFormalP1left} was already put forward in Appendix C of Ref.~\citenum{knizia2013intrinsic}, as a simpler alternative formula for constructing IAOs capable of exactly spanning the occupied space.
However, with the definitions of Ref.~\citenum{knizia2013intrinsic}, this was an \emph{approximation} to Eq.~\eqref{eq:OOplus1mO1mOFormalP1right} (a very good approximation, but still an approximation), while the minor revision explained here (namely replacing  Eq.~\eqref{eq:OOplus1mO1mOFormalP1right} by \eqref{eq:OOplus1mO1mOFormalP1left} and Eq.~\eqref{eq:noto13} by \eqref{eq:noto14}),
this simplification from Eq.~\eqref{eq:OOplus1mO1mOFormalP1left} to Eq.~\eqref{eq:1plusOminusOtFormalP1left} is \emph{exact}.
A similar simplification was also discussed by Janowski;\cite{janowski2014near} however, it was derived under the formal prerequisite that $\lin(\mathcal{B}_2)$ is an exact subspace of $\lin(\mathcal{B}_1)$---a condition which in practice is often violated, and under which also the simplified formula in Appendix C of Ref.~\citenum{knizia2013intrinsic} is exact.
By confirming Eq.~\eqref{eq:Otimes1minusTildeOisZero}, we show that
with the present tweaks in the definitions of intermediate quantities, it is not necessary that $\lin(\mathcal{B}_2)\subseteq \lin(\mathcal{B}_1)$ to achieve exact equivalence of Eq.~\eqref{eq:OOplus1mO1mOFormalP1left} and Eq.~\eqref{eq:1plusOminusOtFormalP1left}.

\subsection{Matrix formulation of the IAO construction}\label{sec:IaoMatrixFormulation}
So far we described the algebraic reasoning behind the IAO construction using the abstract state vector formalism.
For the more practical minded, we here translate the equations into a directly implementable matrix formulation.
To this end, let $\mat C$ denote the $(\dim(\mathcal{B}_1),N_{\rm occ})$-shape occupied orbital coefficient matrix representing the $\{\ket{\oo{i}};\,i\in\{1,\ldots,N_{\rm occ}\}\}$ of Eq.~\eqref{eq:MOcoeffoccB1}, and let $\mat S_{11}$, $\mat S_{12}$, $\mat S_{21}$ ($=\mat S_{12}^\dagger$), and $\mat S_{22}$ denote the indicated overlap matrices between $\mathcal{B}_1$ and $\mathcal{B}_2$, with elements defined as usual, e.g.,
\begin{align}
   [\mat S_{12}]_{\mu\nu} := \psh{\mainbf{\mu}}{\minbf{\nu}}
\end{align}
for $\mu\in\{1,\ldots,\dim(\mathcal{B}_1)\}$, $\nu\in\{1,\ldots,\dim(\mathcal{B}_2)\}$.
[Note that in the the main text of this work, $\mathcal{B}_1$ is represented by a full set of orthonormal molecular orbitals, so $\mat S_{11}$ is an identity matrix; however, we retain the general form in this section].

\textsf{Revised IAO construction:}
We will first describe a matrix formulation of Eq.~\eqref{eq:1plusOminusOtFormalP1left}, which in its abstract state vector form reads
\begin{align}
   \ket{\iao{\rho}} &= \hat P_1\bigl(1 + \hat O - \tOprj\bigr) \ket{\minbf{\rho}}
   \quad\left(\text{for }\ket{\minbf{\rho}}\in\mathcal{B}_2)\right)
   \label{eq:1plusOminusOtFormalP1leftV2}
\end{align}
and provides the proto-versions (not yet orthogonalized) of the revised IAOs resulting from Eq.~\eqref{eq:OOplus1mO1mOFormalP1left} combined with Eq.~\eqref{eq:noto14}.
To express the $\ket{\iao{\rho}}$ numerically, we will represent them with the $\bigl(\dim(\mathcal{B}_1),\dim(\mathcal{B}_2)\bigr)$-shape coefficient matrix $\smash{\mat c^{\rm IAO}}$, which represents the expansion
\begin{align}
   \ket{\iao{\rho}} = \sum_{\mu=1}^{\dim(\mathcal{B}_1)} \ket{\mainbf{\mu}} c^{\rm IAO}_{\mu \rho}\qquad\text{$(\rho\in\{1,\ldots,\dim(\mathcal{B}_2)\})$}.
   \label{eq:IaoExpansionRev1}
\end{align}
To derive the concrete form of $\smash{\mat c^{\rm IAO}}$, first note that Eq.~\eqref{eq:1plusOminusOtFormalP1leftV2} can be reformulated as
\begin{align}
   \ket{\iao{\rho}} &= \hat P_1\ket{\minbf{\rho}} + \hat O \ket{\minbf{\rho}} - \hat P_1 \tOprj \ket{\minbf{\rho}}
   \label{eq:ProtoIao14AbstractSplitUp}
\end{align}
because $\hat P_1\hat O=\hat O$ (Eq.~\eqref{eq:P1oP1vOVP1}).
We now process the individual terms of Eq.~\eqref{eq:ProtoIao14AbstractSplitUp}.
Inserting Eq.~\eqref{eq:nontildeProjectors} for $\hat O$ and Eq.~\eqref{eq:MOcoeffoccB1} for its $\{\ket{\oo{i}}\}$, we find:
\begin{align}
   \hat O \ket{\minbf{\rho}} &= \biggl(\sum_{i=1}^{N_{\rm occ}} \ket{\oo{i}} \bra{\oo{i}}\biggr) \ket{\minbf{\rho}}
\notag\\
   &= \sum_{i=1}^{N_{\rm occ}}\biggl(\,\sum_{\mu=1}^{\dim(\mathcal{B}_1)} \ket{\mainbf{\mu}} C_{\mu i}\biggr) \biggl(\,\sum_{\nu=1}^{\dim(\mathcal{B}_1)}
   C_{\nu i}^* \bra{\mainbf{\nu}}\biggr) \ket{\minbf{\rho}}
\notag\\
   &= \sum_{i=1}^{N_{\rm occ}}\sum_{\mu=1}^{\dim(\mathcal{B}_1)} \ket{\mainbf{\mu}} \underbrace{C_{\mu i}}_{[\mat C]_{\mu i}} \sum_{\nu=1}^{\dim(\mathcal{B}_1)}
   \underbrace{C_{\nu i}^*}_{[\mat C^\dagger]_{i\nu}}\underbrace{\psh{\mainbf{\nu}}{\minbf{\rho}}}_{[\mat S_{12}]_{\nu\rho}}
\notag\\
      &= \sum_{\mu=1}^{\dim(\mathcal{B}_1)} \ket{\mainbf{\mu}} [\mat C \mat C^\dagger \mat S_{12}]_{\mu\rho}.
   \label{eq:ProtoIao14AbstractSplitUpTerm1}
\end{align}
Next, by inserting $\hat P_1$ from Eq.~\eqref{eq:P1P2Abstract}, we get for $\hat P_1\ket{\minbf{\rho}}$:
\begin{align}
   \hat P_1\ket{\minbf{\rho}} &= \biggl(\,\sum_{\mu,\nu=1}^{\dim(\mathcal{B}_1)} \ket{\mainbf{\mu}}[\mat S_{11}^{-1}]^{\mu\nu} \bra{\mainbf{\nu}}\biggr) \ket{\minbf{\rho}}
\notag\\
   &= \sum_{\mu=1}^{\dim(\mathcal{B}_1)} \ket{\mainbf{\mu}} \biggl(\,\sum_{\nu=1}^{\dim(\mathcal{B}_1)}[\mat S_{11}^{-1}]^{\mu\nu} \underbrace{\psh{\mainbf{\nu}}{\minbf{\rho}}}_{[\mat S_{12}]_{\nu\rho}}\biggr)
\notag\\
   &= \sum_{\mu=1}^{\dim(\mathcal{B}_1)} \ket{\mainbf{\mu}} [\mat S_{11}^{-1} \mat S_{12}]_{\mu\rho}.
   \label{eq:ProtoIao14AbstractSplitUpTerm2}
\end{align}
To process $\hat P_1 \tOprj$, we first need a matrix representation of $\mat \SnotoName$ of Eq.~\eqref{eq:Snoto}; this is easily obtained by again inserting Eq.~\eqref{eq:MOcoeffoccB1} for the $\{\ket{\oo{i}}\}$ and Eq.~\eqref{eq:P1P2Abstract} for $\hat P_2$:
\begin{align}
   &\Snoto{i}{j} = \psh{\oo{i}}{\hat P_2 |\oo{j}}
\notag\\
   &= \bra{\oo{i}} \biggl(\,\sum_{\mu,\nu=1}^{\dim(\mathcal{B}_2)} \ket{\minbf{\mu}}[\mat S_{22}^{-1}]^{\mu\nu} \bra{\minbf{\nu}}\biggr) \ket{\oo{j}} 
\notag\\
   &= \sum_{\lambda=1}^{\dim(\mathcal{B}_1)} C_{\lambda i}^* \bra{\mainbf{\lambda}} \sum_{\mu,\nu=1}^{\dim(\mathcal{B}_2)} \ket{\minbf{\mu}}[\mat S_{22}^{-1}]^{\mu\nu} \bra{\minbf{\nu}} \sum_{\kappa=1}^{\dim(\mathcal{B}_1)} \ket{\mainbf{\kappa}} C_{\kappa j} 
\notag\\
   &= \sum_{\lambda,\kappa=1}^{\dim(\mathcal{B}_1)}\sum_{\mu,\nu=1}^{\dim(\mathcal{B}_2)} \underbrace{C_{\lambda i}^*}_{[\mat C^\dagger]_{i\lambda}} \underbrace{\psh{\mainbf{\lambda}}{\minbf{\mu}}}_{[\mat S_{12}]_{\lambda\mu}} [\mat S_{22}^{-1}]^{\mu\nu} \underbrace{\psh{\minbf{\nu}}{\mainbf{\kappa}}}_{[\mat S_{21}]_{\nu\kappa}} C_{\kappa j} 
\notag\\
   &= \bigl[\mat C^\dagger \mat S_{12} \mat S_{22}^{-1} \mat S_{21} \mat C\bigr]_{ij}.
\end{align}
That is, we get
\begin{align}
   \mat \SnotoName = \mat C^\dagger \mat S_{12} \mat S_{22}^{-1} \mat S_{21} \mat C.
   \label{eq:SnotoMatrix}
\end{align}
Using this, we can now insert $\tOprj$ from Eq.~\eqref{eq:TildeOv14} to 
obtain the final term of Eq.~\eqref{eq:ProtoIao14AbstractSplitUp}:
\begin{align}
   &\hat P_1 \tOprj \ket{\minbf{\rho}}
   = \hat P_1 \sum_{j,k=1}^{N_{\rm occ}} \hat P_2\ket{\oo{j}} \InvSnoto{j}{k} \bra{\oo{k}} \hat P_2 \ket{\minbf{\rho}}
\notag\\
   &= \sum_{j,k=1}^{N_{\rm occ}} \sum_{\mu,\nu=1}^{\dim(\mathcal{B}_1)} \ket{\mainbf{\mu}}[\mat S_{11}^{-1}]^{\mu\nu} \underbrace{\bra{\mainbf{\nu}} \hat P_2\ket{\oo{j}}}_{[\mat S_{12}\mat S_{22}^{-1} \mat S_{21}\mat C]_{\nu j}} \InvSnoto{j}{k} \underbrace{\bra{\oo{k}} \hat P_2 \ket{\minbf{\rho}}}_{[\mat C^\dagger \mat S_{12}]_{k\rho}}
\notag\\
   &= \sum_{\mu=1}^{\dim(\mathcal{B}_1)} \ket{\mainbf{\mu}} [\mat S_{11}^{-1} \mat S_{12}\mat S_{22}^{-1} \mat S_{21}\mat C \,\mat \SnotoName^{-1} \mat C^\dagger \mat S_{12}]_{\mu\rho}
   \label{eq:ProtoIao14AbstractSplitUpTerm3}
\end{align}
In this, we used $\hat P_2 \ket{\minbf{\rho}}=\ket{\minbf{\rho}}$ to simplify $\psh{\oo{k} }{\hat P_2|\minbf{\rho}}$.

Collecting the terms from Eqs.~\eqref{eq:ProtoIao14AbstractSplitUpTerm3}, \eqref{eq:ProtoIao14AbstractSplitUpTerm2}, and \eqref{eq:ProtoIao14AbstractSplitUpTerm1} for Eq.~\eqref{eq:ProtoIao14AbstractSplitUp}, and comparing to the formal expansion Eq.~\eqref{eq:IaoExpansionRev1}, we find as expression for the proto-IAO expansion matrix
\begin{align}
   \mat c^{\rm IAO} = \mat S_{11}^{-1} \mat S_{12} + \mat C \mat C^\dagger \mat S_{12} - \mat S_{11}^{-1} \mat S_{12}\mat S_{22}^{-1} \mat S_{21}\mat C \,\mat \SnotoName^{-1} \mat C^\dagger \mat S_{12}.
   \label{eq:ProtoIao14MatrixForm1}
\end{align}
This expression can be directly computed from the occupied orbital matrix $\mat C$ and the indicated overlap matrices.
The formula can be evaluated more efficiently when introducing the intermediate matrices
\begin{align}
   \label{eq:ProtoIao14MatrixForm2IntStart}
   \mat P_{12} := \mat S_{11}^{-1} \mat S_{12}
\qquad
   \mat t_1 := \mat S_{21} \mat C
\qquad
   \mat t_2 := \mat S_{22}^{-1} \mat t_1
\end{align}
In terms of these, we may first rewrite Eq.~\eqref{eq:ProtoIao14MatrixForm1} into
\begin{align}
   \mat \SnotoName &= \mat t_1^\dagger \mat t_2,
\end{align}
and then introduce the intermediate
\begin{align}
   \mat t_3 &= \mat t_2 \mat\SnotoName^{-1} = \left(\bigl(\mat\SnotoName^\dagger\bigr)^{-1}\, \mat t_2^\dagger  \right)^\dagger
   \label{eq:ProtoIao14MatrixForm2IntEnd}
\end{align}
[We formulated $\mat t_3$ in terms of adjoints because linear algebra solvers are normally solve $\mat A \mat X = \mat B$ ($\Leftrightarrow$ $\mat X = \mat A^{-1}\mat B$) rather than $\mat X \mat A = \mat B$ ($\Leftrightarrow$ $\mat X = \mat B \mat A^{-1}$) we would otherwise need].
Using these, we may then rewrite Eq.~\eqref{eq:ProtoIao14MatrixForm1} as
\begin{align}
   \mat c^{\rm IAO} &= \mat S_{11}^{-1} \mat S_{12} + \mat C \mat t_1^\dagger - \mat S_{11}^{-1} \mat S_{12} \mat t_2 \,\mat \SnotoName^{-1} \mat t_1^\dagger,
\notag\\
   &= \mat P_{12} + \mat C \mat t_1^\dagger - \mat P_{12} \mat t_2 \,\mat \SnotoName^{-1} \mat t_1^\dagger
\notag\\
   &= \mat P_{12} + \left(\mat C - \mat P_{12} \mat t_2 \,\mat \SnotoName^{-1}\right) \mat t_1^\dagger
\notag\\
   &= \mat P_{12} + \left(\mat C - \mat P_{12} \mat t_3\right) \mat t_1^\dagger
   \label{eq:ProtoIao14MatrixForm2}
\end{align}
The intermediates Eqs.~\eqref{eq:ProtoIao14MatrixForm2IntStart} to \eqref{eq:ProtoIao14MatrixForm2IntEnd} combined with the last line of Eq.~\eqref{eq:ProtoIao14MatrixForm2} provide an efficient formulation of the construction of $\mat c^{\rm IAO}$ in terms of binary matrix operations (see Fig.~\ref{fig:IaoRevAlgo}).
It may be noted that in this factorization, the only operation in the entire construction which scales as $\mathcal{O}(\dim(\mathcal{B}_1)^3)$ is calculating the Cholesky decomposition of $\mat S_{11}$, which is needed to solve for $\mat P_{12}$ in Eq.~\eqref{eq:ProtoIao14MatrixForm2IntEnd}; however, this decomposition of $\mat S_{11}$ (which is the computational main basis overlap matrix), would very most likely be already present in the host program because at least one of those is typically computed as part of the SCF process (or alternatively a spectral decomposition of $\mat S_{11}$ or a $\smash{\mat S_{11}^{-1/2}}$ matrix, either of which would also work for constructing $\mat P_{12}$).
Note that the three formal matrix inverses in Eq.~\eqref{eq:ProtoIao14MatrixForm2IntStart} and \eqref{eq:ProtoIao14MatrixForm2IntEnd} should never be calculated as actual inverses, but rather expressions like ``$\mat S_{22}^{-1} \mat t_1$'' should be read as ``solve the equation system $\mat S_{22} \mat x = \mat t_1$ for $\mat x$ using a suitable matrix decomposition of $\mat S_{22}$''---in the current case, all formally inverted matrices are symmetric and positive definite, so the systems are most efficiently solved using the (extremely efficient) Cholesky decomposition of the overlap matrices combined with two triangular matrix solves per equation (dpotrf/dtrsm in LAPACK/BLAS terminology).

\begin{figure}
   \centering
   \rule{\columnwidth}{1pt}
   \begin{align*}
   \mat P_{12} &:= \operatorname{solve}\left(\mat S_{11},\, \mat S_{12}\right)
\notag\\
   \mat t_1 &:= \mat S_{12}^\dagger \mat C
\notag\\
   \mat t_2 &:= \operatorname{solve}\left(\mat S_{22},\, \mat t_1\right)
\notag\\
   \mat \SnotoName &:= \mat t_1^\dagger \mat t_2
\notag\\
   \mat t_3 &:= \operatorname{solve}\bigl(\mat\SnotoName^\dagger,\, \mat t_2^\dagger\bigr)^\dagger
\notag\\
   \mat c^{\rm IAO} &:= \mat P_{12} + \left(\mat C - \mat P_{12} \mat t_3\right) \mat t_1^\dagger
   \end{align*}
   \caption{Recommended final numerical algorithm for revised construction of the IAOs, as provided in Eqs.~\eqref{eq:ProtoIao14MatrixForm2IntStart} to \eqref{eq:ProtoIao14MatrixForm2}. 
   This construction is based on Eqs.~\eqref{eq:OOplus1mO1mOFormalP1left} and \eqref{eq:noto14}; see Appx.~\ref{appx:DetailsAndDifferencesOfIaoVariants} for a discussion of the differences to Ref.~\citenum{knizia2013intrinsic}.
   Algorithm inputs are: the $\mathcal{B}_1$ to $\mathcal{B}_2$ overlap matrices $\mat S_{11}$, $\mat S_{12}$, and $\mat S_{22}$, and the occupied molecular orbital cofficient matrix $\mat C$ of Eq.~\eqref{eq:MOcoeffoccB1} (with shape $(\dim(\mathcal{B}_1), N_{\rm occ})$); output is the $(\dim(\mathcal{B}_1),\dim(\mathcal{B}_2))$-shape proto-IAO coefficient matrix $\mat c^{\rm IAO}$ of Eq.~\eqref{eq:IaoExpansionRev1}.
   The linear solves are most efficiently computed with Cholesky decompositions (dpotrf/dtrsm).
   }
   \label{fig:IaoRevAlgo}
   \rule{\columnwidth}{1pt}
\end{figure}

\textsf{Original 2013 IAO construction:}
To compare, we also provide the matrix formulation of the original 2013 IAO construction.
As mentioned, this is obtained from either Eq.~\eqref{eq:OOplus1mO1mOFormalP1left} or Eq.~\eqref{eq:OOplus1mO1mOFormalP1right} when defining the proto-depolarized occupied orbitals $\{\ket{\oto{i}}\}$ via Eq.~\eqref{eq:noto13} (i.e., as $\ket{\noto{i}} := \hat P_1 \hat P_2 \ket{\oo{i}}$) instead of Eq.~\eqref{eq:noto14} (which omits the final $\hat P_1$ projection).

When we again denote the occupied orbital matrix as $\mat C$ and the internal and crossed $\mathcal{B}_1$/$\mathcal{B}_2$ overlap matrices as $\mat S_{11}$, $\mat S_{12}$, $\mat S_{21}\;(=\mat S_{12}^\dagger)$, and $\mat S_{22}$, then a direct translation of Eq.~\eqref{eq:OOplus1mO1mOFormalP1left} for the proto-IAOs into its matrix form yields
\begin{align}
   \mat P_{12} &= \mat S_{11}^{-1} \mat S_{12}
\\ \tilde {\mat C} &= \mathrm{orth}(\mat S_{11}^{-1} \mat S_{12} \mat S_{22}^{-1} \mat S_{21} \mat C,\,\mat S_{11})
\\ \mat c^{\rm IAO} &= \mat C \mat C^T \mat S_{11} \tilde{\mat C} \tilde{\mat C}^T\mat S_{11} \mat P_{12} +
\notag\\&\qquad(\mat 1-\mat C \mat C^T \mat S_1)(\mat 1-\tilde{\mat C} \tilde{\mat C}^T \mat S_{11}) \mat P_{12}
\label{eq:Iao2013NumericalFormulation}
\end{align}
In this, orthonormalization is defined as
\begin{align}
   \mathrm{orth}(\mat C,\,\mat S) := \mat C [\mat C^T \mat S\mat C]^{-1/2},
\end{align}
where $\mat X^{-1/2}$ denotes the matrix inverse square root (any other orthogonalization would also work and produce identical results).
These are equivalent to the formulas in Appendix C of Ref.~\citenum{knizia2013intrinsic}.
As in the last subsection, the proto-IAOs described by $\mat c^{\rm IAO}$ are not yet orthogonal, and still need to be orthogonalized if orthogonal IAOs are desired.
Eq.~\eqref{eq:Iao2013NumericalFormulation} can still be factorized; however, both the formal complexity and the complexity of the numerical terms will be higher than in Fig.~\ref{fig:IaoRevAlgo}, because as mentioned, the transition from Eq.~\eqref{eq:OOplus1mO1mOFormalP1left} to Eq.~\eqref{eq:1plusOminusOtFormalP1left} is not exact in this case.

It should be mentioned, however, that either formulation of the IAOs will be trivial in terms of computational cost when compared to a full SCF calculation---even with the most efficient of non-hybrid DFT programs.
Additionally, both versions produce numerically near indistinguishable results in all cases we have investigated.
The preference for the revised IAO version is therefore mainly from a formal nature, but we expect that it is unlikely to play a major role except for special applications like analytic gradients, where simpler formulas are much preferable.

\subsection{Discussion of possible choices in the formal IAO construction}
\label{appx:DetailsAndDifferencesOfIaoVariants}
      There are multiple different sensible ways of defining the polarization-free occupied orbitals $\{\ket{\oto{i}}\}$ introduced in Eq.~\eqref{eq:MOcoeffoccB2}, which in turn determine the projector $\tOprj$ of Eq.~\eqref{eq:tildeProjectors}.
      In the original article\cite{knizia2013intrinsic}, Eq.~\eqref{eq:noto13} was chosen, which is
      \begin{align}
         \ket{\noto{i}} &:= \hat P_1 \hat P_2 \ket{\oo{i}}.\label{eq:noto13x}
      \end{align}
      In contrast, here we presented derivations using Eq.~\eqref{eq:noto14},
      \begin{align}
         \ket{\noto{i}} &:= \hat P_2 \ket{\oo{i}},\label{eq:noto14x}
      \end{align}
      as a more elegant alternative (which actually has been discussed in the IAO/IBO reference implementation (ibo-ref.py) of the IAO method since 2014, but has not been published in an academic context.
      The revision was prompted by an article of Janowski\cite{janowski2014near}, who employed Eq.~\eqref{eq:noto14x} under the premise that $\mathcal{B}_2$ is exactly spanned by $\mathcal{B}_1$).
      
      The differences between Eq.~\eqref{eq:noto13x} and Eq.~\eqref{eq:noto14x} are very subtle, because they involve differences in handling the degree to which the minimal basis $\mathcal{B}_2$ cannot be expressed in terms of the full basis $\mathcal{B}_1$.
      For the same reason, they have almost no bearing on the numerical results obtained:
      as far as we are aware of, differences between results produced by both variants are negligible for all practical purposes.
      Additionally, while the amount of numerical work differs for the final formulas of both choices (Appx.~\ref{sec:IaoMatrixFormulation}), evaluating either of them is \emph{so} cheap compared to even the fastest non-hybrid DFT or other SCF methods, that it is hard to imagine a usage case where the gain in computational efficiency afforded by Eq.~\eqref{eq:noto14x} could actually matter.
      
      Nevertheless, these choices do imply some formal differences, which can easily confuse.
      Additionally, in some special applications (e.g., analytical gradient formulas) the minor formal differences might have noticeable consequences on program complexity.
      We therefore point them out and discuss them inside this section, but mostly isolate it from the rest of the text to reduce potential for distraction by details of ultimately very little importance.

      One such subtle difference appears when comparing Eq.~\eqref{eq:OOplus1mO1mOFormalP1left} to Eq.~(2) of Ref.~\citenum{knizia2013intrinsic}, which as already mentioned, would read as Eq.~\eqref{eq:OOplus1mO1mOFormalP1right} if translated into the current notation.
      While formally different, for the concrete definition of the depolarized orbitals $\ket{\oto{i}}$ used in Ref.~\citenum{knizia2013intrinsic} (given in Eq.~\eqref{eq:noto13x}), not only the $\ket{\oo{i}}$ lie inside the span of $\mathcal{B}_1$ (as here), but the $\ket{\oto{i}}$ do so, too.
      As a consequence, with that definition \emph{both} $\hat O$ and $\tOprj$ are projectors into subspaces of $\lin(\mathcal{B}_1)$; 
      therefore, we have not only $\hat P_1\hat O = \hat O\hat P_1 = \hat O$,
      but also
      \[ \hat P_1\tOprj = \tOprj\hat P_1 = \tOprj \qquad\text{(with $\{\ket{\oto{i}}\}$ from Eq.~\eqref{eq:noto13x})}\]
      (that is, the operator $\hat P_1$ acts as identity within the subspaces spanned by the $\{\ket{\oo{i}}\}$ and $\{\ket{\oto{i}}\}$, because both of them already lie completely inside $\lin(\mathcal{B}_1)$ from the outset).
      If we expand \eqref{eq:OOplus1mO1mOFormalP1right}, we can therefore see that if Eq.~\eqref{eq:noto13} is used to construct the $\{\ket{\oto{i}}\}$, the only term actually affected by $\hat P_1$ is the $1 1 \hat P_1\ket{\minbf{\rho}}$ term originating from $(1-\ldots)(1-\ldots)\hat P_1\ket{\minbf{\rho}}$ in the virtual part---and for this term it does not matter whether $\hat P_1$ stands to the left or right, because the identity operator $1$ commutes with everything.
      In summary, this means that for the original 2013 definition of the $\ket{\oto{i}}$ (Eq.~\eqref{eq:noto13}), the choice of either Eq.~\eqref{eq:OOplus1mO1mOFormalP1left} or Eq.~\eqref{eq:OOplus1mO1mOFormalP1right} does not matter, because both formulas provide mathematically \emph{exactly} identical results.
      This is no longer the case with the $\ket{\oto{i}}$ resulting from Eq.~\eqref{eq:noto14}, though, because with these orbitals $\tOprj$ does not necessarily project into an exact subspace of $\mathcal{B}_1$ (unless $\mathcal{B}_1$ happens to span $\mathcal{B}_2$ \emph{exactly}), and therefore no longer commutes with $\hat P_1$.
      
      As also mentioned, with the definition Eq.~\eqref{eq:noto13x}, the transition from Eq.~\eqref{eq:OOplus1mO1mOFormalP1left} to Eq.~\eqref{eq:1plusOminusOtFormalP1left} is not exact anymore, and as outlined in Appx.~\ref{sec:IaoMatrixFormulation}, the numerical complexity is higher.
      
      So why was Eq.~\eqref{eq:noto13x} originally chosen in preference to using Eq.~\eqref{eq:noto14x} directly?
      After all, the latter not only turns out to be formally and numerically favorable after a closer look, but actually also is in closer alignment with the spirit of the IAO construction described in Appx.~\ref{sec:RationalizationOfTheIaoFormula}.
      The reason for this was, unfortunately, not particularly good:
      the method appeared simpler to implement (particularly in the concrete Fortran framework used) if more quantities were expressed in terms of the regular computational basis $\mathcal{B}_1$; and combined with the fact that $\hat P_1$ would in practice be an almost-identity operator in any case, with little effect regardless of where it is placed, GK made the unfortunate decision to just depolarize the occupied orbitals via Eq.~\eqref{eq:noto13x} (because it obviously accomplishes the removal of polarization constributions from the occupied orbitals and it produces an orbital matrix $\mat {\tilde C}$ of compatible format to $\mat C$), but without thinking particularly deeply about it---and then never revisited the matter before Ref.~\citenum{knizia2013intrinsic} was published because it seemed inconsequential.
      In retrospect, it would have been preferable to either use Eq.~\eqref{eq:noto14x} directly; or, alternatively, if all quantities \emph{should} be expressed in terms of $\mathcal{B}_1$, to just project the reference free-atom orbitals onto $\mathcal{B}_1$ first (making them an exact subspace), rather than using either \eqref{eq:noto13x} or \eqref{eq:noto14x} directly.
      That would also result in Eq.~\eqref{eq:1plusOminusOtFormalP1left} being exact, and could be obtained from the present derivation by simply not treating the $\mathcal{B}_2$ functions as raw basis functions, but instead considering them as basis expansions as already done with the $\mathcal{B}_1$ and $\mathcal{B}_2$ functions in the main text.

%

\end{document}